\documentclass[lettersize,journal]{IEEEtran}
\usepackage{marginnote} 
\usepackage{amsmath,amsfonts}
\makeatletter
\newcommand{\removelatexerror}{\let\@latex@error\@gobble}
\makeatother
\usepackage{algorithm}
\usepackage{array}
\usepackage[caption=false,font=normalsize,labelfont=sf,textfont=sf]{subfig}
\usepackage{textcomp}
\usepackage{stfloats}
\usepackage{url}
\usepackage{verbatim}
\usepackage{graphicx}
\usepackage{cite}
\usepackage{tabularx}
\usepackage{multirow}
\usepackage{graphicx, wrapfig}
\usepackage{booktabs}
\usepackage{amsmath} 
\usepackage{algorithmicx} 
\usepackage{algpseudocode}
\usepackage{stfloats}
\usepackage{xcolor}
\usepackage{makecell}
\usepackage{adjustbox}
\usepackage{tikz}
\usepackage{graphicx} 
\usepackage{longtable}
\usepackage[numbers,square]{natbib}
\usepackage{xcolor}
\usepackage[colorlinks=true, urlcolor=blue, linkcolor=blue, citecolor=blue, allcolors=blue]{hyperref}
\usepackage{etoolbox} 
\makeatletter
\def\NAT@citex[#1][#2]#3{%
	\ifNAT@swa\else\if*#2*\else\NAT@cmt#2\fi
	\fi
	\if\relax#1\relax\else\NAT@open\fi
	\hyper@natlinkstart{\@citeb\@extra@b@citeb}#3\hyper@natlinkend
	\if\relax#1\relax\else\NAT@close\fi
}
\renewcommand\NAT@open{\textcolor{blue}{[}}
\renewcommand\NAT@close{\textcolor{blue}{]}}
\def\eqref#1{%
	\textup{%
		\hyperref[#1]{
			\textcolor{blue}{(}%
			\ref*{#1}%
			\textcolor{blue}{)}%
		}%
	}%
}
\makeatother
\AtBeginEnvironment{thebibliography}{\hypersetup{urlcolor=black}}
\newcommand{\circled}[1]{%
	\tikz[baseline=(X.base)]{%
		\node[inner sep=0pt, circle, draw,  text=black, minimum size=0.9em] (X) {\normalfont\scriptsize\bfseries #1};%
	}%
}

\begin{document}
\title{Metronome: Efficient Scheduling for Periodic Traffic Jobs with Network and Priority Awareness}
\author{Hao Jiang,~\IEEEmembership{Student Member,~IEEE,} Meng Qin,~\IEEEmembership{Member,~IEEE,} \\Ruijie Kuai,~\IEEEmembership{Student Member,~IEEE,} Dandan Liang,~\IEEEmembership{Member,~IEEE,} Yue Gao,~\IEEEmembership{Fellow,~IEEE}
	% <-this % stops a space
\thanks{This work was supported in part by the Major Key Project of Pengcheng Laboratory (PCL) under Grant PCL2024A03-1, in part by the National Science Foundation
	of China under Grant 62201310, and in part by the National Science Foundation of China under Grant 62201304. \textit{(Corresponding author: Dandan Liang.)}}
\thanks{Hao Jiang, Ruijie Kuai are with the School of Electronic Information and Electrical Engineering, Shanghai Jiao Tong University, Shanghai 200240 and also with PCL, Shenzhen 518055, China (e-mail: haojiang@sjtu.edu.cn; kuairuijie@sjtu.edu.cn).}
\thanks{Meng Qin and Dandan Liang are with the Department of Strategic and Advanced Interdisciplinary Research, PCL, Shenzhen 518055, China (e-mail: qinm01@pcl.ac.cn; liangdd@pcl.ac.cn).}
\thanks{Yue Gao is with the Department of Strategic and Advanced Interdisciplinary Research, PCL, Shenzhen 518055 and also with the Institute of Space Internet, Fudan University, Shanghai 200433, China (e-mail: gao.yue@fudan.edu.cn).}
}


\maketitle

\begin{abstract}
With the rapid growth in computing power demand, cloud native networks have emerged as a promising solution to address the challenges of efficient resource coordination, particularly in coping with the dynamic fluctuations of network bandwidth in  clusters.
We propose Metronome, a network-aware and priority-aware scheduling mechanism for cloud native networks. 
This mechanism is designed to support jobs that exhibit periodic traffic patterns and dynamic bandwidth demands, particularly in the context of distributed training.
Specifically, Metronome employs a time-division multiplexing approach  that leverages job traffic characteristics to construct an elastic network resource allocation model, enabling efficient bandwidth sharing across multiple jobs.
In addition, it incorporates a multi-objective optimization strategy, jointly considering latency and job priorities to achieve globally optimal as well as dynamic resource allocation.
Finally, Metronome adapts to the dynamic environment by monitoring the cluster and performing reconfiguration operations.
Extensive experiments with 13 common machine learning  models demonstrate that Metronome can enhance cluster resource utilization while guaranteeing service performance. 
Compared with the existing Kubernetes scheduling mechanisms across multiple scenarios, Metronome reduces job completion time by up to 19.50\% while improving average bandwidth utilization by up to 23.20\%.

\end{abstract}

\begin{IEEEkeywords}
Job scheduling, periodic traffic patterns,  cloud native networks, network-aware.
\end{IEEEkeywords}

\section{Introduction}
\label{1}
\IEEEPARstart{T}{he} convergence of Artificial Intelligence as a dominant cloud workload \cite{CNCF}, \cite{impossiblecloud2024cloud}, and the pervasive adoption of the cloud native paradigm are poised to make computing power a ubiquitous utility \cite{sui2024large}.
In this evolving technological ecosystem, cloud native networks have emerged as critical infrastructure for resource coordination, in which the efficient network resource allocation plays a pivotal role in guaranteeing computational service performance.
Particularly for communication-intensive applications such as distributed training and federated learning jobs \cite{gad2024communication}, \cite{chen2025mobility}, mitigating resource allocation mismatches between computational and network capacities through efficient scheduling remains a critical challenge in cloud native networks \cite{yu2019cybertwin}, \cite{yu2023cybertwin}.

\begin{figure}[H]
	\centering
	\captionsetup[subfloat]{
		font=small,
		labelfont=bf,
		textfont=normalfont,
		justification=centering
	}
	\subfloat[Agnostic-based   scheduling]{\label{1-sc1}\includegraphics[height=0.14\textwidth]{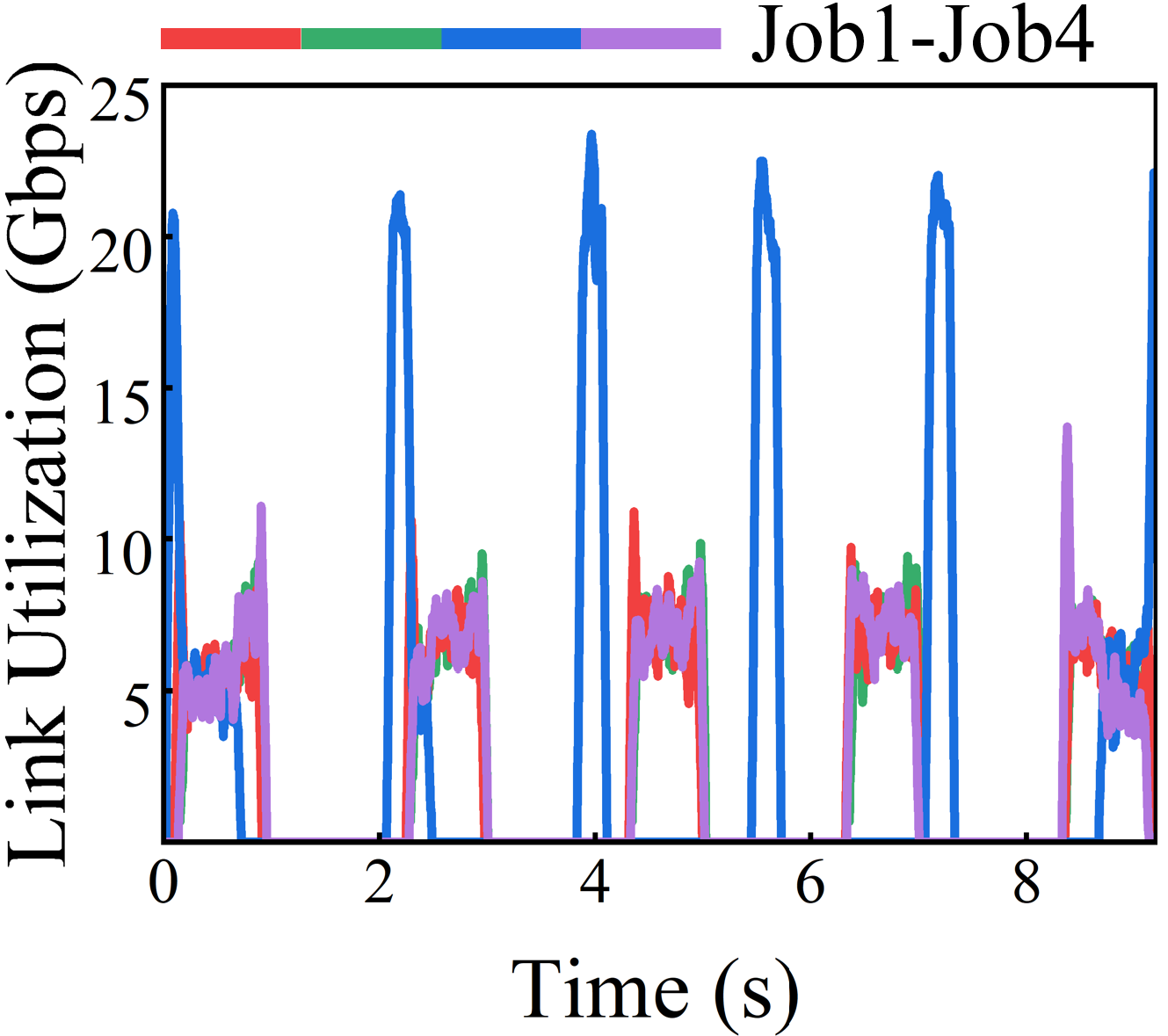}}
	\hspace{0em} 
	\subfloat[ \hspace{-0.7em} Exclusive-based  scheduling]{\label{1-sc2}\includegraphics[height=0.14\textwidth]{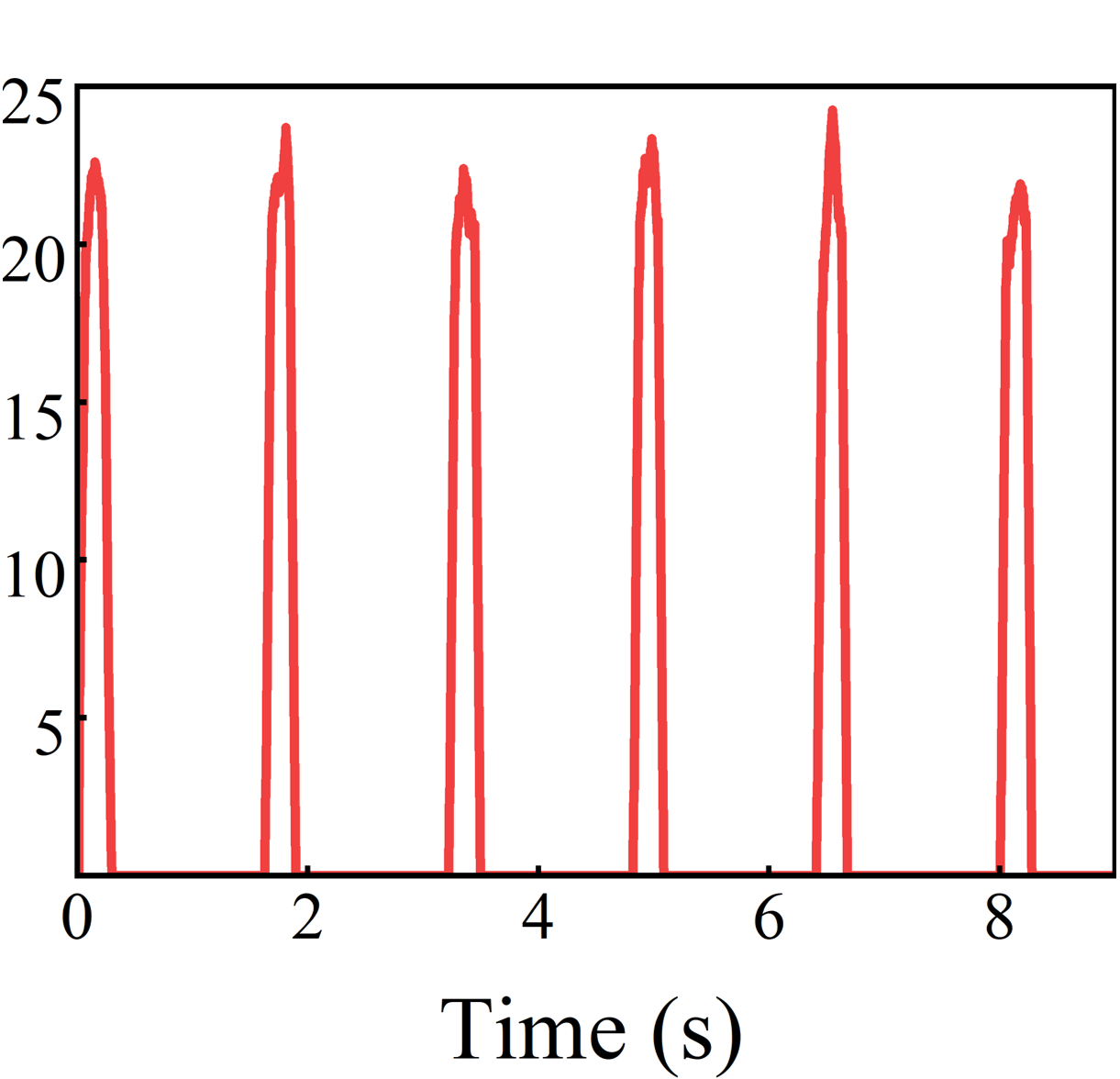}}
	\hspace{0em}  
	\subfloat[TDM-based  scheduling]{\label{1-sc3}\includegraphics[height=0.14\textwidth]{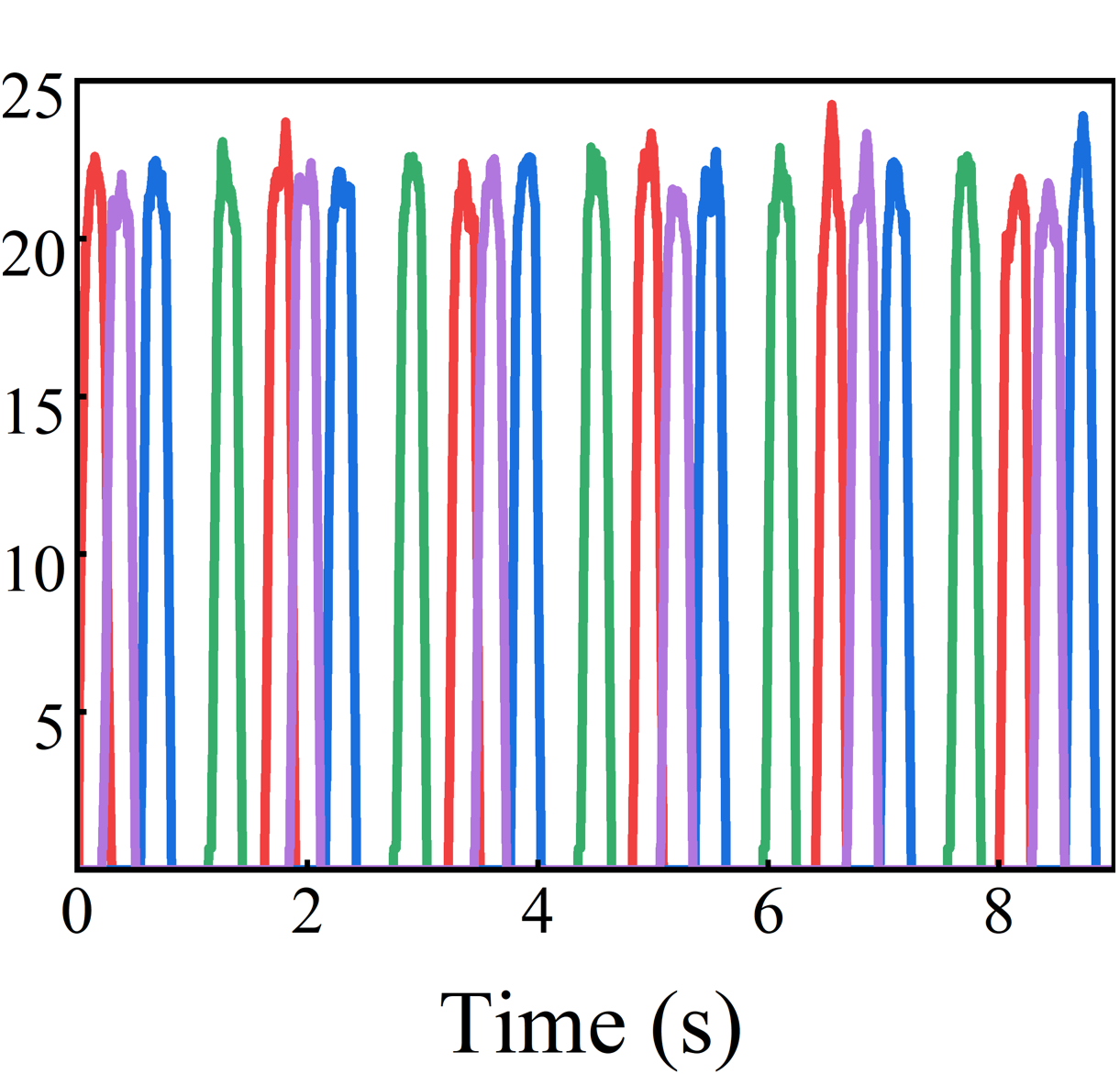}}	
	\caption{Comparison of link utilization across different approaches.}
	\label{1-sc}
\end{figure}

This challenge is particularly pronounced in multi-tenant GPU clusters, where  resources are shared among distinct tenants \cite{straggler}. Consequently, the traffic patterns of each job are affected by cluster itself and other jobs running concurrently \cite{cao2024crux}. However, by treating jobs as black boxes, existing schedulers fail to adapt to the distinct traffic patterns of periodic jobs, particularly in distributed training scenarios.
In this scenario, each iteration consists of a local computation phase (e.g., forward pass) and a synchronization phase that generates a large amount of traffic between workers.  Furthermore, the interdependent nature of these computation and communication phases gives rise to a well-defined, periodic ``on-off" communication pattern  \cite{Cassini}, \cite{SkeletonHunter}, as shown in Fig. \ref{1-sc}.

\vspace{-2ex}

A concrete example of this mismatch is shown in Fig. \ref{1-sc1}, where four distributed training jobs  share a link without bandwidth-aware scheduling \cite{kubernetes_scheduler}, \cite{themis2}, leading to competition for bandwidth resources.
The resulting communication contention causes delayed flows, which stall the subsequent  computations, thereby slowing down the entire training process and reducing link utilization.
Data from Alibaba indicates that communication contention impacts up to 36.3\% of jobs in multi-tenant environments \cite{cao2024crux}. 
Currently, network-aware schedulers guarantee training performance by employing exclusive allocation of bandwidth resources  (Fig. \ref{1-sc2}) \cite{lai2023delay}, \cite{toka2021ultra}. 
This model strictly constrains the allocatable bandwidth on a link, which means the allocatable amount is derived solely by subtracting the bandwidth consumed by existing jobs from the link's predefined total capacity. However, these existing approaches lead to job rejection when bandwidth is insufficient.
In addition, network resource utilization decreases substantially as links are idle during the computation phases of exclusively scheduled jobs.
Hence, neither the agnostic-based nor the exclusive-based approach can simultaneously optimize resource utilization and service performance.

This paper proposes Metronome, a network-aware and priority-aware scheduling mechanism.  
It leverages the distinct traffic patterns of different jobs to jointly optimize bandwidth resource allocation across two dimensions: quantity and time evolution.
Based on traffic pattern awareness, Metronome comprises a customized scheduler for the periodic jobs.
The scheduler jointly considers latency and priority indicators to enable temporal bandwidth allocation by employing time division multiplexing (TDM), while also achieving effective spatial resource utilization through dynamic allocation.
Metronome concurrently accepts four jobs while achieving performance comparable to that of exclusive scheduling scenarios, as shown in Fig. \ref{1-sc3}.
We conceptualize this approach as two-dimensional bandwidth scheduling, which gives rise to what we call two-dimensional bandwidth resources.

Conclusively, the main contributions in this paper are summarized as the following:

$\bullet$ \textbf{Scheduler mechanism based on TDM. }We propose a three-stage optimization approach to  enhance the cluster's link utilization while maintaining service performance guarantees. First, Metronome interleaves the communication phases of different jobs to maximize resource sharing. Second, Metronome proactively avoids congested links during scheduling while ensuring compact deployment of interdependent components. Finally, Metronome reserves cushion slots to mitigate subsequent contention. 

$\bullet$ \textbf{Priority-based  monitoring mechanism.} 
Maintenance of interleaving requires all jobs to confine their communications within their assigned intervals. To this end, Metronome monitors the status of deployed jobs to facilitate dynamic adaptation.
When communication contention occurs due to drift, readjustment operations are automatically triggered. 
Given that phase shifting is inherently a relative adjustment, Metronome resolves conflicts by having low priority jobs continuously adapt their communication phases, thereby allowing high priority jobs to maintain uninterrupted execution.
This approach ensures that high priority jobs experience performance comparable to that achieved with exclusive network resource allocation.

$\bullet$ \textbf{Implementation based on cloud native platform.}
We integrate Metronome into Kubernetes (K8s) \cite{burns2016borg} to enable its flexible deployment across cloud native environments.
To the best of our knowledge, this is the first implementation of TDM-based bandwidth resource allocation on a cloud native platform.
Given the computational complexity of the scheduling algorithm, we propose an innovative offline recomputation mechanism where the scheduler generates feasible solutions while delegating optimal scheme computation to a subsequent controller for offline processing.

$\bullet$ \textbf{Mechanism evaluation based on distributed training jobs.} 
We conduct extensive practical evaluations of Metronome using 13 popular Machine Learning (ML) models on our experimental platform. 
The results demonstrate that with the proposed mechanism, the completion time of high priority jobs deviates by no more than 2\% from the contention-free ideal scenario, confirming effective priority preservation.
Additionally, when compared with the existing K8s scheduling mechanisms across multiple scenarios, Metronome can reduce job completion time by up to 19.50\% while improving the average bandwidth utilization by up to 23.20\%.

\begin{table}[t]
	\centering
	\setlength{\abovecaptionskip}{0pt}
	\setlength{\belowcaptionskip}{10pt}
	\caption{Frequently used notations\label{t1}}
	\begin{tabularx}{\columnwidth}{@{} l >{\raggedright\arraybackslash}X @{}}
		\toprule
		\textbf{Symbol} & \textbf{Description} \\
		\midrule		
		$ \mathcal{L}, \mathcal{N}  $ &Set of links and nodes in cluster\\
		$ \mathcal{W} $  & Set of ML training workloads\\	
		$ \mathcal{J}_w $ & Set of jobs of workload \\	
		$ \mathcal{P}_{w,j} $  & Set of tasks of job $j$ of workload $w$ \\
		$ P_{w,j}(n)  $  &Number of tasks of job $j$ of workload $w$ on node $n$\\
		$\bar{\mathcal{P}}_l$ &Set of tasks   shared link $l$ \\		
		$\bar{\mathcal{P}}_l(n) $ &Set of tasks   shared host link $l$ of node $n$ \\
		${D}_w$  & Whether workload $w$ is deployed, ${D}_w \in \{0, 1\}$\\
		${D}_{w,j}$  & Whether job $j$ of workload $w$ is deployed,  ${D}_{w,j} \in \{0, 1\}$\\
		$B_l$  &Bandwidth capacity of link $ l $\\	
		$B_l(n) $  &Bandwidth capacity of host link $l$ of node $n$\\	
		$B^{max}$  &Maximum bandwidth capacity of links\\	
		$ R^{{s}}(n) $  & Total amount of resource  $s$  of node $ n $
		\\	
		$ L^{{s}}(n)  $ &Allocatable resource $s$ of node $n$ \\		 	  	
		$ t_p $&Period of task $ p $\\
		$d_p $&Communication duty cycle of task $ p $, $ d_p \in [0,1] $\\
		$ r_p^{{s}} $  &
		Request of Resource $s$ for task $p$ 
		\\	
		$ \theta_{l, p} $& Rotation angle on link $l$ for task $p$ \\
		$  p^{h}_l$&Task $p$ of the highest-priority job that is utilizing link  $l$\\
		$ \tau_{x,y} $ &The network latency matrix, $ \tau_{x,y} $ indicates the latency between nodes $x$ and $y$ (if $x = y$, 	$ \tau_{x, y} = 1$ ) \\
		$\nu_w$ & Job dependencies in workload $w$, if $(a,b)  \in \nu_w$(where $a,b \in \mathcal{J}_w $), it denotes a dependency between jobs $a$ and $b$\\
		\bottomrule
	\end{tabularx}
\end{table}

This paper is organized as follows. §\ref{2} formulates the system model as a multi-objective optimization problem. §\ref{3} details the design of our proposed mechanism, Metronome. The extensive testbed evaluation is presented in  §\ref{4}. 
We then discuss the additional design considerations and limitations in §\ref{5}, followed by a review of the related work in §\ref{6}.
Finally,  §\ref{7} concludes the paper. 

\section{System Modeling And Problem Formulation}
\label{2}
In this section, we first describe the notations in §\ref{2-1}, then develop the system model based on the geometric abstraction of the traffic patterns in §\ref{2-2}, and finally propose a three-stage optimization model in §\ref{2-3} to improve the cluster's link utilization while guaranteeing service performance.  

We employ distributed training jobs as the fundamental modeling unit in our framework, where job completion time serves as the primary metric for service performance evaluation. 
To structure this framework, we adopt a hierarchical system architecture with a top-down granularity structure consisting of three levels: workload, job, and task.
In this paradigm, users submit workloads that may encapsulate multiple distributed training jobs (such as hyperparameter optimization, HPO). Each job is fulfilled through the coordinated execution of numerous parallel tasks. 
Meanwhile, tasks will be scheduled to nodes and may share communication links between nodes with other tasks.

\subsection{Notations Description}
\label{2-1}
The major notations in this work are summarized and explained in Table \ref{t1}.

\subsection{System Modeling visa Geometric Abstraction}
\label{2-2}
To abstract the traffic pattern of a group of tasks $p$ sharing link $l$, we unify the iteration time of tasks to obtain the Least Common Multiple (LCM)  of period $T_l=\underset{p \in \bar{\mathcal{P}}_l}{\mathrm{LCM}}  (t_1,t_2,...,t_p) $. 
Following a similar approach to that of Cassini \cite{Cassini}, we then abstract the  traffic pattern of task $p$ into a circle whose perimeter is equal to period $T_l$. 

In this representation, for a  task with period $t_p$, the number of times its communication phase repeats on the circle is denoted by $mul_p$. For task $p\in \bar{\mathcal{P}_l}$, the communication duration per  iteration is $m_p = t_p \cdot d_p$, and the corresponding angle, denoted as $\alpha_p$, is given by:
\begin{equation}
	\label{eq1}
	\alpha_p=\frac{2 \pi}{T_l} \cdot m_p=\frac{2 \pi}{mul_p \cdot t_p} \cdot m_p=\frac{2 \pi}{mul_p} \cdot d_p .
\end{equation}

Assume the task initiates its communication phase at time $t = 0$.
Then, for each task, $mul_p$  units are placed on the circle, each containing the same ratio of communication and computation phases.  For  task $p\in\bar{\mathcal{P}}_l$, the angle span of its communication phase  is given as follows:

 \begin{equation}
 	\label{eq2}
 	\mathrm{Comm}_p=\bigcup_{i=0}^{mul_p-1}\left[\frac{2 \pi i}{mul_p}, \frac{2 \pi i}{mul_p}+\alpha_p\right),
 \end{equation}

\noindent and the i-th communication phase can be expressed as:

\begin{equation}
	\label{eq3}
	\mathrm{Comm}_p^{ (i) }=\left[\frac{2 \pi i}{mul_p}, \frac{2 \pi i}{mul_p}+\alpha_p\right), i=0,1,...,mul_p-1 .
\end{equation}

Thus, the total bandwidth demand of all tasks at angle $\theta$ on the circle of the link $l$ can be obtained as:

\begin{equation}
	\label{eq4}
	S_l (\theta) =\sum_{p \in P_l} r_p^{\text{BW}} \cdot \mathbf{I}_{\operatorname{Comm}_p}\left (\theta - \theta_{l, p}\right) ,
\end{equation}

\noindent where  $\theta_{l, p}$ represents the rotation angle in the geometric abstraction of task  $p$'s traffic pattern, while $\mathbf{I}_{\operatorname{Comm}_p}(x) $ is a characteristic function that holds the value of 1  when $x$ belongs to the communication phase of task $p$ and 0 otherwise.
Subsequently, we will adjust these  rotation angles to interleave the communication phases of tasks.   Eq. \eqref{eq4} can be employed for the subsequent computation of link utilization under the given rotational offset scheme.

\subsection{Problem Formulation and Solution}
\label{2-3}
In this section, we implement multi-stage optimization to improve resource utilization while guaranteeing service performance.
As cluster managers, we prioritize maximizing cluster utilization. Therefore, the primary optimization objective  is to  maximize average bandwidth utilization by alleviating communication contention.
When multiple optimal scheduling schemes exist, all solutions optimal in the first stage are carried forward to the second stage as candidates.
In the second stage, the optimizer selects the candidate that minimizes network latency between dependent tasks,   achieving the most compact schedule possible, which in turn ensures high service performance for users.
This prioritization aligns with empirical observations that bandwidth constraints typically become binding before latency limits in distributed training environments \cite{CNCF}, \cite{datamove}, \cite{themis}.
These first two stages collectively determine optimal task placement across nodes.
The third stage will maximize the minimum communication interval between tasks. 
This design reduces the probability of future contention caused by imperfect task synchronization, thereby ensuring the persistence of performance gains.
In this context,  applying multi-stage optimization is reasonable, as the objectives have differing priorities.

Considering the heterogeneity of bandwidth resources, the primary optimal metric is the average bandwidth utilization (avg. BW util.) \cite{themis}, which can be expressed as:

\begin{equation}
	\label{eq5}
	\Gamma = \sum_{l \in \mathcal{L}} \frac{B_l \cdot \xi_l }{B^{\max }}/ |\mathcal{L}|,
\end{equation}

\noindent where $|\mathcal{L}|$ represents the number of links, and $\xi_l$ is the link utilization of link $l$, which is defined as:

\begin{equation}
	\label{eq6}
	\xi_l=\dfrac{\int_0^{2 \pi} \min \left (S_l (\theta), B_l\right)  d \theta}{\int_0^{2 \pi} B_l d \theta},
\end{equation}

\noindent where $S_l (\theta)$  can be derived from Eq. \eqref{eq4}. Based on the above definition, if the total bandwidth demand exceeds the bandwidth capacity of the link  at any angle $\theta$, both $\xi_l$ and $\Gamma$ will decrease. This corresponds to a state of communication contention, which results in congestion.

From a cluster perspective, the objective of the second phase can be reformulated as minimizing the sum of the latency between all dependent tasks.
We define the sum of the latency of the workload $w \in \mathcal{W}$ as $\Lambda_{w}$, and the second optimization metric  can be expressed as:

\begin{equation}
	\label{eq7}
	{\Lambda}= {\sum_{w \in \mathcal{W}} {\textstyle\Lambda_{w}} },
\end{equation}

\noindent where $\Lambda_{w}$ can be further separated into:

\begin{equation}
	\label{eq8}
	\begin{split}
		\Lambda_{w} &= 
		\sum_{ (a, b)  \in \nu_w} \sum_{\substack{x, y \in \mathcal{N} }} \tau_{x, y} {P}_{w, a} (x)  {P}_{w, b} (y) \\
		&\quad +  \sum_{j \in \mathcal{J}_w} \sum_{\substack{x, y \in \mathcal{N} \\ x \leq y}} \tau_{x, y} {P}_{w, j} (x)  {P}_{w, j} (y) .
	\end{split}
\end{equation}

\noindent The first term of Eq. \eqref{eq8} denotes the total latency between dependent tasks across different jobs. On the other hand, for ML workloads, tasks within a training job inherently exhibit dependencies due to mutual communication requirements. 
Thus, the second term, which uses ordered node pairs to avoid duplicate calculations, indicates the total latency between dependent tasks within the same job.

The objective of the third stage is to maximize the minimum communication interval between contending tasks. This optimization metric    $\Psi$ is defined as:

\begin{equation}
	\label{eq9}
	\begin{aligned}
		\Psi = 
		& \min \left\{ \mathop{\mathbf{Distance}} \Bigl ( 
		\operatorname{Comm}_{s}^{ (i) } + \theta_{l,{s}},\, 
		\operatorname{Comm}_{t}^{ (i') } + \theta_{l,{t}} 
		\Bigr)\right\}, \\
		&\forall i \in [0, {mul}_{s} - 1],\ \forall i' \in [0, {mul}_{t} - 1],
	\end{aligned}
\end{equation}

\noindent where $s$ and $t$  are two contending tasks ($s, t \in \bar{\mathcal{P}}_l$),  which implies that  the combined bandwidth requirements of the two tasks on the shared link $l$  are at least equal to the  bandwidth capacity of the link.
The $\mathbf{Distance}$ function is expressed as $\min(|\phi - \psi|, 2\pi - |\phi - \psi|)$, where $\phi$ and $\psi$ are the angles at the midpoint of the intervals. 
Optimizing the communication interval between tasks ensures a sufficient cushion separating the communication phases, thereby reducing the probability of future contention,especially when communication drift occurs over time.

Based on the above abstractions and optimization objectives, our three-stage optimization is modeled as follows:

{\allowdisplaybreaks
	\begin{align}
		\underset{\text{1st}}{\max}\, \Gamma 
		&\mathrel{\boldsymbol{\rightarrow}} 
		\underset{\text{2nd}}{\min}\, \Lambda 
		\mathrel{\boldsymbol{\rightarrow}} 
		\underset{\text{3rd}}{\max}\, \Psi 
		\label{eq10} \\  
		\text{s.t.}\quad 
		& \forall w \in \mathcal{W}, \forall j \in \mathcal{J}_w: 
		 {D}_w \leq {D}_{w, j};
		\label{eq11} \\  
		& \forall w \in \mathcal{W}, \forall j \in \mathcal{J}_w:
		 {D}_{w, j} \leq \frac{\sum_n P_{w, j}(n) }{|\mathcal{P}_{w, j}|} \leq 1;
		\label{eq12} \\  
		& \forall n \in \mathcal{N}, s \in \{\text{CPU}, \text{MEM}, \text{GPU}\}: \nonumber \\
		& \quad \sum_{w \in \mathcal{W}}
		\sum_{j \in \mathcal{J}_w}
		\sum_{p \in \mathcal{P}_{w,j}} P_{w,j}(n)  \cdot r_p^{s} \leq R^{s} (n);		
		\label{eq13} \\  
		&  \forall l \in \mathcal{L},\forall n \in \mathcal{N},\forall w \in \mathcal{W},\forall j \in \mathcal{J}_w,
		  \forall p \in \mathcal{P}_{w,j}: \nonumber \\
		& \quad r_p^{\text{BW}} \leq B_l (n),  if \ P_{w,j}(n) > 1;
		\label{eq14} \\  
		& \forall l \in \mathcal{L}, \forall p \in \bar{\mathcal{P}}_l: 
		  \theta_{l, p} \in \left[0, \frac{2\pi}{\mathrm{mul}_p}\right);
		\label{eq15} \\  
		& \forall l \in \mathcal{L}: 
		 \theta_{l, p^{h}_l} = 0;
		\label{eq16} \\ 
		& \forall e, f \in \mathcal{L}, \forall w \in \mathcal{W},\ \forall j \in \mathcal{J}_w,  \nonumber\\
		& \forall s \in (\bar{\mathcal{P}}_{e} \cap \mathcal{P}_{w,j}),\ 
		\forall t \in (\bar{\mathcal{P}}_{f} \cap \mathcal{P}_{w,j})
		 :\nonumber\\
		& \quad \theta_{e,s} = \theta_{f,t}.
		\label{eq17}  
	\end{align}
}

Eqs. \eqref{eq11}-\eqref{eq12} illustrate the All-or-Nothing characteristic of distributed training scheduling. 
Eq. \eqref{eq11}  states that a workload is deployed only when all its constituent jobs are deployed, while Eq. \eqref{eq12} specifies that a job is deployed only when all its constituent tasks are deployed.

Eqs. \eqref{eq13}-\eqref{eq14} reflect the resource limitations of the nodes. Eq.  \eqref{eq13}  indicates that the total CPU, memory (MEM), and GPU resource requirements for each node $n$ must not exceed the node capacity. 
Bandwidth is another critical resource. Since communication-intensive workloads such as distributed training place extremely high demands on the network, the most advanced networks have approximately achieved an oversubscription ratio of 1:1  \cite{alibabaHPN}, \cite{jiang2024megascale}. This means that in an ideal scenario, if there is no communication contention on the host links, there should also be no contention on links between switches. 
Therefore, we simplify the problem by only considering the bandwidth capacity of host links, but we can obtain more cases by adding additional constraints. In summary, considering the two-dimensional scheduling of bandwidth resources, Eq. \eqref{eq14}  indicates that the bandwidth demand of any task $p$ on  node $n$ must not exceed the bandwidth capacity of the node (host link).

Eqs. \eqref{eq15}-\eqref{eq17}   address the restrictions on the task rotation angles. 
For task $p$, its communication phase recurs with a period of $2\pi/mul_p$. Therefore, Eq. \eqref{eq15} minimizes the search space to avoid duplicate results.
Since the rotation is essentially a relative operation, Eq. \eqref{eq16}  specifies for  task $p$ of the highest priority job on link $l$ that no rotate operation is required, thereby ensuring the service performance of that job.
Finally, Eq. \eqref{eq17} signifies that when tasks  $s$ and  $t$ belong to the same job, they perform synchronized communication with each other, which results in highly aligned traffic patterns and equal rotation angles.

It is worth noting that an additional constraint is added to ensure the optimal value of the previous stage is kept while executing the current stage. 
After the first optimization stage is complete, an additional constraint $ \Gamma = \Gamma^*$ is introduced. $\Gamma^*$ represents the maximum average bandwidth utilization obtained in the first stage. 
Subsequently, after the second stage, additional constraints $ P_{w,j}(n)  = {P}_{w,j}^* (n) $ are enforced, indicating that the optimization in the third stage is performed under the assumption that all tasks have already been assigned to the nodes determined in the second stage.
 
\section{Metronome Scheduling Mechanism: Design and Implementation}
\label{3}
\begin{figure*}[t]
	\centering
	\includegraphics[width=16.4cm]{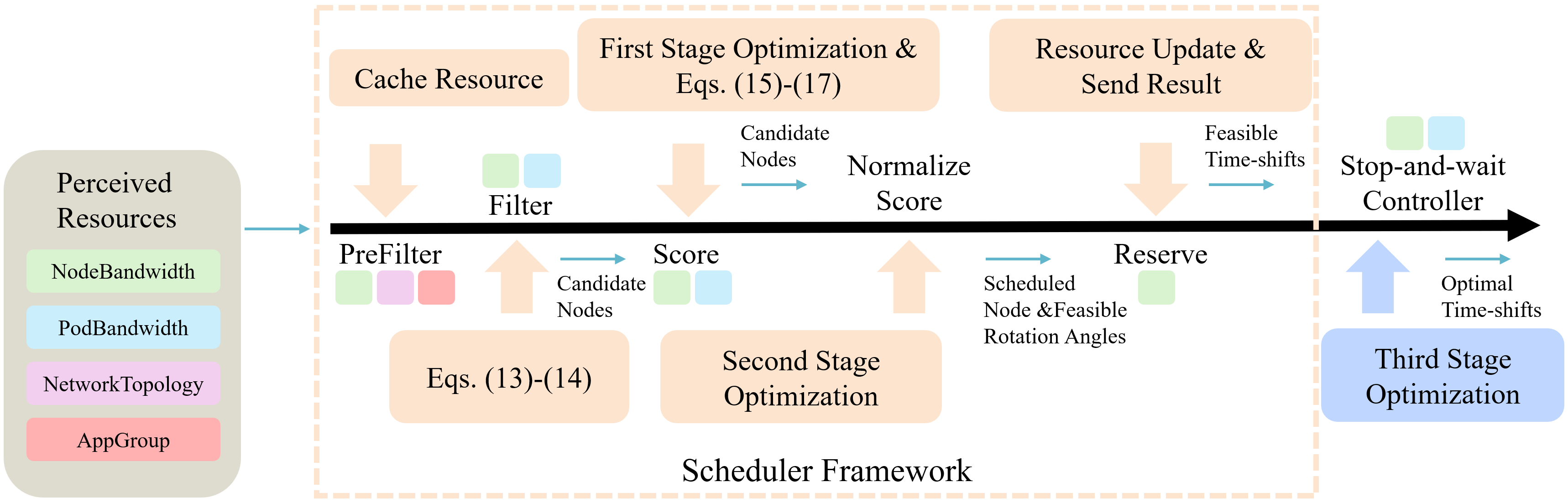}
	\caption{Workflow of Metronome mechanism.} 
	\label{2-workflow}
\end{figure*}

This section presents the design and implementation of the Metronome scheduling mechanism. 
The corresponding workflow is illustrated in Fig. \ref{2-workflow}. The proposed mechanism builds upon the foundational optimizations discussed in §\ref{2}, with the objective of facilitating enhanced resource awareness through the integration of additional Custom Resources (CRs). 
The corresponding colors in the figure denote the CRs (as we describe in §\ref{3-1}) involved at each phase.
The mechanisms for the scheduler and the stop-and-wait controller are detailed in §\ref{3-2} and §\ref{3-3}, respectively.
Finally, we  pull it all together for a comprehensive discussion in §\ref{3-4}.

In K8s, the jobs within the workload can correspond to resources such as Deployments, while tasks can correspond to pods. Pod is the unit of scheduling, and we denote the pod during the scheduling process as $p_{wait}$.

\subsection{CRs For Awareness Mechanism  }
\label{3-1}
We introduce four Custom Resource Definitions  (CRDs)  to allow Metronome to recognize network resources and  dependencies between pods. 
Each CRD operates in conjunction with its dedicated operator, forming an integrated unit where the resource definition and its management logic are tightly coupled.
The NodeBandwidth CRD provides Metronome with visibility into the bandwidth capabilities of cluster nodes and the pods deployed on them. For nodes, the default bandwidth value is derived from the network interface speed. This default value can be manually adjusted to account for exclusive scheduling or reserved resources.
PodBandwidth CRD enables Metronome to be aware of the two-dimensional bandwidth resource required by the pod. 
NetworkTopology  and  AppGroup CRDs basically leverage Diktyo's proven models \cite{diktyo} to handle latency between nodes and dependencies between pods.

\subsection{Metronome Scheduler}
\label{3-2}
Compared with the default scheduler, Metronome benefits from the extensible architecture of cloud native platforms.  
It introduces custom scheduling logic at the PreFilter, Filter, Score, Normalize Score, and Reserve extension points within the K8s scheduling framework (v0.26.7 \cite{plugin}). The framework successfully provides an effective solution to the optimization problem.\footnote{For Eqs. \eqref{eq11}-\eqref{eq12}, we reused the logic of the Coscheduling scheduler plugin  \cite{co}  without significant changes, so it is not shown in Fig.  \ref{2-workflow} and will not be elaborated further.}
The pseudocode of the scheduler process in Metronome is shown in \textbf{Algorithm} \ref{ag1}. Each extension point of the scheduler is discussed below.

$\bullet$ \textbf{PreFilter (lines 1-3).} 
At the PreFilter  phase, we design an operation to calculate the latency score for each node in advance and to cache resources. 

Metronome employs the \textsc{CalculateLatencyScore} function to derive a latency score $\Delta n$ for each node. First, it retrieves the pods that have dependencies with pod $p_{wait}$. 
In addition to these, Metronome automatically treats all pods within the same job as dependent. This is because inherent interdependencies arise from the communication required for synchronization within the job.
Then, for each candidate node, it assumes that pod $p_{wait}$ is deployed on it. The score for that node is computed as the sum of the latency between pod $p_{wait}$  and all its dependent pods that have already been deployed.
If the total latency is 0, it indicates that the pod has not explicitly declared the bandwidth requirement  (termed  a \texttt{LowComm} pod, whose communication does not need to be guaranteed)  or that pod $p_{wait}$  does not have a dependent pod deployed. 
In this scenario, Metronome calculates the average latency between the candidate node and all nodes in the cluster as the latency score.

Finally, Metronome caches the resources obtained at this phase. This allows for later retrieval, improving scheduling efficiency and avoiding redundant operations.

$\bullet$ \textbf{Filter (lines 4-13).} 
At the Filter phase, we employ additional judgment logic to filter out candidate nodes that do not meet the requirements, with a particular focus on bandwidth resources. This step reduces the number of nodes to be scored and improves scheduling efficiency.

First, according to Cassini \cite{Cassini}, if the placement causes a dependency loop, Metronome will filter out that node. A simple example of a dependency loop is as follows: job $a$ competes with job $b$ on link $e$, job $b$ competes with job $c$ on link $f$, and job $c$ competes with job $a$ on link $g$. 
Such a loop prevents the subsequent assigning of global offset.

Furthermore, Metronome filters out nodes whose allocatable resources, such as CPU, MEM, and GPU, are insufficient to meet the requirement of pod $p_{wait}$. 
For bandwidth resources, as specified in Eq. \eqref{eq14}, Metronome directly compares the pod's bandwidth requirements with the bandwidth capacity of the node.

$\bullet$ \textbf{Score (lines 14-16).} 
At the Score phase, we propose a scoring method for the bandwidth resources of each candidate node to determine its scheduling priority. Furthermore, for each candidate node, the feasible rotation angles are calculated assuming that the pod $p_{wait}$ is scheduled on it.

Metronome first determines whether pod $p_{wait}$ is a \texttt{LowComm} pod. If so, the early return operation is executed. 
If there are no deployed pods on the candidate node, or if the aggregate bandwidth consumption of all pods on the  node (including pod $p_{wait}$) does not exceed the bandwidth capacity, there will be no communication contention when scheduling the pod $p_{wait}$ on  it.  In this case, Metronome will revert to exclusive bandwidth scheduling and perform the return operation.
The above early return operations will set the node score to 100 (the perfect score) and set rotation angles $\vec{Ro}_{n}$ to null. These operations will also conclude the scoring phase and set the \texttt{EarlyReturn} flag to 1,  which will later be used to skip the third optimization stage.

If no early return occurs (i.e., the \texttt{EarlyReturn} flag remains 0), Metronome adopts the LCM of all pod periods as a base period $T_l$.
While in real-world environments pod periods are often not exact multiples of each other and their traffic patterns can be variable, using an excessively large LCM period would significantly complicate the scheduling calculation and interleaving process.
To address this problem, two thresholds are introduced. First, a threshold \texttt{$G_T$} is established. 
If the difference between the multiples of any two pod periods is within \texttt{$G_T$}, a common period can be derived by averaging their multiples for simplification.
Second, a  threshold \texttt{$E_T$} is defined for cases where the difference exceeds \texttt{$G_T$} but remains below \texttt{$E_T$}. In such scenarios, idle time is intentionally injected into the low priority pod's computation phase. 
This adjustment enables the pods to maintain an approximately periodic relationship and is motivated by two key insights:

$\bullet$  Introducing a controlled idle delay can interleave the communication phases between pods, and this delay is typically less than the communication contention delay;

$\bullet$ Injecting idle time also reduces the communication duty cycle for low priority pods, thereby decreasing the probability of future contention,  as our evaluation in §\ref{4-2-3} will show. 
Consequently, it  guarantees the service performance of high priority jobs more efficiently.

\begin{figure}[!htbp]
	\small
	\begin{algorithm}[H]
		\caption{Scheduling logic in Metronome}
		\label{ag1}
		\begin{algorithmic}[1] 
			\algnotext{EndIf}      
			\algnotext{EndFor}		
			\Require $\mathcal{N}, L^{s} (n), B_l(n), P_{w,j}(n),\bar{\mathcal{P}}_l (n),\tau_{x, y},\nu_w, r_{p}^{s}, p_{wait}$
			\Ensure $\vec{Shifts},\texttt{SkipPhaseThree},\bar{\mathcal{P}}_l (n^*) $
			
			\ForAll{$n \in \mathcal{N}$}  
			\State $\begin{aligned}[t]
				\Delta_{n} \gets &\textsc{CalculateLatencyScore} \\
				& (P_{w,j}(n), n, \tau_{x, y}, \nu_w,  p_{wait}) 
			\end{aligned}$			
			\EndFor 
			\State \textsc{CacheResource}\\
			\Comment{\parbox[t]{\dimexpr\linewidth-\algorithmicindent}{Eqs.  (13)-(14) }}
			\ForAll{$n \in \mathcal{N}$}
			\If {\textsc{HasDependencyCycle}}
			\State $\mathcal{N} \gets \mathcal{N} \setminus \{n\}$
			\State \textbf{continue} 
			\EndIf
			\If{$\exists s \in \{CPU, MEM, GPU\} : $ \\
				\hspace*{1.2em}$L^{s} (n)  < r_{p_{wait}}^{s}$ \\
				\hspace*{1.2em}\textbf{or } $B_l (n)  < r_{p_{wait}}^{BW}$}
			\State $\mathcal{N} \gets \mathcal{N} \setminus \{n\}$  
			\State \textbf{continue} \\
			\EndIf		
			\EndFor
			\Comment{\parbox[t]{\dimexpr\linewidth-\algorithmicindent}{1st Opt. \& Eqs.  (15)-(17) }}
			\ForAll{$n \in \mathcal{N}$}
			\State $\begin{aligned}[t]
				&Score_{n},\vec Ro_{n},\texttt{EarlyReturn} \gets\\
				&\textsc{ScoreAndFindFeasibleRotation}\\
				& ( B_l (n), \bar{\mathcal{P}}_l (n), n, r_{p}^{s}, p_{wait}  ) 
			\end{aligned}$ 
			\EndFor			
			\State $MaxScore \gets \textsc{FindMaxScore} (Score_{n}) $
			\State $Candidates \gets \textsc{FindMaxScoreNode} (Score_{n}) $
			\If{$|\mathit{Candidates}| \neq 1$} \\
			\Comment{\parbox[t]{\dimexpr\linewidth-\algorithmicindent}{2nd Opt.}}			
			\ForAll{$n \notin \mathit{Candidates}$}
			\State $Score_{n} \gets 0$
			\EndFor
			\ForAll{$n \in \mathit{Candidates}$}				
			\State $\Delta_{n} \gets  \textsc{NormalizeScore} (\Delta_{n}) $
			\If{\textsc{NotLowCommPod} $(p_{wait})$ } 			
			\State $Score_{n} \gets \Delta_{n}$ 
			\Else
			\State $Score_{n} \gets 100 - \Delta_{n}$ 
			\EndIf
			\EndFor
			\EndIf
			\State $n^* \gets \textsc{FindMaxScoreNode} (Score_{n}) $						
			\State $\bar{\mathcal{P}}_l (n^*)  \gets \bar{\mathcal{P}}_l (n^*)  \cup \{p_{wait}\}$
			\State \textsc{Update} $(\bar{\mathcal{P}}_l (n^*) ) $ 			
			\If{$MaxScore \neq 100$ \textbf{or} $\bar{\mathcal{P}}_l (n^*)  = 2$ \\ 
				\hspace*{1.2em}\textbf{or} $\texttt{EarlyReturn} = 1$}
			\State \texttt{SkipPhaseThree} $\gets 1$
			\Else
			\State \texttt{SkipPhaseThree} $\gets 0$
			\EndIf						
			\If {$\vec{Ro}_{n^*} \neq \text{null}$}
			\State $\vec{Shifts} \gets \textsc{ConversionToDelay} (\vec{Ro}_{n^*}) $			\\			
			\Comment{\parbox[t]{\dimexpr\linewidth-\algorithmicindent}{Send to stop-and-wait controller}}				
			\State \textsc{Send} $(\vec{Shifts},\texttt{SkipPhaseThree},\bar{\mathcal{P}}_l (n^*) ) $
			\EndIf			
		\end{algorithmic}		
	\end{algorithm}	
\end{figure}

Additionally, Metronome abstracts the traffic patterns between pod $p_{wait}$ and the pods already deployed on the node based on the base period $T_l$. 
Subsequently, Metronome exhaustively enumerates all possible rotation angle schemes for subsequent scoring.
This is achieved by enumerating all combinations generated from an angular discretization (precision: \texttt{$Di$-$Pre$}) and then rotating the abstraction circles of the pods accordingly.
Since rotation is an inherently relative operation, the reference pod is excluded from this process. Thus, the high priority pod is selected for this role.
If multiple pods share the same priority level, the one deployed earlier is chosen, allowing earlier jobs to complete sooner.
Currently, only two priority levels  (high and low)  are defined, with priorities assigned via pod labels. Nevertheless, Metronome can be readily adapted to accommodate alternative priority determination logic. 

Finally, the bandwidth scoring equation for every node $n$ is defined as follows:

\begin{equation}
	\label{eq18}
	\mathit{Score_n} = 100 - \frac{Excess}{B_l (n)  \cdot\texttt{$Di$-$Pre$}}\\,
\end{equation}

\noindent where $Excess$ is defined as the sum of the excess, over each angular range (determined by \texttt{$Di$-$Pre$}), between the total bandwidth demand of all pods within that range and the bandwidth capacity of the node. 
A perfect  score of 100 indicates that pod $p_{wait}$ is compatible with the pods already deployed on the node. 
A lower score signifies communication contention, with the score decreasing as the conflict duration increases.
This scoring process is computationally intensive. 
Therefore, at this phase, the algorithm traverses the offset scheme until it encounters the first interval with the perfect  score. It then selects this interval and returns the scheme corresponding to its middle index.
This approach yields a locally optimal feasible solution, while the stop-and-wait controller handles all computations. 
In the worst case, when no offset scheme achieves full compatibility, the scheme with the highest score is selected.

$\bullet$ \textbf{Normalize Score (lines 17-29).}
At the Normalize Score  phase, we employ the proposed normalization algorithm to select the final scheduled node $n^*$ for pod $p_{wait}$.
 
Metronome first determines whether the candidate node with the highest score is unique, in which case no further action is required. Otherwise, the nodes with the highest score will be further scored for latency. 
In this case, the previously cached latency scores are normalized.  Let $\Delta_{\text{min}}$ and $\Delta_{\text{max}}$ be the minimum and maximum values of all cached $\Delta_{n}$. The normalization equation is defined as follows:

\begin{equation}
	\label{eq19}
	{\Delta}_{n} = 
	\begin{cases}
		\displaystyle
		100 - \left\lfloor 100 \cdot \frac{\Delta_{n} - \Delta_{\text{min}}}{\Delta_{\text{max}} - \Delta_{\text{min}}} \right\rfloor, & \Delta_{\text{max}} \neq \Delta_{\text{min}} \\
		100 -  (\Delta_{n} - \Delta_{\text{min}}) , & \Delta_{\text{max}} = \Delta_{\text{min}}
	\end{cases}.
\end{equation}

\noindent The equation above performs a reverse mapping of the scores. If pod $p_{wait}$ is a \texttt{LowComm}  pod, a further reverse conversion is conducted. Nodes with higher score are favored. 
This operation enables scheduling dependent pods as compactly as possible (2nd Opt.) when multiple nodes satisfy the bandwidth requirement (1st Opt.). Otherwise, if the pod has no dependencies, the node with the lowest average latency is selected to avoid congestion. Whereas for \texttt{LowComm} pods, the node with the worst network condition is chosen.

$\bullet$ \textbf{Reserve (lines 30-40).} 
At the Reserve phase, we clean up the scheduler's state and incorporate the logic to transmit the scheduling outcome to the stop-and-wait controller, making the scheduler ready for the next pod.

At this phase, Metronome records the node selected for the pod and updates the corresponding CR. In our pseudocode, the selected node is denoted as $n^*$,  and in practice, Metronome updates the NodeBandwidth field of the CR (represented by $\bar{\mathcal{P}}_l (n^*)$).

Furthermore, the \texttt{SkipPhaseThree} flag is set to 1 under three cases to skip the third optimization stage: (1) the \texttt{EarlyReturn} flag is set to 1; (2)  the highest score obtained during the Score phase is not a perfect score; (3) the scheduled node contains only two pods (including pod $p_{wait}$). The first two cases correspond to scenarios with either no contention or unavoidable contention, respectively, while the last case implies that the local optimum is already the global optimum. Hence, the third optimization stage is skipped in all these situations.

Finally, if the offset scheme is not empty, two-dimensional bandwidth scheduling has been performed. Rotation angles in the scheme are  converted to the time-shifts using the equation $\vec{Shifts} = \vec{Ro}_{n^*}/\texttt{$Di$-$Pre$} \cdot T_l$. These values, along with other relevant messages (\texttt{SkipPhaseThree}, $\mathcal{P} _l (n^*)$), are then sent to the stop-and-wait controller.

\subsection{Metronome Stop-and-wait Controller}
\label{3-3}
The stop-and-wait controller is developed with the dual objective of yielding the final offset scheme and ensuring continuous regulation.

$\bullet$ \textbf{Global offset.}
The Metronome stop-and-wait controller receives information from the scheduler. However, the result from the scheduler is only the offset scheme for a single link. When jobs traverse multiple links and compete with different jobs on different links, the stop-and-wait controller must  perform a global offset adjustment to align the time-shifts.
Metronome adopts Cassini \cite{Cassini} approach to traverse the affinity graph. 
The primary difference is that, whereas Cassini randomly selects a reference job, Metronome designates the job with the highest priority as the reference for offset calculations.

$\bullet$ \textbf{Offline recalculation.}
The stop-and-wait controller verifies whether  rotation angles need recalculation based on the \texttt{SkipPhaseThree} flag. 
If this flag is set to 0, an offline recalculation must be performed.
The recalculation logic aligns with the Score phase of the scheduler, but differs in that it enumerates all schemes to find the  optimal rotation angles. This exhaustive search is feasible because the set of offset schemes is finite, which guarantees that an optimal solution must exist. Furthermore, we can narrow the search range to reduce the computation time, as the optimal scheme is expected to be located at the median index within the perfect score interval.  The rationale behind this assertion is detailed below.

Considering the scenario with two contending pods, we observe that the score function behaves periodically with respect to the phase rotation of one pod.
This periodicity arises because all abstracted communication phases start from 0 (as mentioned in §\ref{2-2}) and the phase of the higher priority pod is fixed.
Consequently, rotating the phase of the other pod will periodically cause the score to fluctuate from the minimum to the maximum and back again.
As a result, the optimal rotation angle is chosen from the middle indices of each perfect score interval. This principle extends to scenarios with more pods. By sequentially traversing the offset schemes, we can hold the phases of all but one contending pod fixed at their optimal positions. Then, rotating the phase of the final pod reduces the problem to the two-pod scenario described above.

Consequently, the search range is limited to the middle indices of all perfect score intervals, and the offset scheme that maximizes the minimum communication interval is deemed the optimal. The offline recalculation mechanism described above allows the scheduler to quickly return the feasible rotation angles, even if calculating the  optimal rotation angles  (3rd Opt.)  takes a significant amount of time. 
This ensures that the continuous monitoring mechanism operates normally during the initial stage, although not optimal before the scheme is updated through recalculation, as the maximum differences observed in metrics are presented in §\ref{4-3}.

$\bullet$ \textbf{Continuous regulation based on priority.}
The controller first executes the initial traffic pattern rotation for the pods. Given that the communication phase of jobs may drift due to noise and other events \cite{Cassini}, it is also necessary to continuously monitor each pod to ensure it respects the time-shifts. When communication contention arises due to drift, readjustment operations are implemented. Within a fixed time window of 10, if a pod's iteration time exceeds the baseline iteration time by a factor of \texttt{$A_T$} for more than \texttt{$O_T$} occurrences, the stop-and-wait controller will execute a pause operation on low priority pods to correct their communication phase. High priority pods do not require any action. 
The baseline iteration time is defined as the ideal duration without communication contention.

For distributed training jobs, when network congestion occurs or is resolved, the communication duty cycle can change. 
Additionally, in certain scenarios, the batch size  may change during training, altering the duty cycle of the job  \cite{pollux}, \cite{zheng2023shockwave}.
Therefore, upon detecting changes in pod traffic patterns or frequent readjustment triggers, the stop-and-wait controller will sequentially inspect relevant pods' traffic patterns. 
When necessary, it will update the corresponding CRs and recalculate the rotation angles.

\subsection{Metronome Mechanism}
\label{3-4}
\begin{figure}[htbp]
	\centering
	\includegraphics[width=8.8cm]{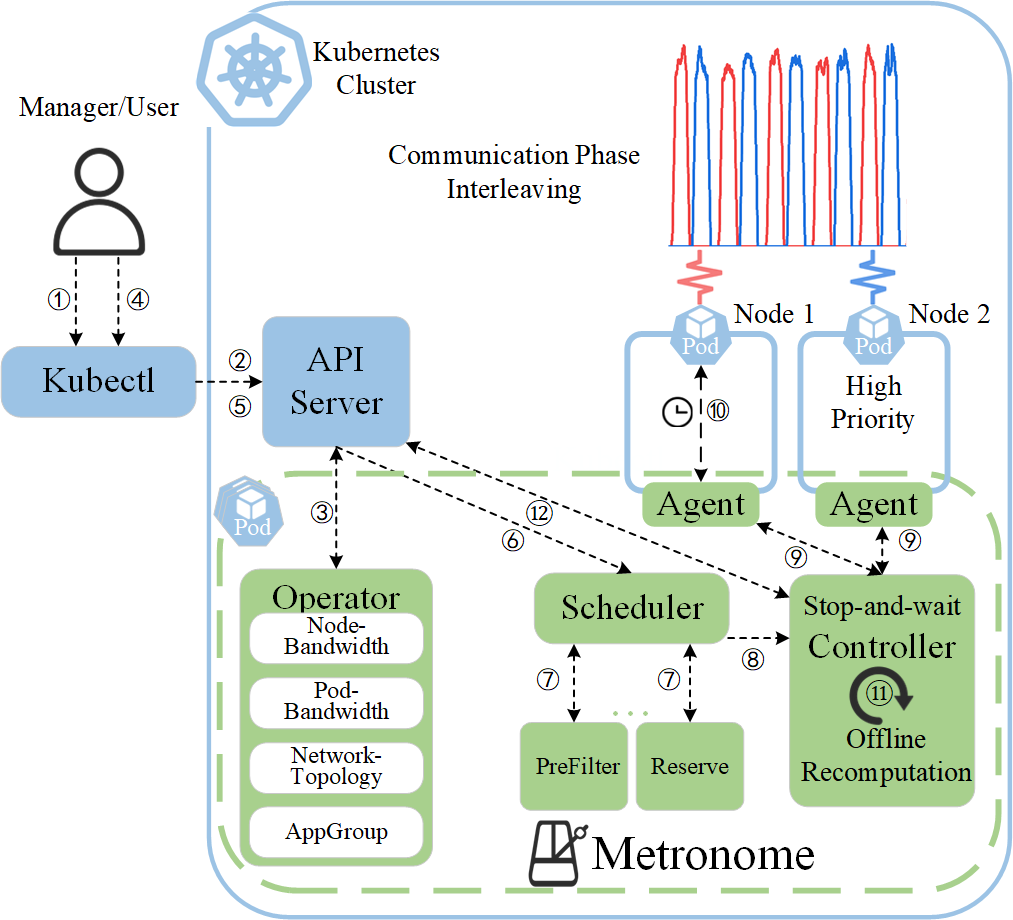}
	\caption{Metronome architecture.} 
	\label{3-metronome}
\end{figure}

The foundation of the Metronome mechanism, based on the design discussed above, is formed by three integral components (highlighted in green in Fig. \ref{3-metronome}) : 
(1) a set of four CRDs and their corresponding operators;  (2) a scheduler enhanced with an additional plugin registered at multiple extension points; (3) a stop-and-wait controller accompanied by monitoring agents deployed on each node. Agents are deployed as DaemonSets in K8s, while operators, the controller, and the scheduler are deployed as Deployments. 
Our mechanism requires no specific support from switches or network interface controllers and does not require any changes to the congestion control protocol.
A complete process in Metronome comprises three phases: steps  \raisebox{0.2ex}{\circled{1}}-\raisebox{0.2ex}{\circled{3}},  \raisebox{0.2ex}{\circled{4}}-\raisebox{0.2ex}{\circled{8}}, and  \raisebox{0.2ex}{\circled{9}}-\raisebox{0.2ex}{\circled{12}}, as shown in Fig. \ref{3-metronome}. The fundamental description of these phases  is provided  in §\ref{3-1}, §\ref{3-2}, and §\ref{3-3}, respectively.

\textbf{Phase 1 (\raisebox{0.2ex}{\circled{1}}-\raisebox{0.2ex}{\circled{3}}).} 
On the cluster side, the manager  is responsible for submitting the NodeBandwidth and NetworkTopology CRs;
on the user side, the user is responsible for providing the AppGroup and PodBandwidth CRs \raisebox{0.2ex}{\circled{1}}.
The API server registers resources \raisebox{0.2ex}{\circled{2}}  and notifies the corresponding operator. The operator then updates the CRs  (e.g., by adding a timestamp field) \raisebox{0.2ex}{\circled{3}}. For \texttt{LowComm} jobs, users do not need to submit the above resources, and Metronome will automatically add the special label to the submitted pod.
 
\textbf{Phase 2 (\raisebox{0.2ex}{\circled{4}}-\raisebox{0.2ex}{\circled{8}}).} Users specify the Metronome scheduler through fields and submit workload requests \raisebox{0.2ex}{\circled{4}} to complete resources registration \raisebox{0.2ex}{\circled{5}}. At this point, the scheduler discovers the pod  \raisebox{0.2ex}{\circled{6}}  and executes additional scheduling logic through extension points based on CRs, returning the selected node \raisebox{0.2ex}{\circled{7}}. If necessary, it also notifies the stop-and-wait controller with relevant information, including feasible time-shifts \raisebox{0.2ex}{\circled{8}}.

\textbf{Phase 3 (\raisebox{0.2ex}{\circled{9}}-\raisebox{0.2ex}{\circled{12}}).} The stop-and-wait controller continuously monitors the traffic patterns between pods based on the global feasible offset scheme. 
When a traffic pattern deviation between pods leads to communication contention, the controller will instruct the agent \raisebox{0.2ex}{\circled{9}}  to perform readjustment operations on low priority pods \raisebox{0.2ex}{\circled{10}}. 
If the offset scheme is not optimal, the controller will execute an offline recalculation and then update the scheme \raisebox{0.2ex}{\circled{11}}. Finally, the controller could update the CRs corresponding to the pods whose traffic patterns have changed \raisebox{0.2ex}{\circled{12}}.
	
As a scheduling mechanism, Metronome operates under the assumption that all necessary two-dimensional bandwidth resource requirements are specified at the time of job submission.
In practice, however, users may initially only provide a bandwidth demand value.
Metronome then performs operations similar to existing studies \cite{cao2024crux}, \cite{Cassini}, \cite{themis2}, \cite{pollux},  by executing multiple iterations of each job in an environment without contention to profile its traffic pattern and update the bandwidth requirement.

\section{Evaluations}
\label{4}
This section presents a comprehensive evaluation of the Metronome, systematically comparing its performance with other scheduling mechanisms. §\ref{4-1} details the experimental methodology and configuration. Subsequently, §\ref{4-2} analyzes the performance gains achieved by Metronome over existing schedulers across multiple metrics. To further probe the robustness of these improvements, we assess the mechanism's adaptability by independently varying the bandwidth requirements of pods and latency parameters, complemented by extended-duration trials that confirm the persistence of observed gains. Furthermore, §\ref{4-3}. employs ablation studies to evaluate the overall performance of Metronome. Finally, the rationale for threshold selection is validated in §\ref{4-4} and a comparative analysis of scheduler execution times is presented in §\ref{4-5}.

\subsection{Methodology and Setup}
\label{4-1}
\begin{figure}[htbp]
	\centering
	\includegraphics[width=8.8cm]{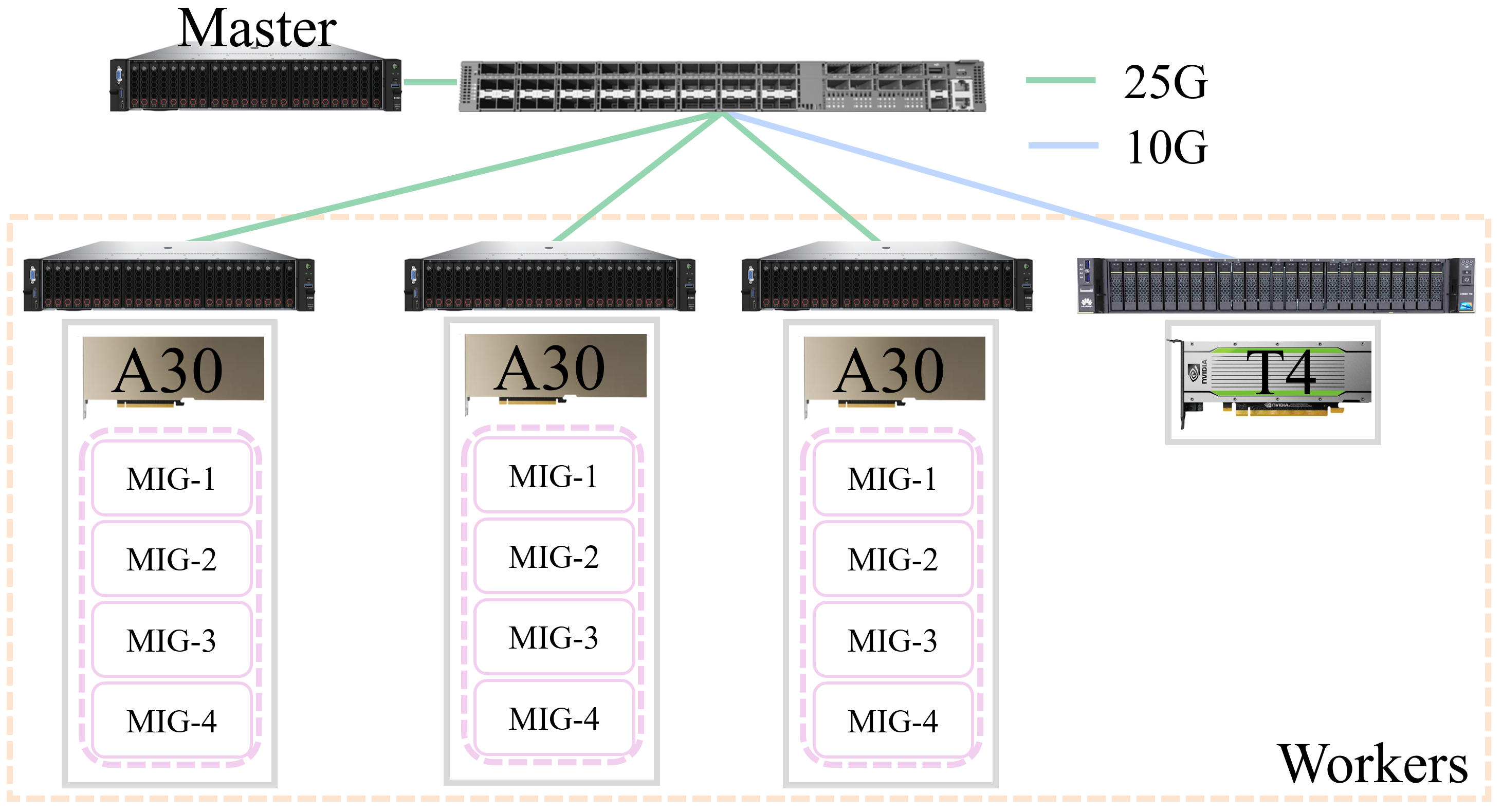}
	\caption{ Testbed environment: a K8s cluster with heterogeneous devices.} 
	\label{4-heterogeneous}
\end{figure}

$\bullet$ \textbf{Topology and setup.}
We build a K8s cluster with heterogeneous devices  (1 master, 4 workers)  to demonstrate the gains of  Metronome. Fig.  \ref{4-heterogeneous} shows the cluster configuration and physical topology. Three worker nodes  (32-core CPU, 1 TB memory, 4 TB SSD)  are equipped with an NVIDIA A30 GPU  (24 GB HBM2 memory). We utilize Multi-Instance GPU  technology to virtualize each A30 GPU into four  logical  GPUs, and the nodes connect to the switch with 25Gbps links. One worker node  (20-core CPU, 32 GB memory, 6 TB HDD)  is equipped with a Tesla T4  (16 GB GDDR6 memory), which is connected to the switch with 10 Gbps link. 
This diverse GPU type and bandwidth experimental testbed effectively demonstrates the complex heterogeneous environment in large modern multi-tenant clusters. 

The server runs Ubuntu 24.04 LTS. The K8s cluster version is v1.27, and we use Kubeflow v1.9  to deploy distributed training jobs.
The relationships between the additional variables involved in the scheduling mechanism and the resource fields are illustrated in Table \ref{t2}.

\begin{table}[!htbp]
	\setlength{\abovecaptionskip}{0pt}
	\setlength{\belowcaptionskip}{10pt}
	\setlength{\tabcolsep}{7.7pt} 
	\caption{ The relationship between the variables and the resource fields\label{t2}}  
	\centering
	\resizebox{8.8cm}{!}{
		\begin{tabular}{@{}p{2.2cm}p{5.7cm}@{}}
			\toprule
			\textbf{Variables} & \textbf{Resource fields} \\
			\midrule
			$\texttt{LowComm}$ & Pod.spec.labels.low\_comm\\
			$P_{w,j}(n), \bar{\mathcal{P}}_l(n)  $ & NodBandwidth.status.pods\\ 
			$B_l$ & NodBandwidth.spec.bandwidth\\
			$t_{p}$ & PodBandwidth.spec.period \\
			$d_{p}$ & PodBandwidth.spec.duty\_cycle \\
			$r_{p}^{BW}$ & PodBandwidth.spec.bandwidth \\
			$\tau_{x,y}$ & NetworkTopology.spec.weights...costlist* **\\
			$\nu_w$ & AppGroup.spec.workloads**\\

			\bottomrule
			\addlinespace[0.5ex]
			\multicolumn{2}{@{}p{\columnwidth}}{
				\footnotesize 
				\parbox{\dimexpr\linewidth}{%
					* For ease of presentation, intermediate fields have been omitted. \\
					** Reuse the native AppGroup and the modified NetworkTopology CRDs from Diktyo.%
				}
			} 
		\end{tabular}
	}
\end{table}

$\bullet$ \textbf{Scheduling mechanisms.}
With the same resource quota  (5G memory, 5-core CPU), we evaluated the performance of the following existing scheduling mechanisms:

	Default (De): The default scheduler in K8s \cite{kubernetes_scheduler}.
	
	Diktyo (Di): The network-aware scheduling mechanism designed by Ghent University and IBM Research \cite{diktyo}. It mainly considers the latency  during scheduling, intending to minimize the latency between dependent pods.
	We modify the mechanism to automatically identify dependencies, both between and within jobs.
	This modification is specifically designed to handle the characteristics of distributed training, aligning with Metronome's capability to enable a fair performance comparison.
	
	Ideal (Id): The ideal scheduler that runs each training job on a dedicated cluster, thus eliminating any communication contention.
	
	Metronome (Me): The scheduling mechanism proposed in this paper. 
	In our evaluation, the thresholds are configured as follows:  \texttt{$O_T$} = 5, \texttt{$A_T$} = 110\%, \texttt{$G_T$} = 5 ms, \texttt{$E_T$} = 10\% of the low priority job's period, and \texttt{$Di$-$Pre$} = 72.  The value for \texttt{$Di$-$Pre$} is chosen to be consistent with the setting in  Cassini \cite{Cassini}, and the rationale for other threshold selections is verified in §\ref{4-4}.
	
	Exclusive bandwidth scheduling: A general class of schedulers that allocate exclusive bandwidth to pods, as implemented in  \cite{lai2023delay}, \cite{toka2021ultra}.

\begin{table}[!htbp]
	\centering
	\setlength{\abovecaptionskip}{0pt}
	\setlength{\belowcaptionskip}{10pt}
	\setlength{\tabcolsep}{7.7pt} 
	\renewcommand{\arraystretch}{1} 
	\caption{ML models used in our experiments\label{t3}}
	\resizebox{8.8cm}{!}{
		\begin{tabular}{@{}p{3.4cm}p{1.4cm}p{1.5cm}p{1.3cm}@{}}
			\toprule
			\textbf{Name}&\textbf{Type}& \textbf{Parallelization Strategy}& \textbf{Training Strategies}\\
			\midrule				
			\makecell[l]{VGG model family \\ \footnotesize{VGG11 VGG16 VGG19}} & Vision&DP&FT\&Pre\\
			\makecell[l]{ResNet model family \\ \footnotesize{ResNet18 ResNet50 ResNet152}} & Vision&DP&FT\&Pre\\[-1.3ex]	 
			WideResNet101&  Vision&DP&FT\\			
			GoogLeNet&  Vision&DP&FT\\		
			DenseNet201&  Vision&DP&Pre\\			
			AlexNet&  Vision&DP&Pre\\					
			GPT-1&  Language&MP&Pre\\		
			GPT-2&  Language&MP&Pre\\			
			BERT&  Language&MP&Pre\\		
			\bottomrule
		\end{tabular}
	}
\end{table}

$\bullet$ \textbf{Workloads.}
We conduct experiments with 13 popular ML models: VGG \cite{vgg} (VGG11, VGG16, VGG19), ResNet \cite{resnet} (ResNet18, ResNet50, ResNet152), WideResNet101 \cite{wideresnet}, GoogLeNet \cite{googlenet}, DenseNet201 \cite{densenet}, AlexNet \cite{alexnet}, GPT-1 \cite{gpt1}, GPT-2 \cite{gpt2}, BERT \cite{bert}. The training strategy includes pre-training (Pre)  and fine-tuning (FT). 
We employ data parallelism (DP) with PyTorch's DistributedDataParallel (DDP) for computer vision models, while utilizing Microsoft's DeepSpeed for hybrid model parallelism (MP) in language models.
We made minor revisions to DDP and DeepSpeed to enable them to report information about iteration time to the stop-and-wait controller. Specific details of used models are provided in Table \ref{t3}.

\begin{figure}[h]
	\small
	\centering
	\includegraphics[width=8.8cm]{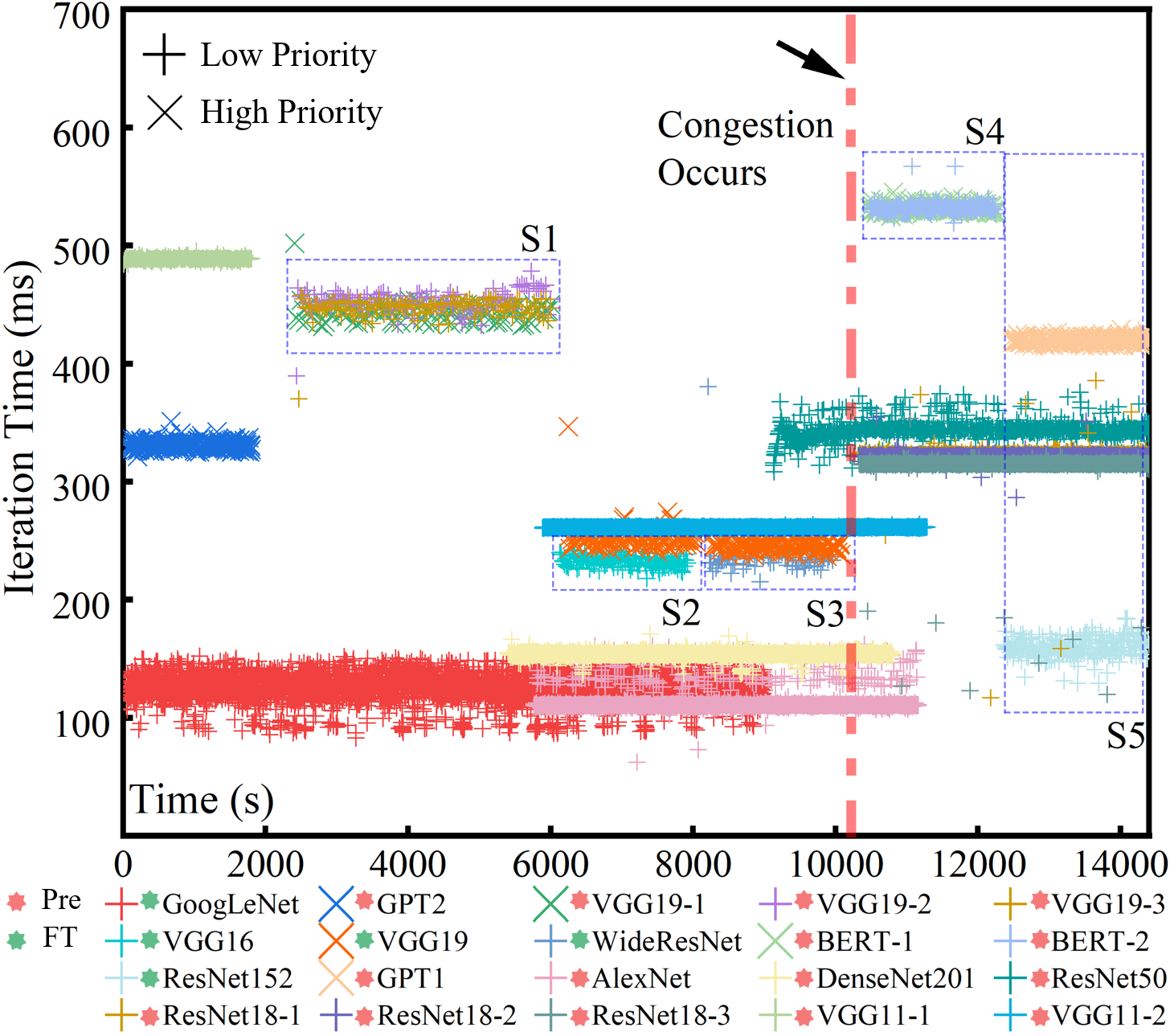}
	\caption{Time series of training jobs and their iteration time.} 
	\label{5-time}
\end{figure}

\begin{figure*}[t]
	\centering
	\captionsetup[subfloat]{
		font=footnotesize,       
		labelfont=bf,         
		textfont=normalfont,   
		justification=centering  
	}
	\subfloat[S1]{\label{6-traffic1}\includegraphics[width=0.19\textwidth]{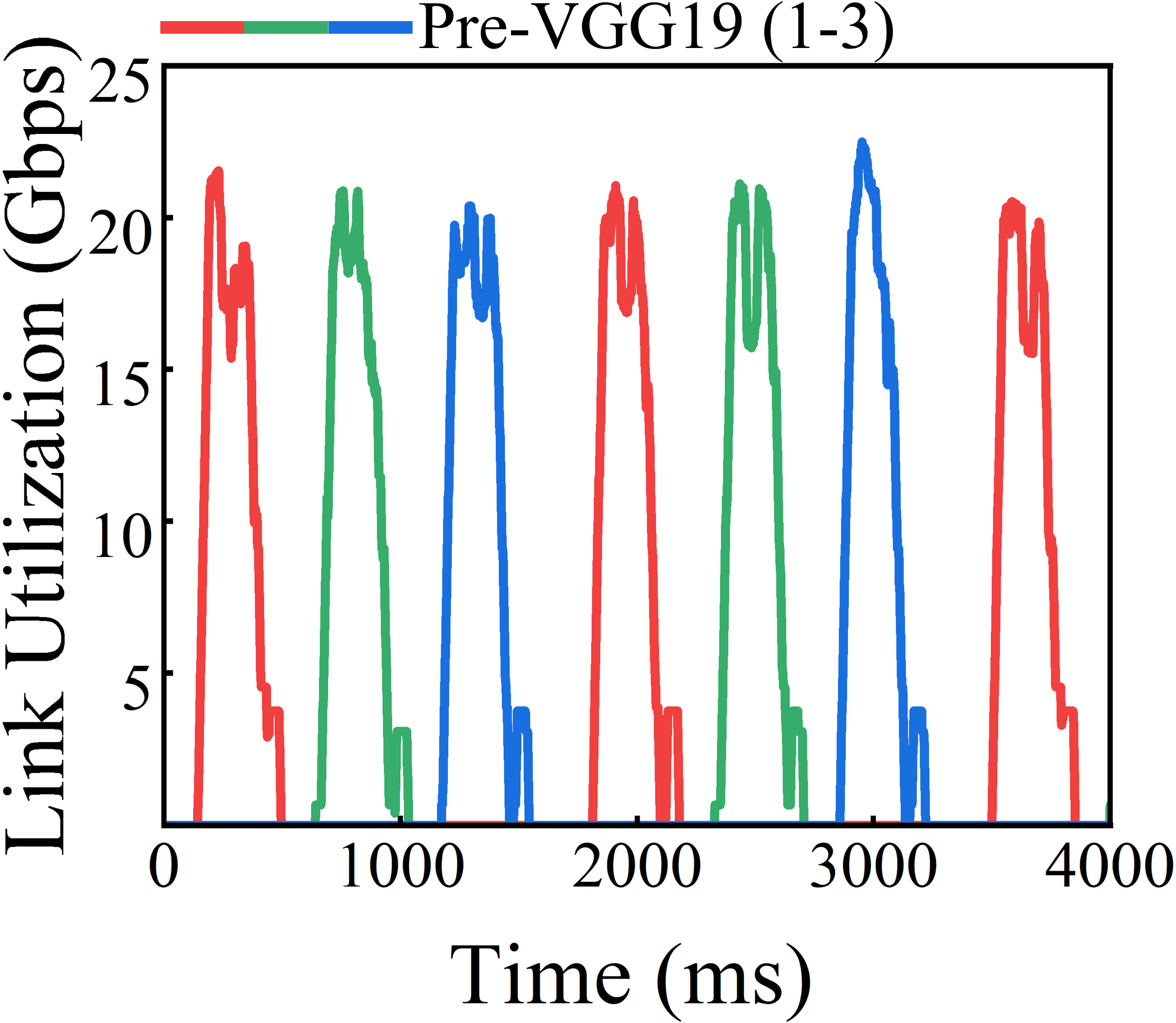}}
	\hfill
	\subfloat[S2]{\label{6-traffic2}\includegraphics[width=0.19\textwidth]{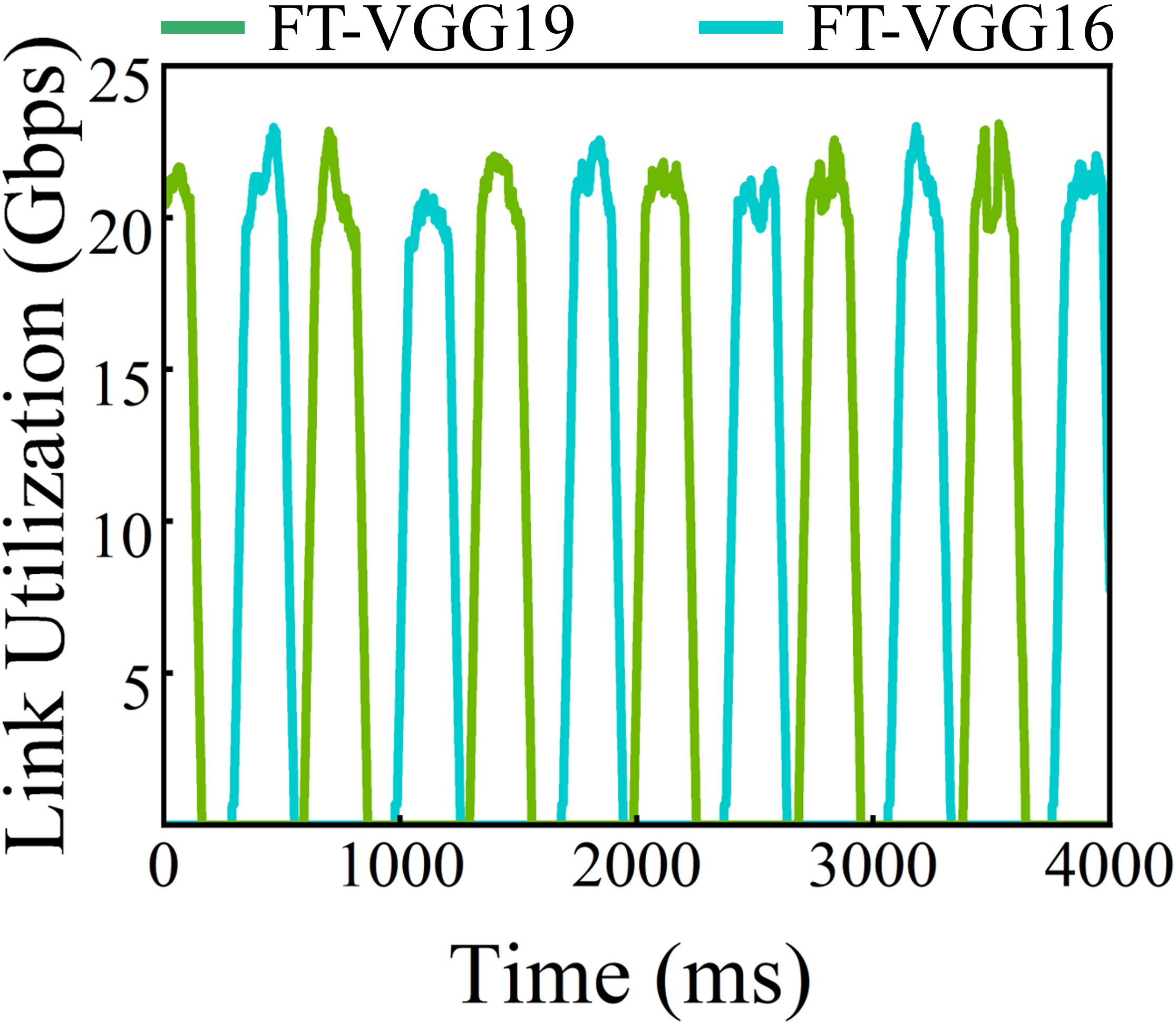}}
	\hfill
	\subfloat[S3]{\label{6-traffic3}\includegraphics[width=0.19\textwidth]{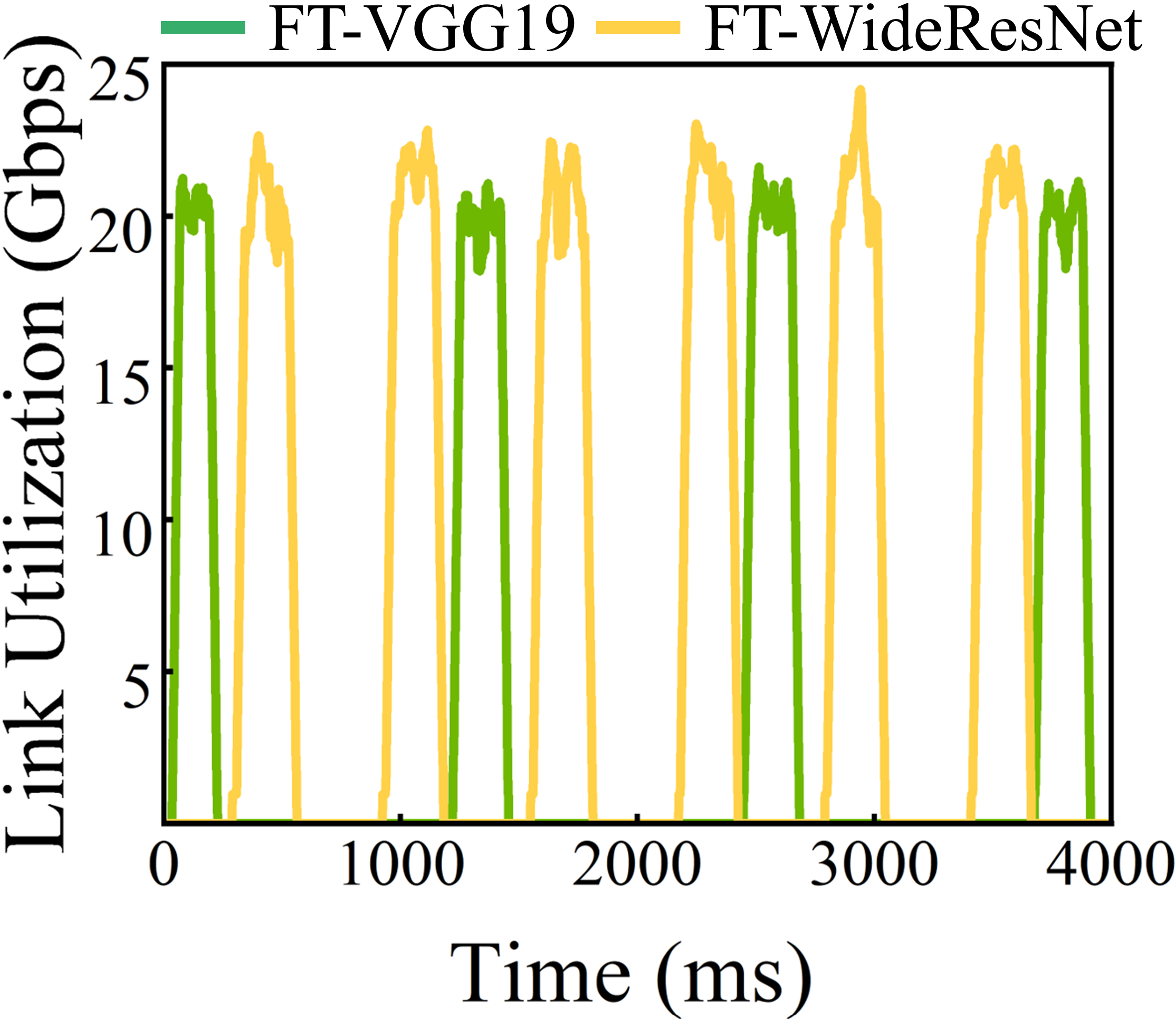}}
	\hfill
	\subfloat[S4]{\label{6-traffic4}\includegraphics[width=0.19\textwidth]{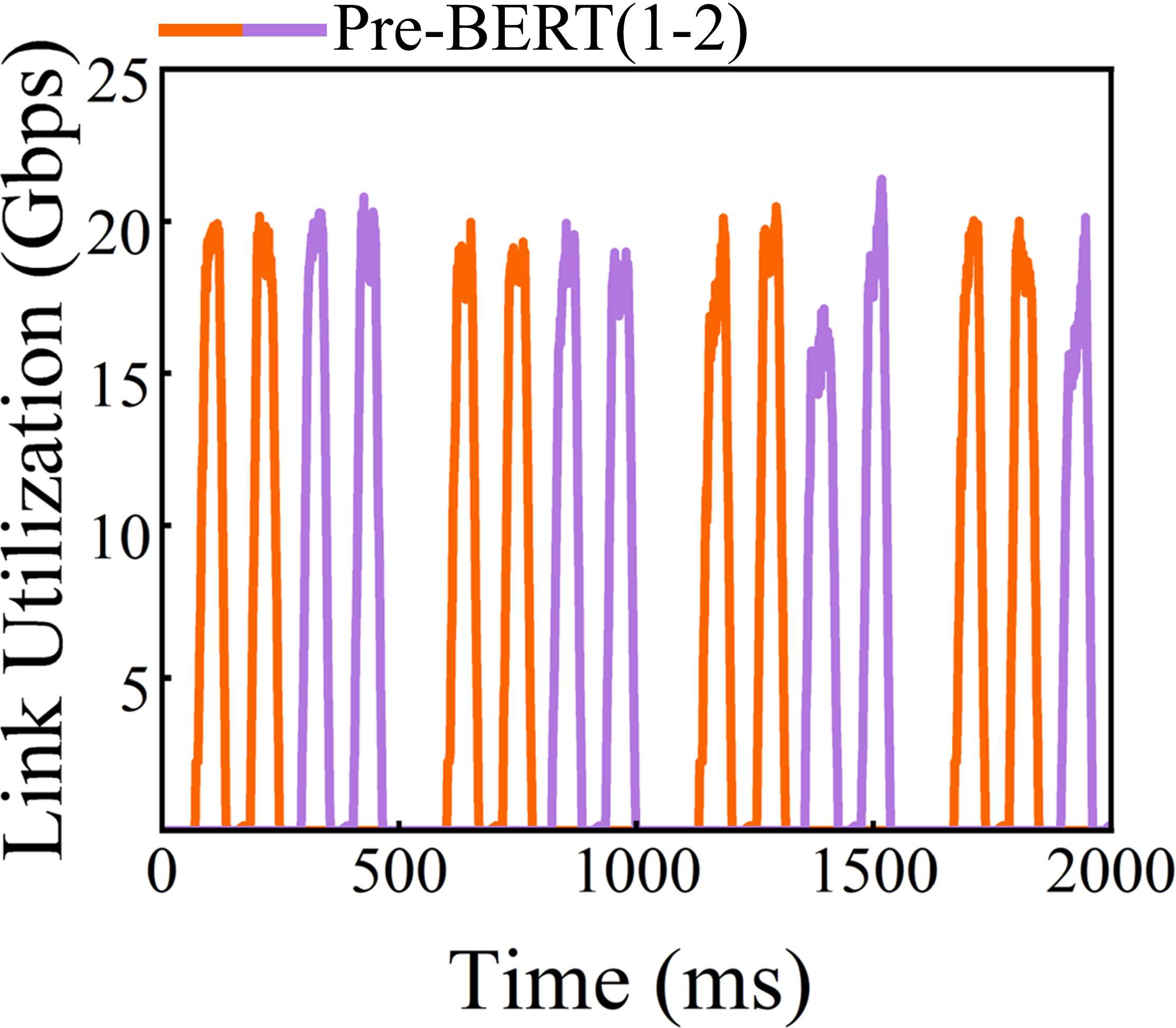}}
	\hfill
	\subfloat[S5]{\label{6-traffic5}\includegraphics[width=0.19\textwidth]{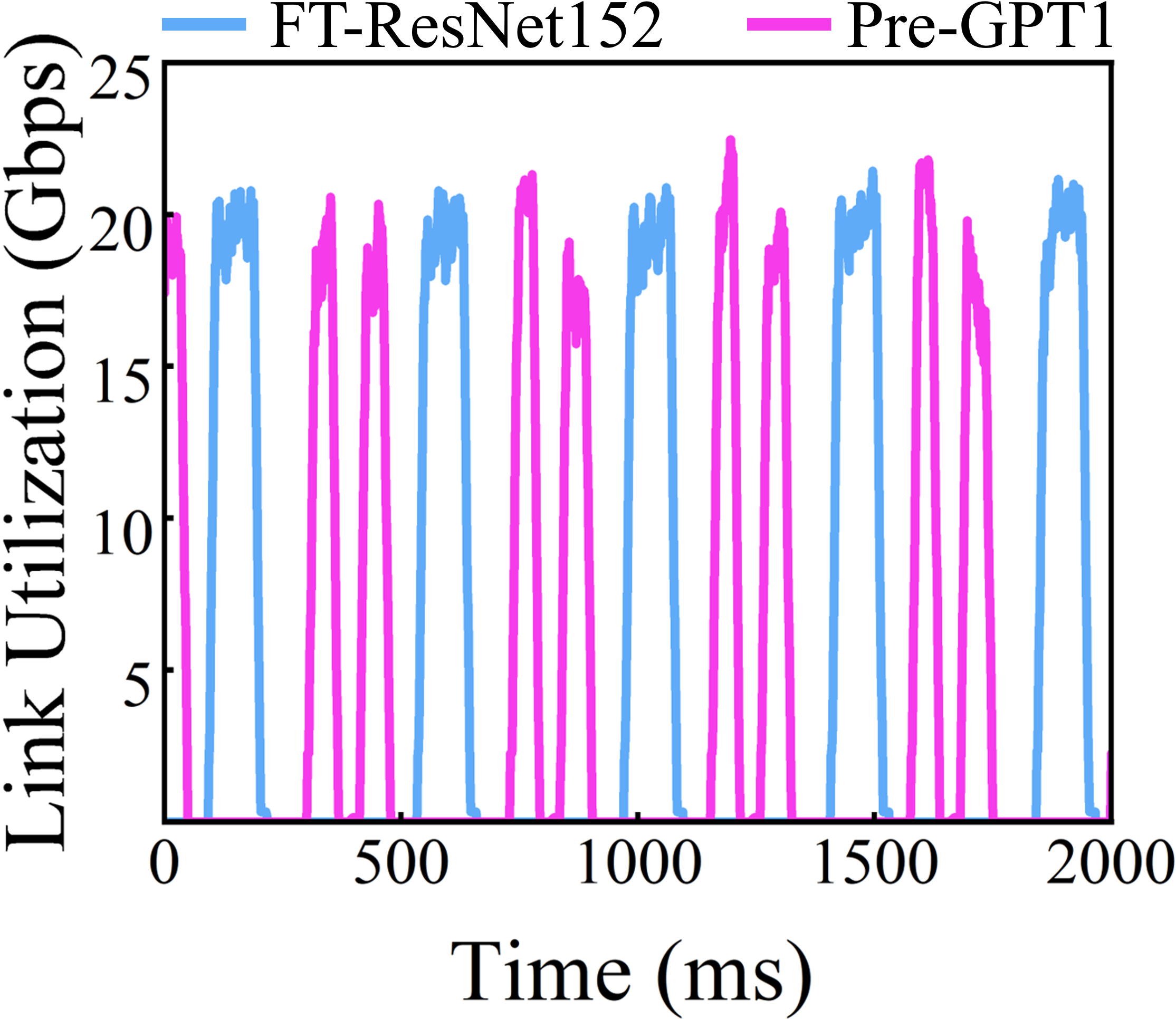}}
	\caption{Traffic patterns of the jobs in snapshot.}
	\label{6-traffic}
\end{figure*}

$\bullet$ \textbf{Traces.}
We use the Gavel workload generator  developed by Stanford University and Microsoft Research \cite{narayanan2020heterogeneity} to create a trace with a total duration of 4 hours. It defines the job arrival sequence, priorities, and individual job durations ranging from 0.5 to 1.5 hours.
The load parameter is defined as the average fraction of GPUs that are serving active jobs in the cluster. Excluding the startup phase, the cluster operates at a load in excess of 60\%  (reaching a maximum of 85\%).
Fig. \ref{5-time} plots the time series of training jobs and their iteration times under ideal conditions. To simulate congestion caused by link failures or unregulated traffic, we use iPerf3 to introduce  a congested node with a high latency link. 
Fig.  \ref{5-time} also highlights five snapshots (S1-S5)  for detailed evaluation in subsequent experiments. 
Each snapshot comprises a set of jobs competing on links, with their traffic patterns illustrated in Fig. \ref{6-traffic} and the detailed composition provided in Table \ref{t4}.

\begin{table}[t]
	\centering 
	\setlength{\abovecaptionskip}{0pt}
	\setlength{\belowcaptionskip}{10pt}
	\setlength{\tabcolsep}{7.7pt} 
	\caption{Composition of ML Models in the snapshot\label{t4}}
	\resizebox{8.8cm}{!}{
	\begin{tabular}{@{}p{0.2cm} p{2.6cm} p{5cm}@{}}
			\toprule
			\textbf{ID} & \textbf{Contending Jobs} & \textbf{Description} \\
			\midrule 
			S1 & Pre-VGG19 $\times$ 3 & DP HPO training job $\times$ 3 \\        
			S2 & FT-VGG16 \linebreak +  FT-VGG19$^\star$ &  DP training jobs \\  
			S3 & FT-WideResNet101  \linebreak +  FT-VGG19$^\star$ & DP  training jobs,  with a ratio of 2:1 for the period after idle time injection\\  
			S4 & Pre-BERT $\times$ 2 & Exist the congested link, DP HPO training job $\times$ 2\\        
			S5 & FT-ResNet152 \linebreak +  Pre-GPT-1$^\star$ &Exist the congested link,  DP and MP training jobs exist simultaneously \\      
			\bottomrule
			\multicolumn{3}{@{}p{\columnwidth}}{\footnotesize $^\star$ represents high priority jobs.  Unless stated otherwise, jobs deployed earlier have higher priority.}
		\end{tabular}
	}
\end{table}

$\bullet$ \textbf{Metrics.}
We evaluated from both the cluster (manager) and the user perspectives. Considering the heterogeneous bandwidth, the cluster side evaluated the average bandwidth utilization and the total completion time  (TCT)  of the trace, while the user side assessed the average time per 1,000 iterations of the job. TCT is a global metric that summarizes the entire trace, whereas link utilization and iteration time are fine-grained metrics measured at each snapshot. We utilized the ethtool counters in the Linux kernel to gather network interface communication information and parse PyTorch logs to retrieve iteration time and TCT. 
Unless stated otherwise, the experiment is run in triplicate, each replicate lasting for 0.5 hours, and all metrics are averaged.

\subsection{Performance Analysis}
\label{4-2}
We now separately evaluate the scheduler performance for snapshots and TCT. 
For distributed training jobs with extremely high network demand, the baseline configuration often requires pod bandwidth allocation to match the full  bandwidth capacity of the node. In that case, exclusive scheduling results in only one job being executed per snapshot, while all others are rejected. Although it can achieve the optimal job training speed under ideal conditions, the average job acceptance rate across  all snapshots is below 50\%. 
This result stands in sharp contrast to two-dimensional bandwidth scheduling, which achieves full acceptance while also delivering training speeds that approach the ideal conditions, as will be demonstrated in the following experiments. 
Therefore, exclusive scheduling will not be considered in further experiments.

\subsubsection{Performance Evaluation Based on Snapshots}
\label{4-2-1}
\begin{figure}[!htbp]
	\centering
	\captionsetup[subfloat]{
		font=footnotesize,      
		labelfont=bf,          
		textfont=normalfont,   
		justification=centering 
	}	
	\subfloat[S1-L-Distribution]{\label{7-dist1}\includegraphics[height= 0.20\textwidth]{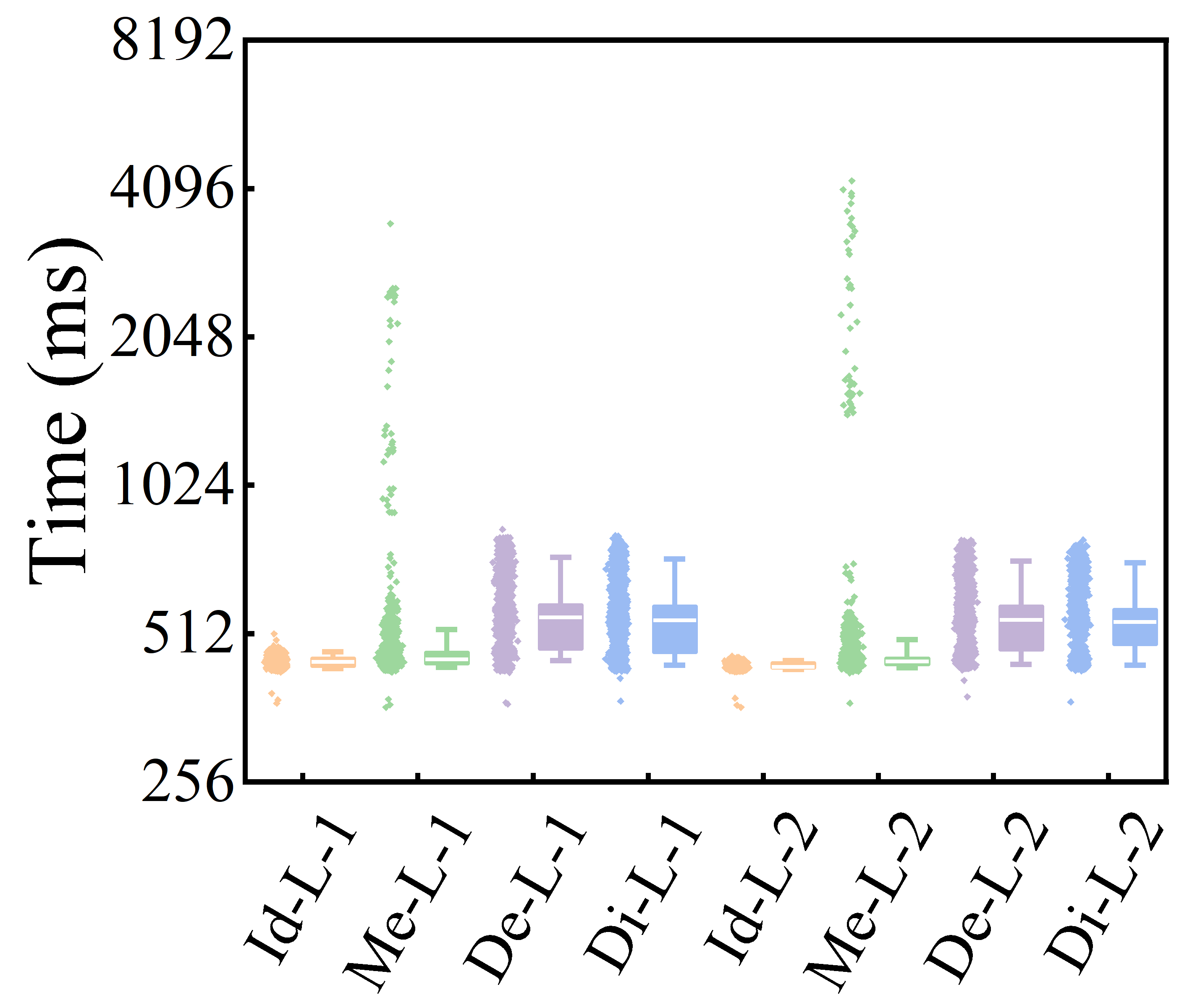}}
\hspace{0.1cm}
	\subfloat[S2-L-Distribution]{\label{7-dist2}\includegraphics[height=0.20\textwidth]{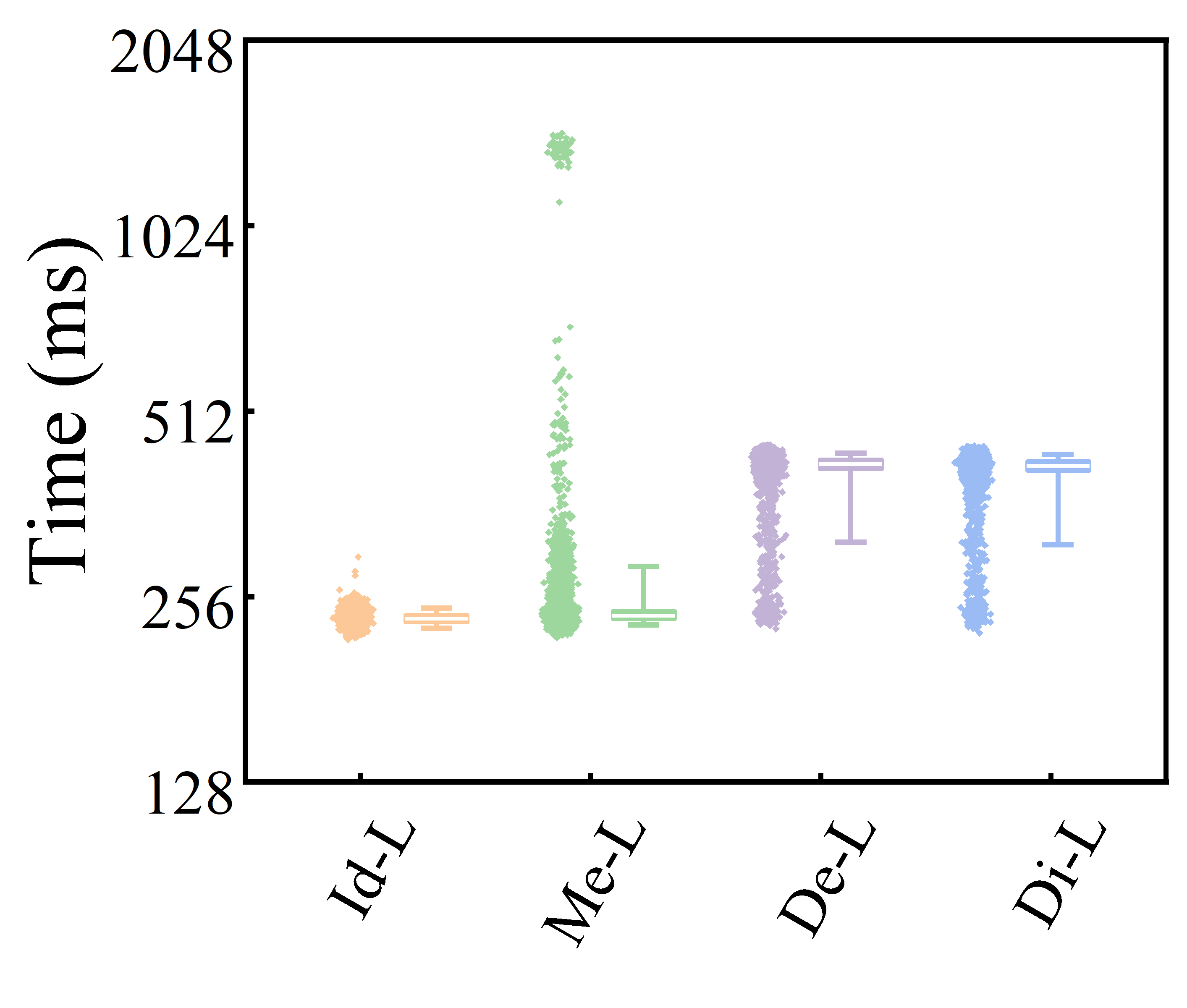}}
\hspace{0.1cm}
	\subfloat[S3-L-Distribution]{\label{7-dist3}\includegraphics[height=0.2\textwidth]{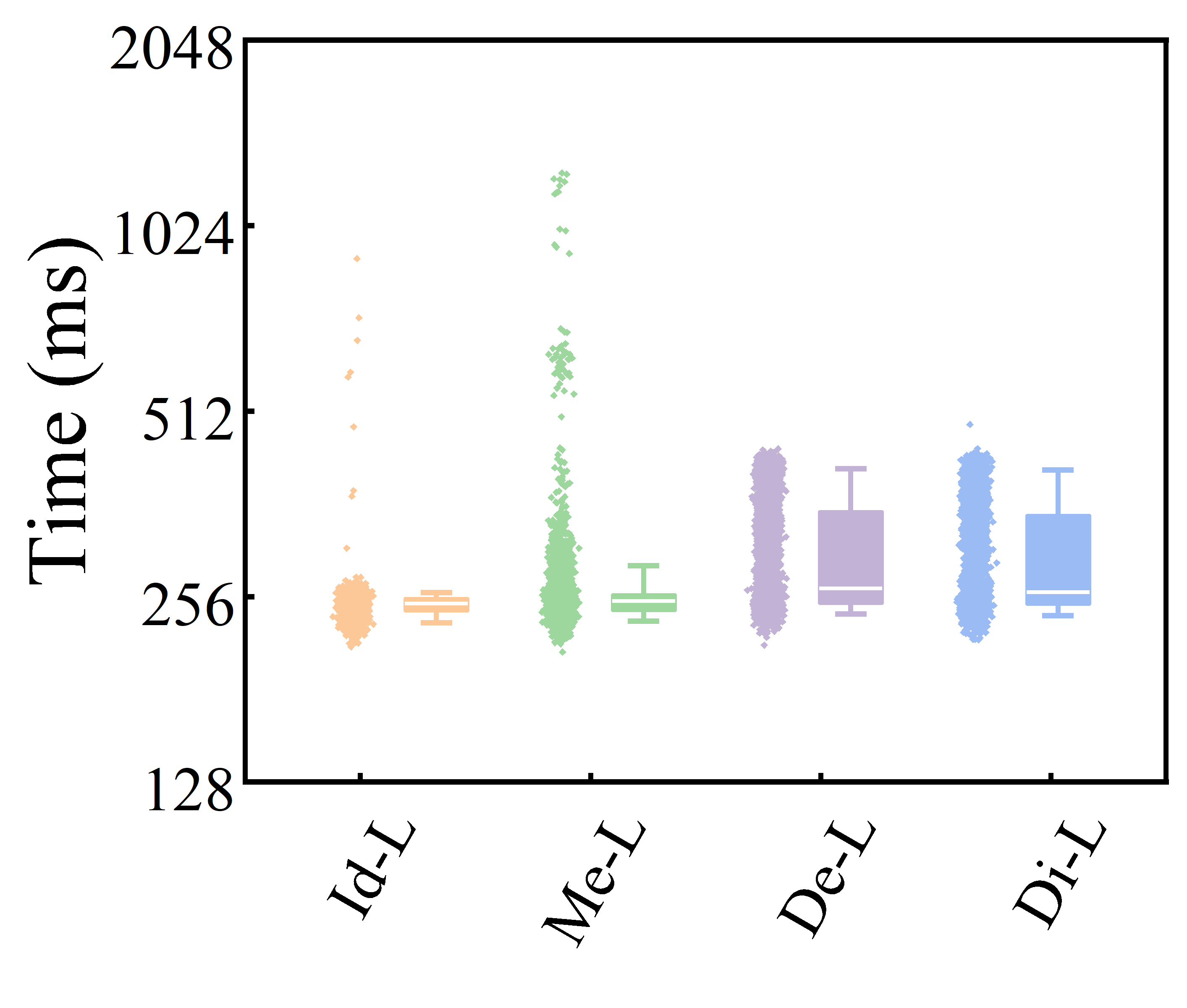}}
\hspace{0.1cm}
	\subfloat[S4-L-Distribution]{\label{7-dist4}\includegraphics[height=0.2\textwidth]{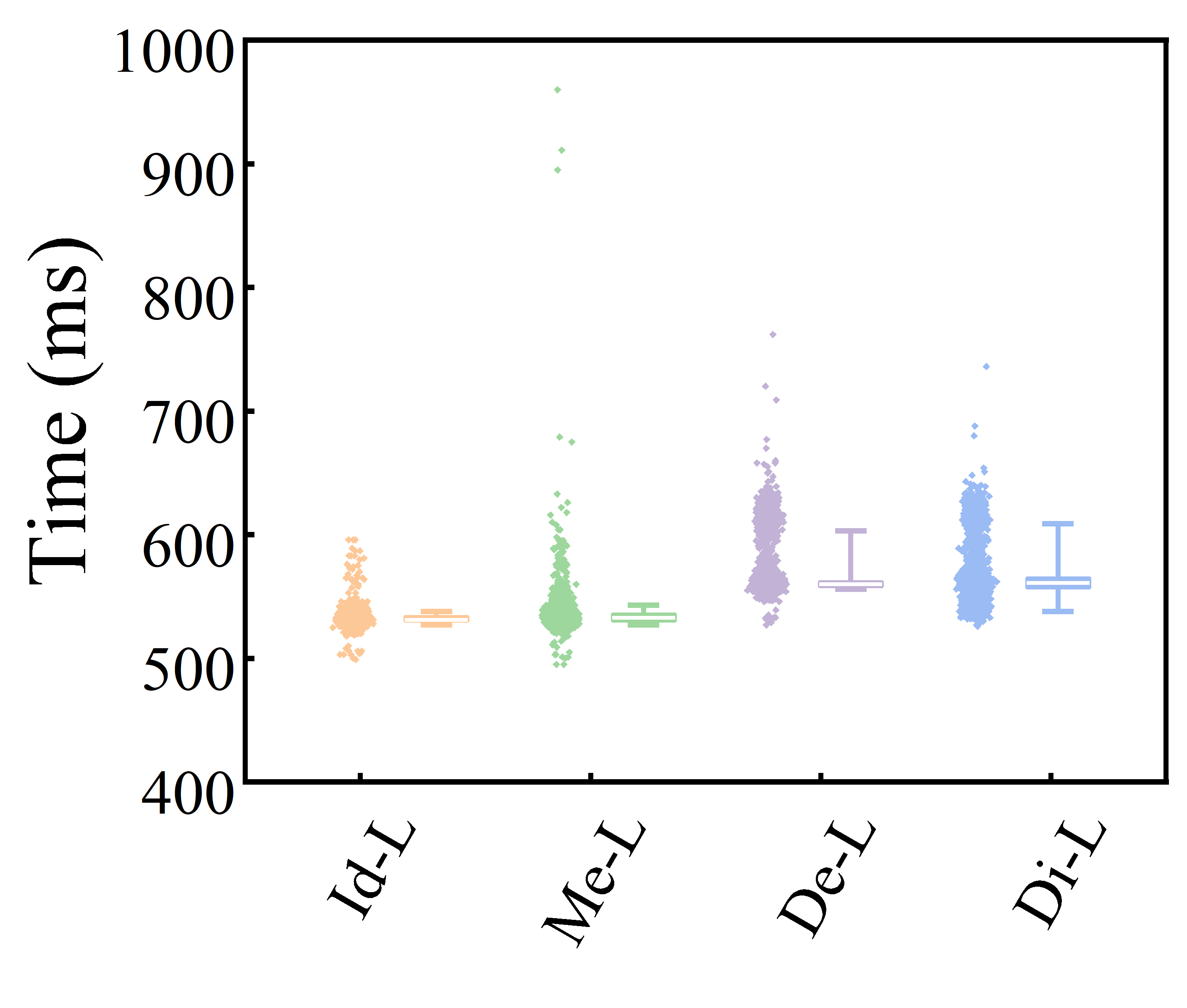}}
\hspace{0.1cm}
	\subfloat[S5-L-Distribution]{\label{7-dist5}\includegraphics[height=0.2\textwidth]{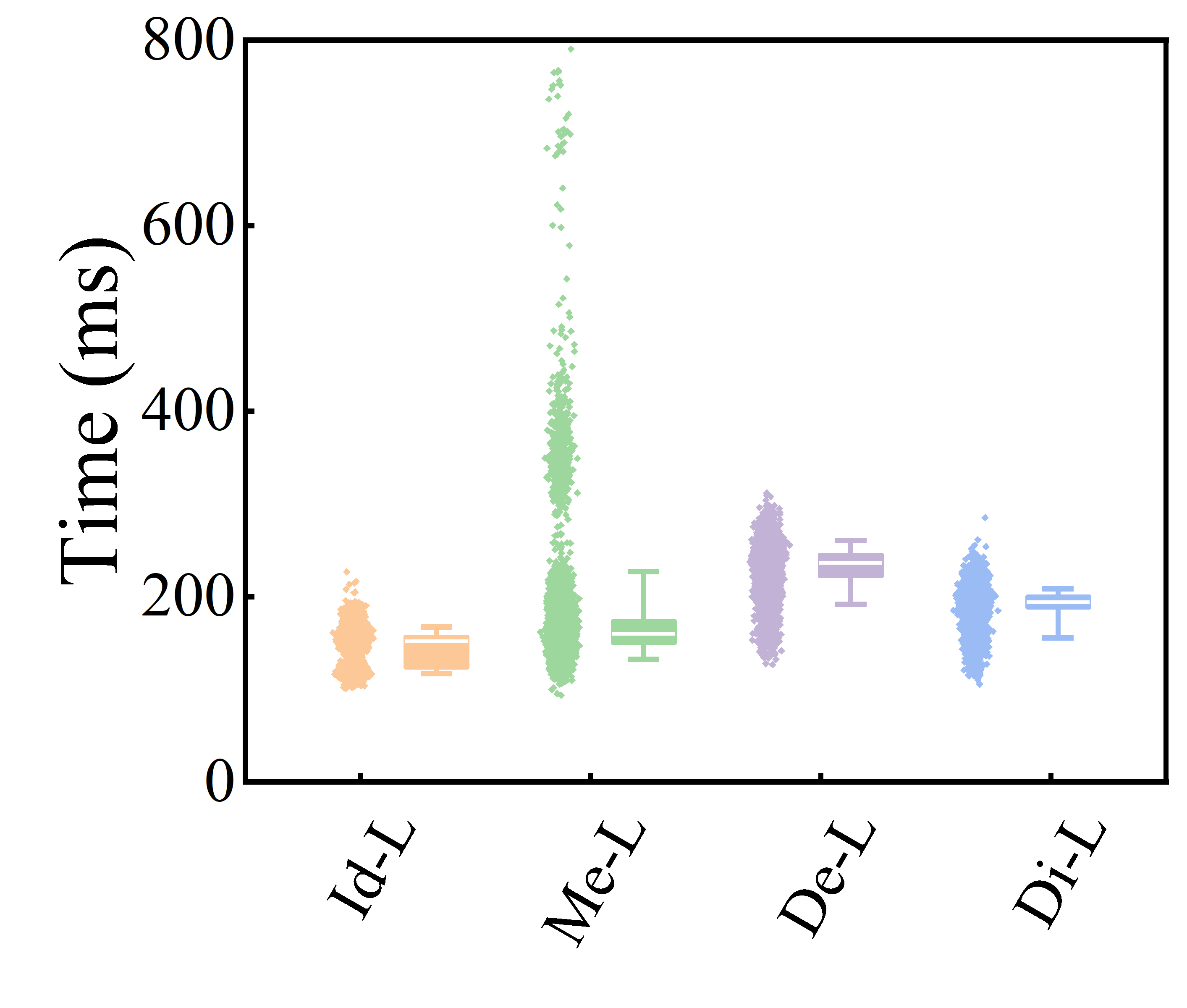}}
\hspace{0.1cm}
	\subfloat[S1-H-Distribution]{\label{7-dist6}\includegraphics[height=0.2\textwidth]{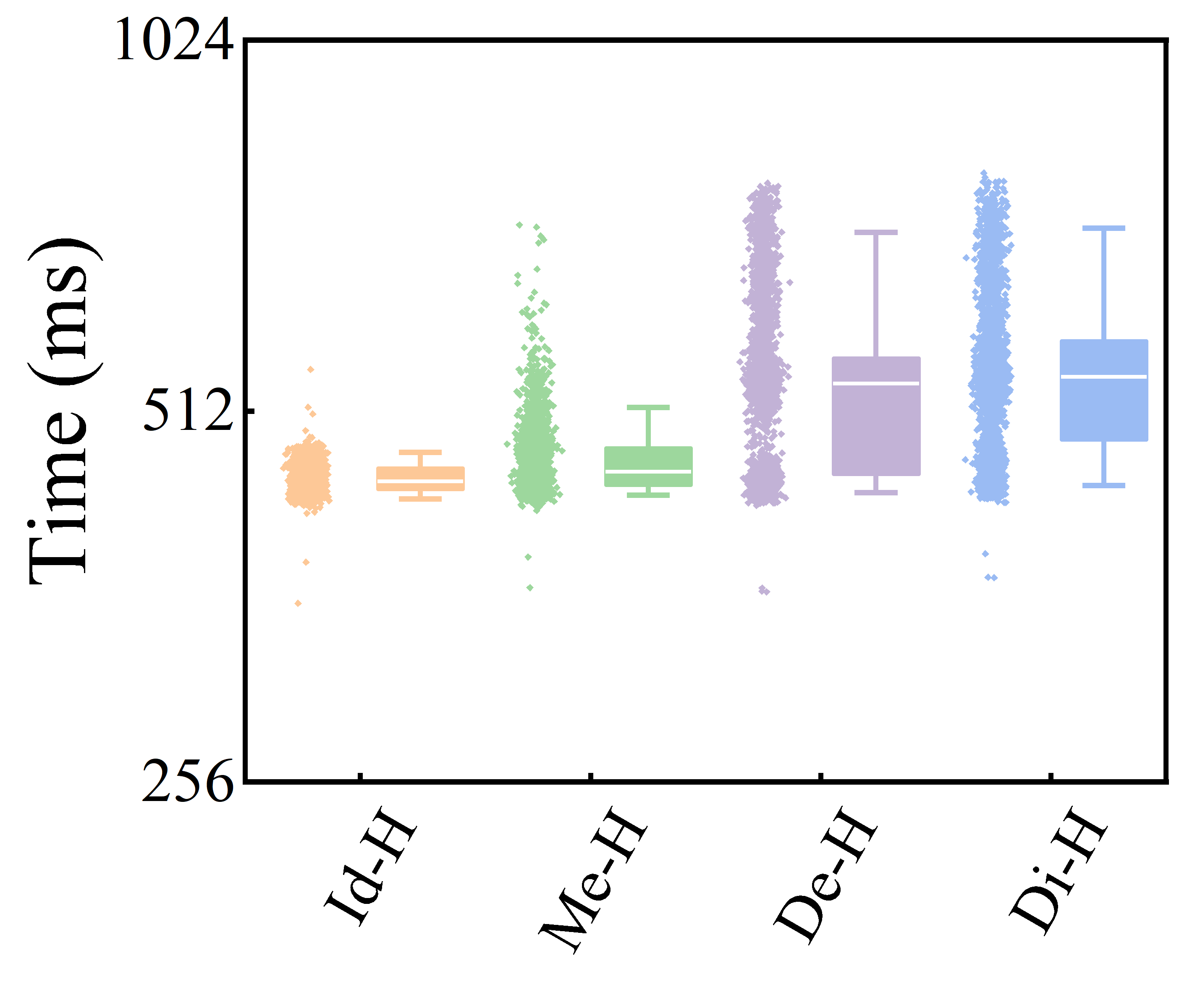}}
\hspace{0.1cm}
	\subfloat[S2-H-Distribution]{\label{7-dist7}\includegraphics[height=0.2\textwidth]{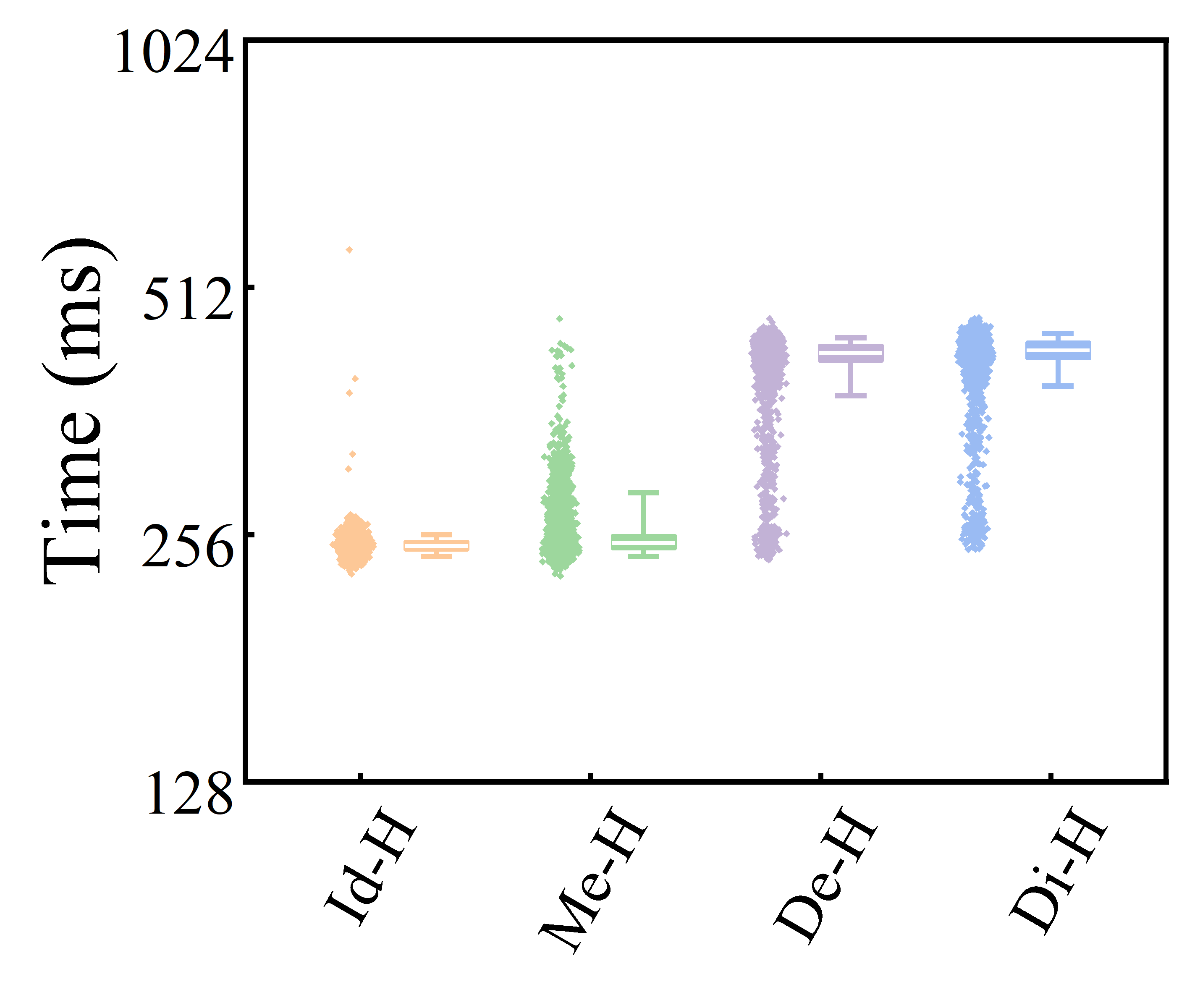}}
\hspace{0.1cm}
	\subfloat[S3-H-Distribution]{\label{7-dist8}\includegraphics[height=0.2\textwidth]{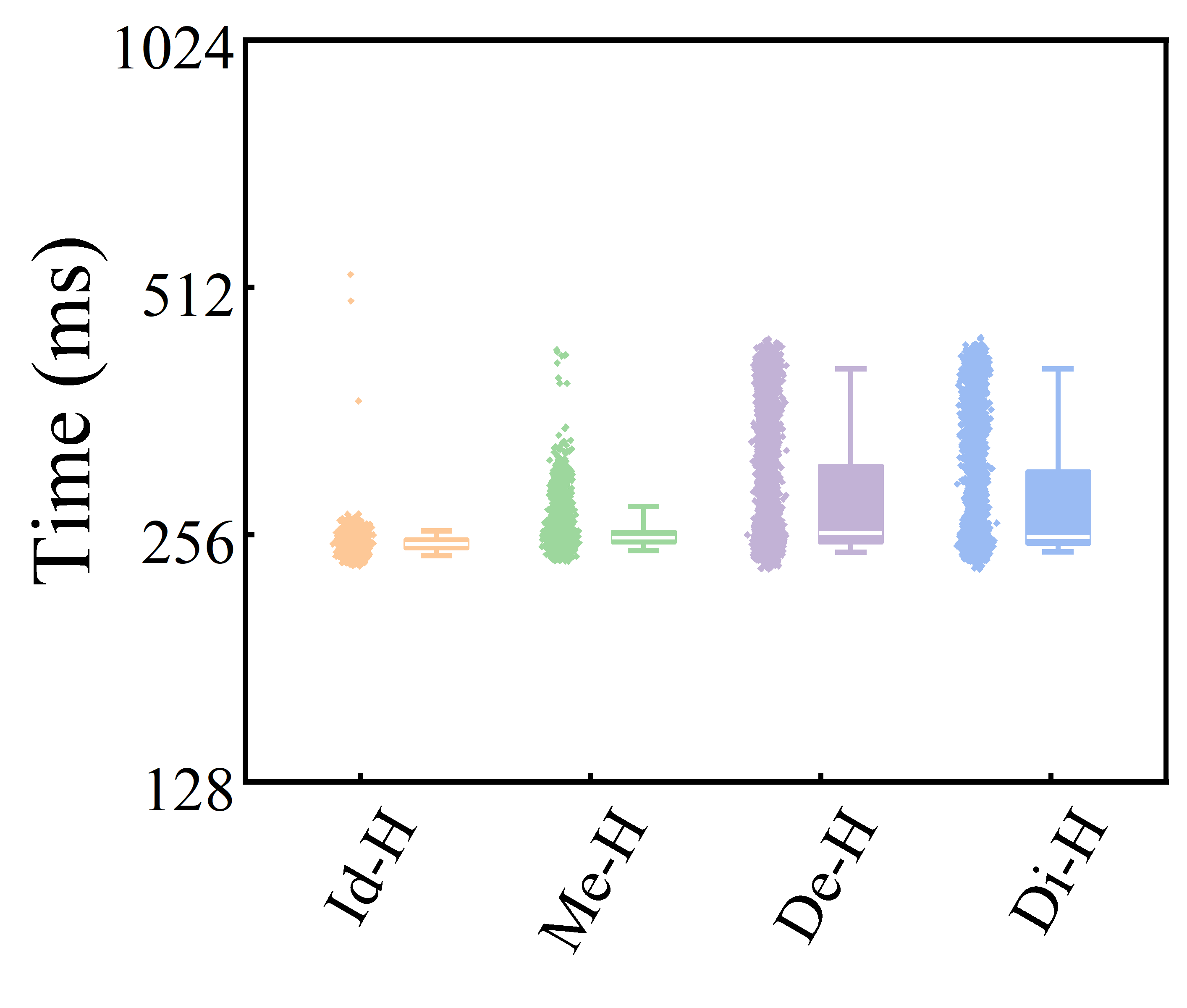}}
\hspace{0.1cm}
	\subfloat[S4-H-Distribution]{\label{7-dist9}\includegraphics[height=0.2\textwidth]{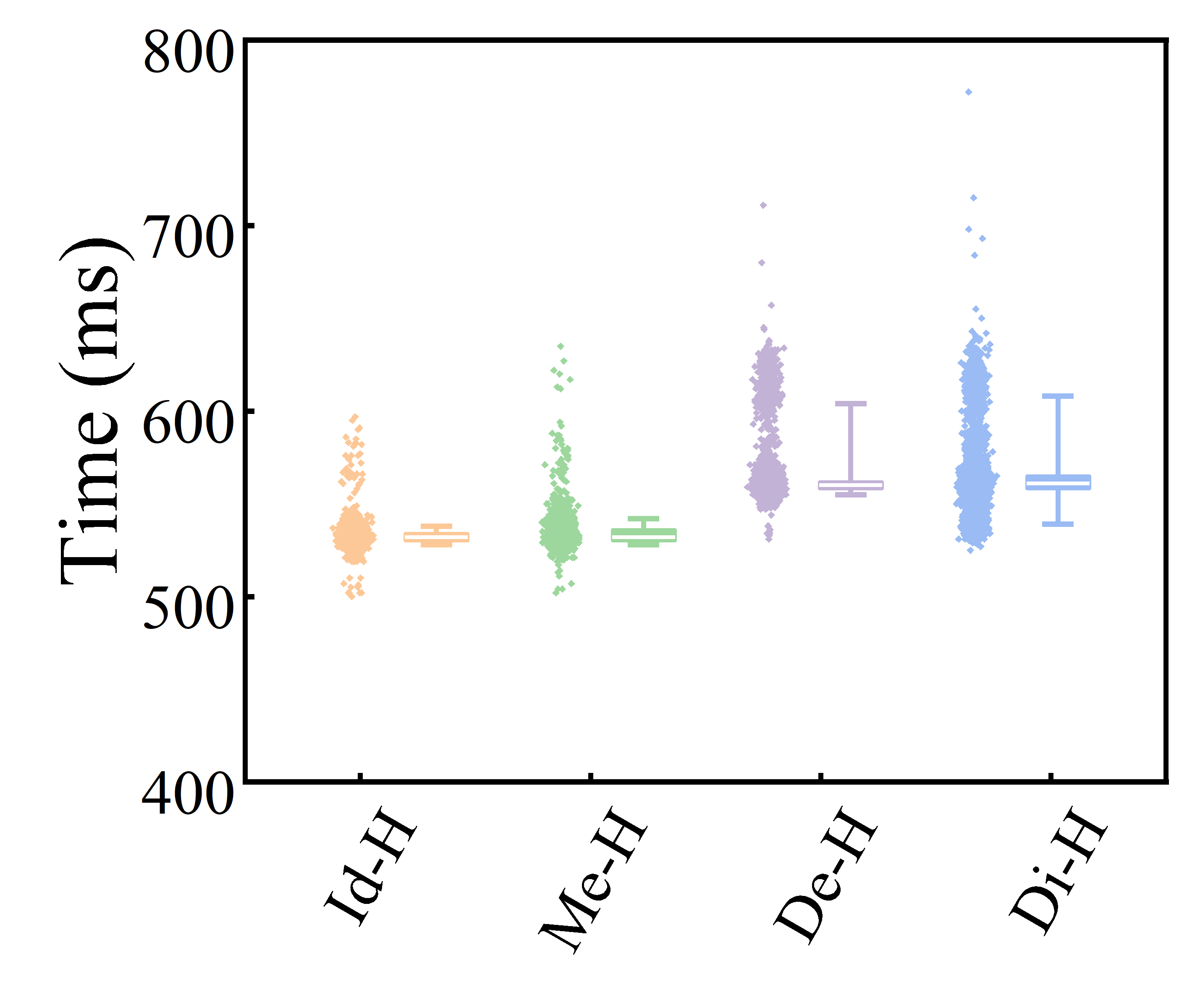}}
\hspace{0.1cm}
	\subfloat[S5-H-Distribution]{\label{7-dist10}\includegraphics[height=0.2\textwidth]{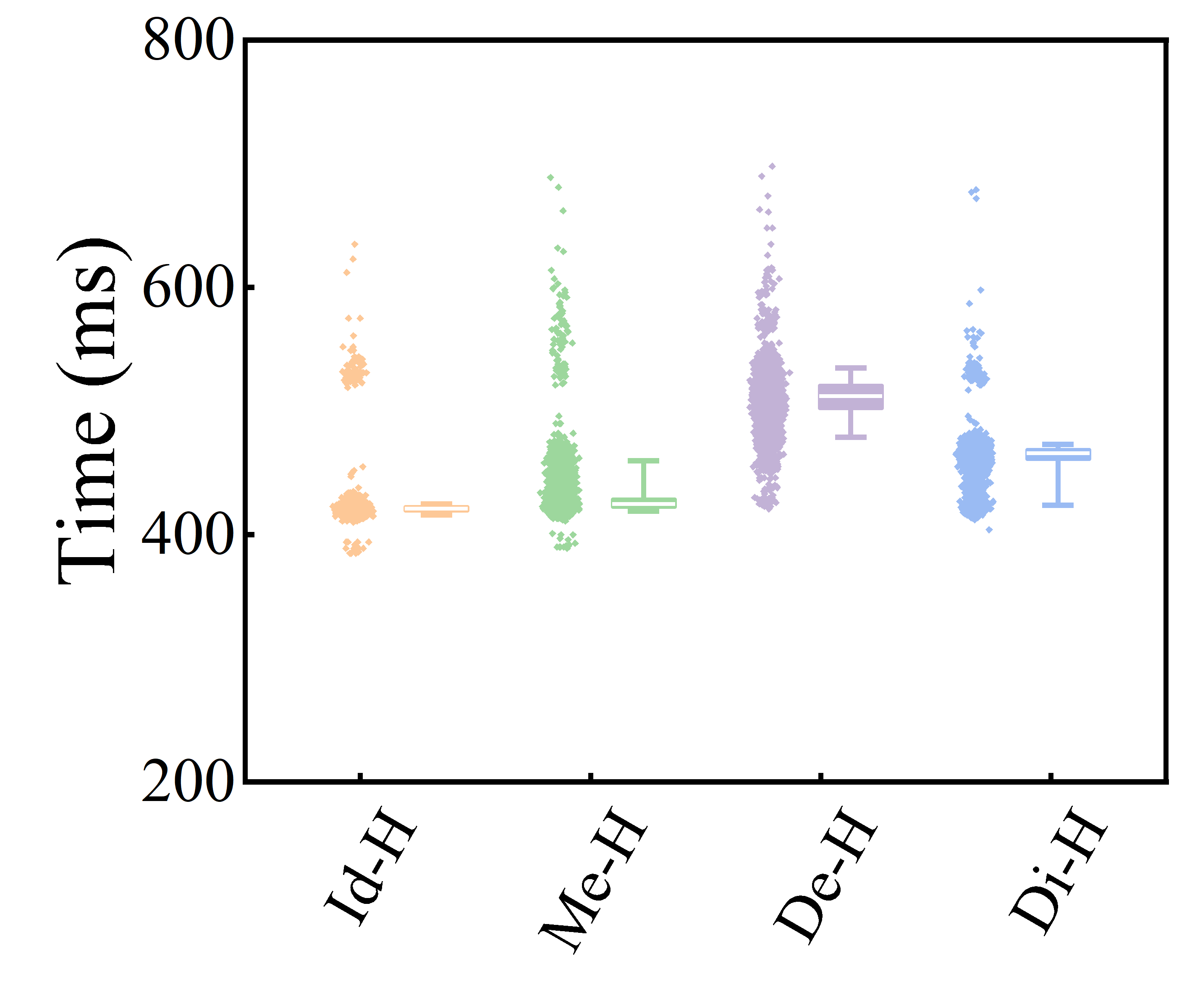}}	
	\caption{Job iteration time distribution under different schedulers.}
	\label{7-dist}
\end{figure}

\begin{figure*}[htbp]
	\centering
	\captionsetup[subfloat]{
		font=footnotesize,      
		labelfont=bf,         
		textfont=normalfont,   
		justification=centering 
	}
	\subfloat[S1-Average]{\label{8-av1}\includegraphics[width=0.19\textwidth]{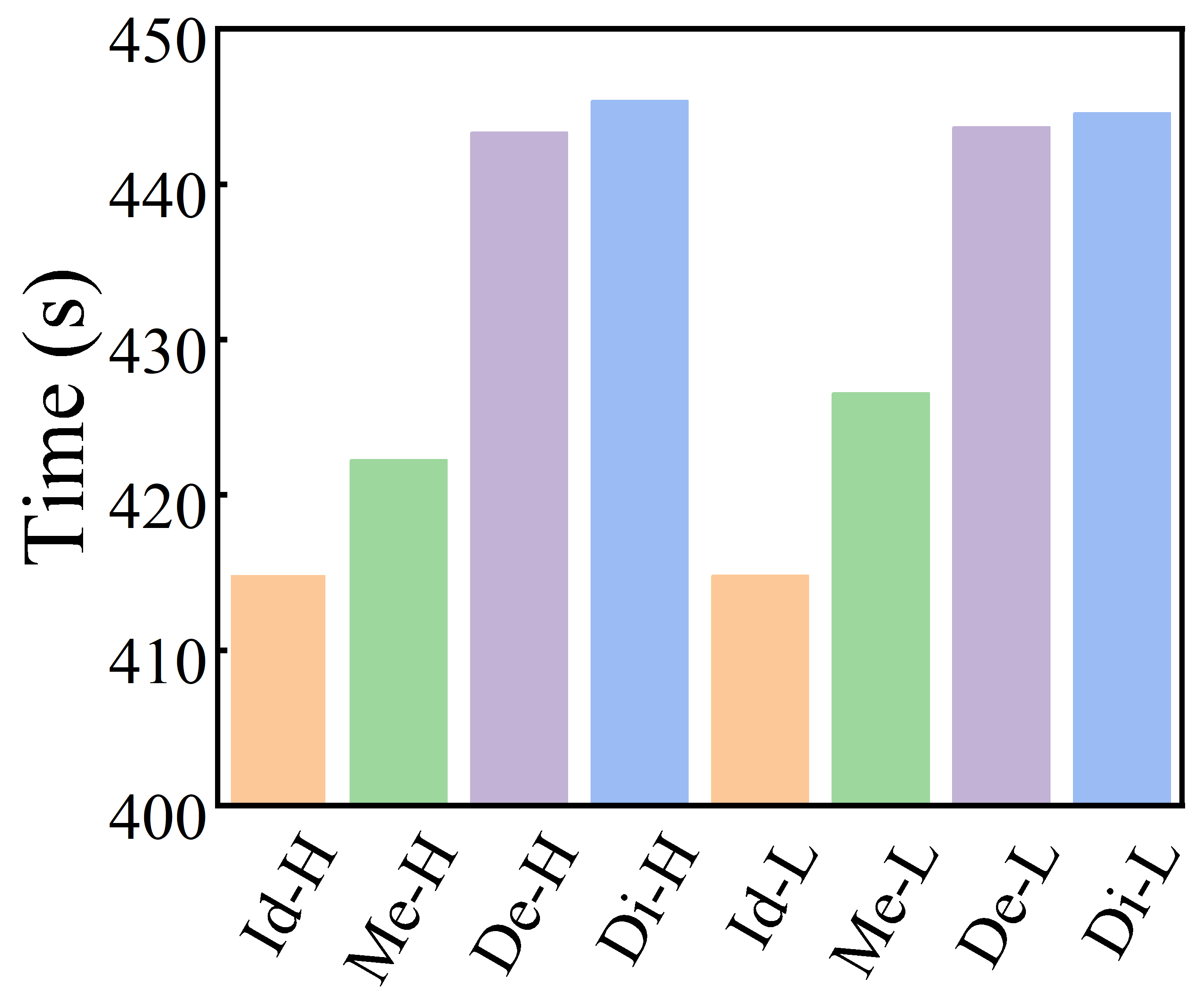}}
	\hfill
	\subfloat[S2-Average]{\label{8-av2}\includegraphics[width=0.19\textwidth]{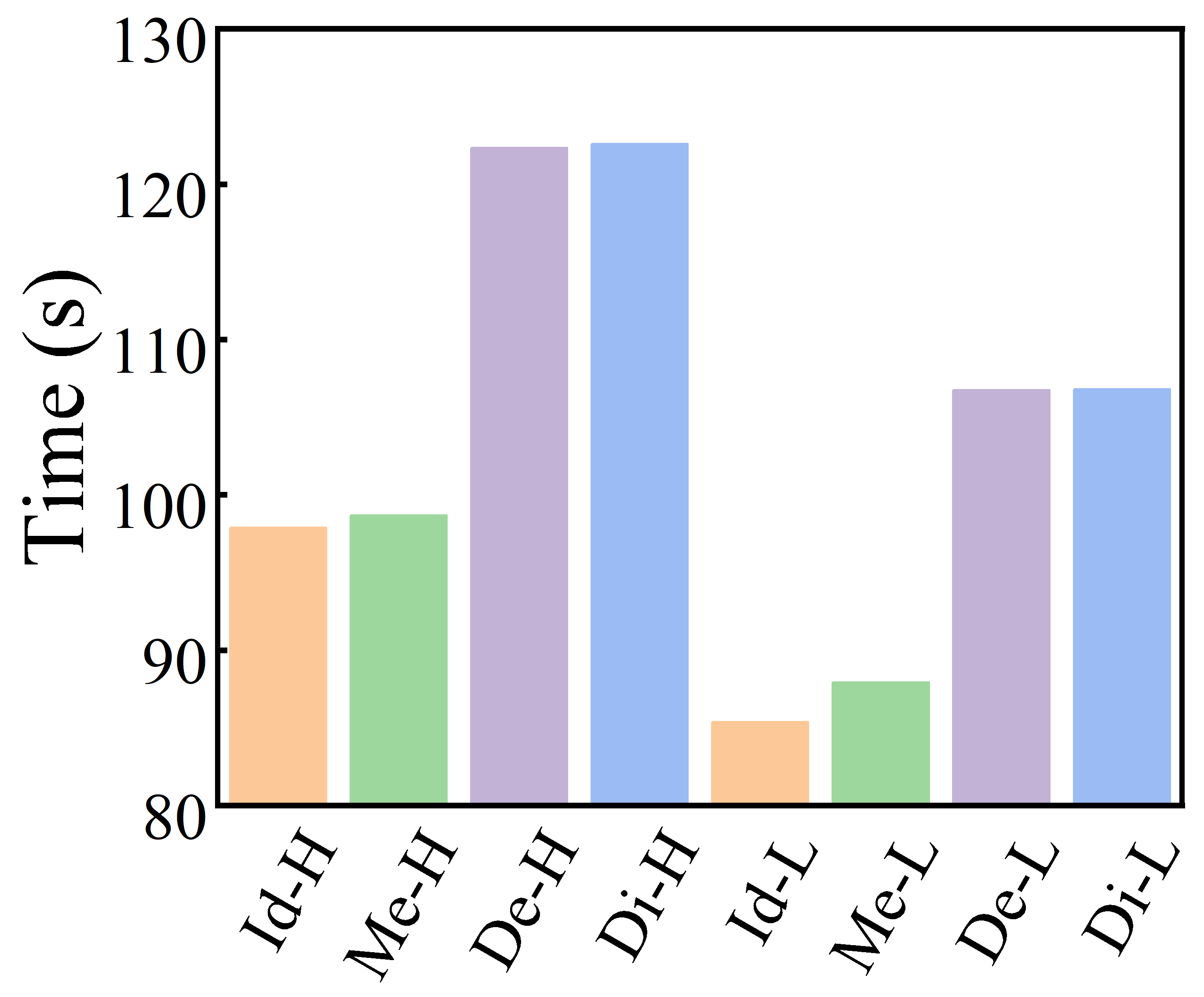}}
	\hfill
	\subfloat[S3-Average]{\label{8-av3}\includegraphics[width=0.19\textwidth]{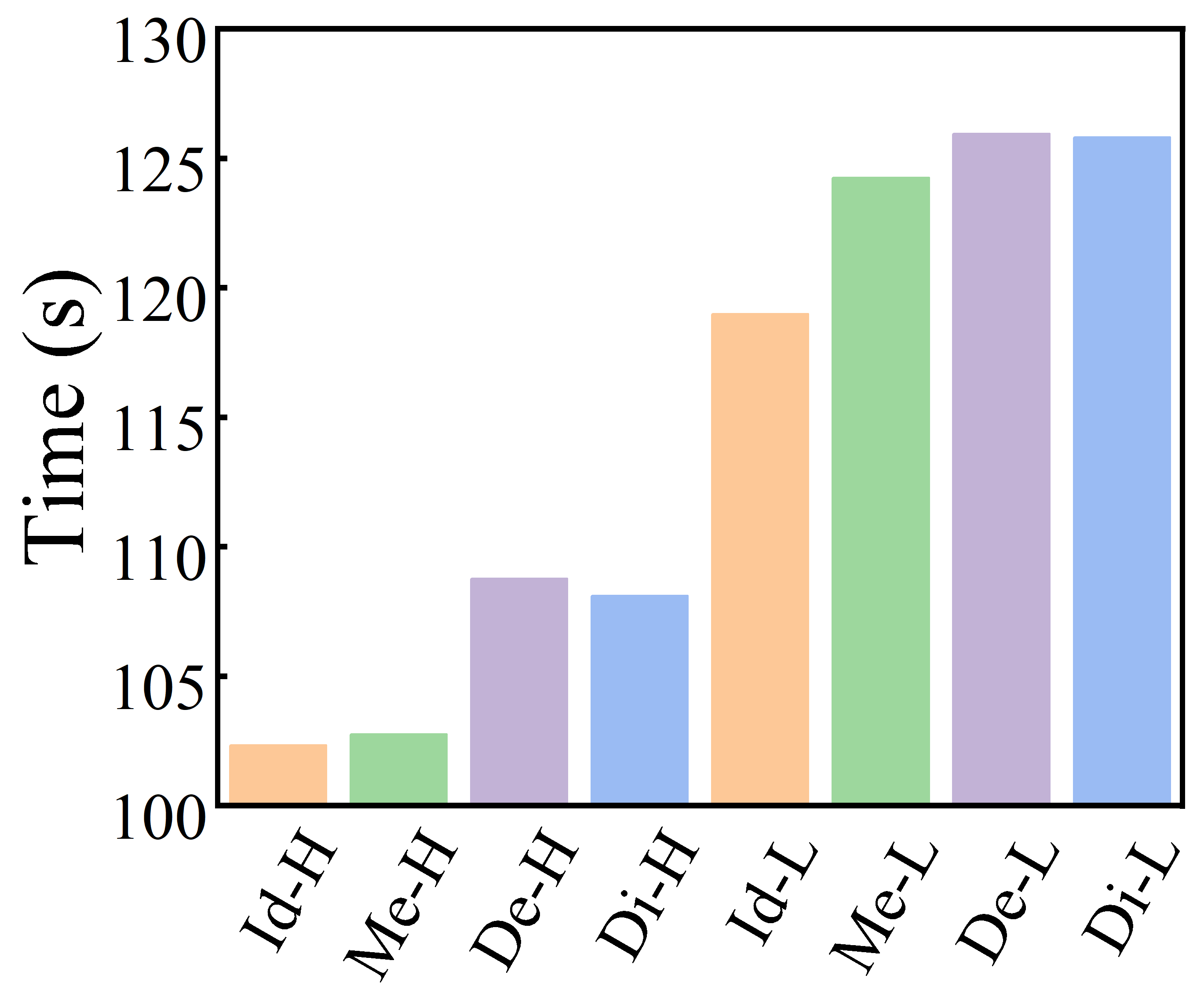}}
	\hfill
	\subfloat[S4-Average]{\label{8-av4}\includegraphics[width=0.19\textwidth]{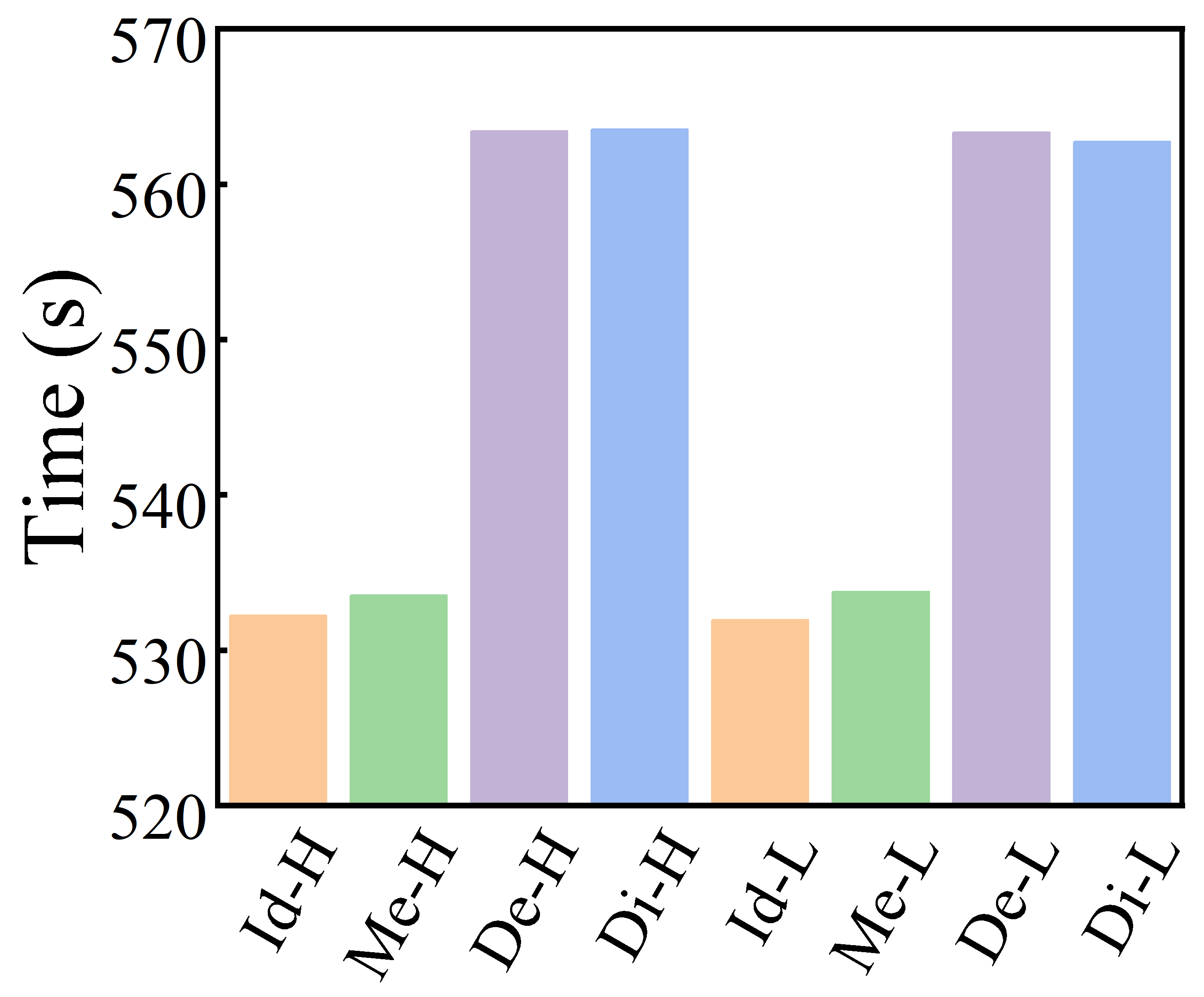}}
	\hfill
	\subfloat[S5-Average]{\label{8-av5}\includegraphics[width=0.19\textwidth]{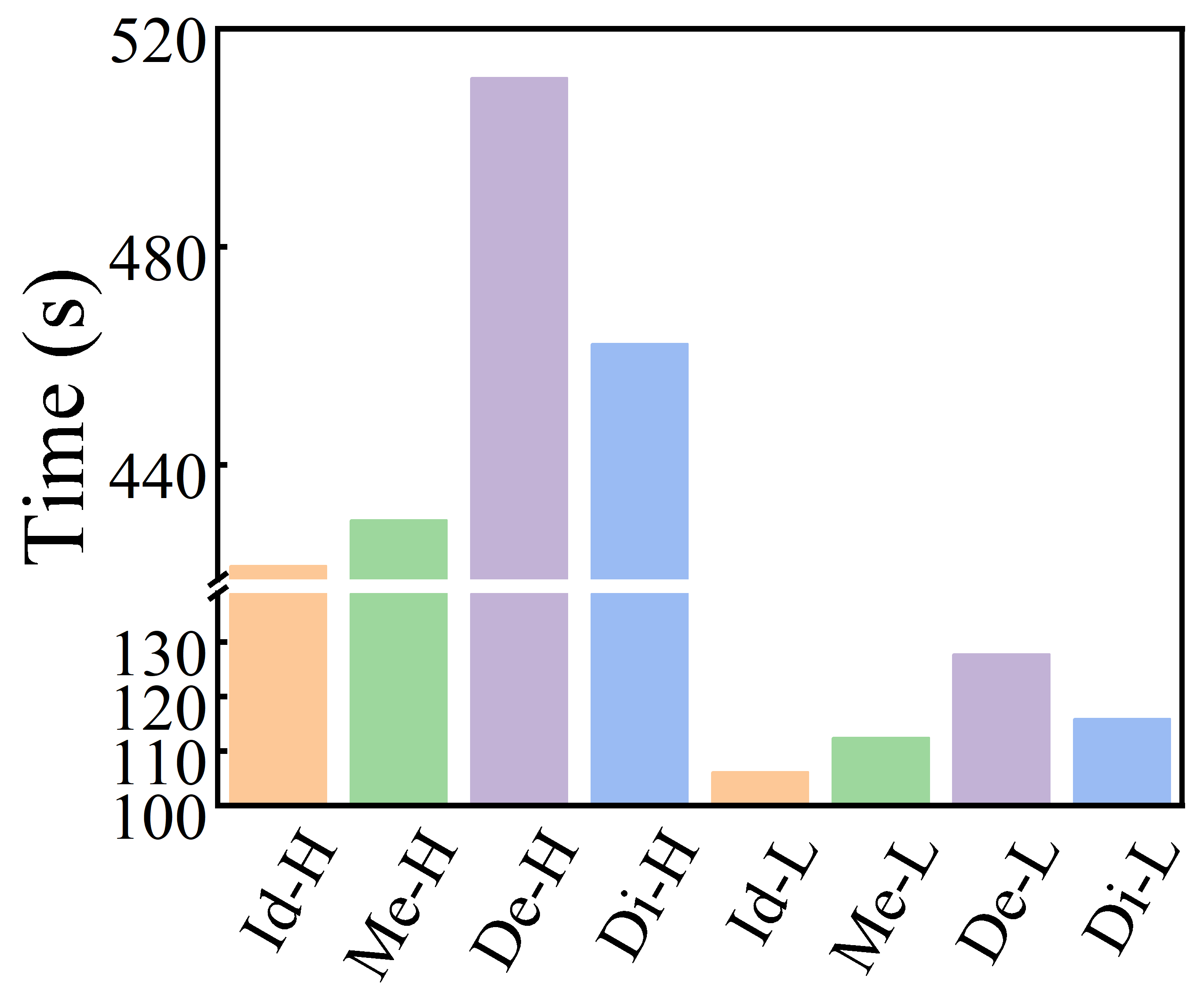}}
	\caption{Average time per 1,000 iterations under different schedulers.}

	\label{8-av} 
\end{figure*}

First, we evaluate the iteration time of jobs in each snapshot, as shown in Fig. \ref{7-dist} and Fig. \ref{8-av},  where `L' denotes low priority jobs and `H' represents high priority jobs.  

The box plots (Fig. \ref{7-dist1}-\ref{7-dist10}) illustrate the iteration time distribution of jobs. The box itself spans the interquartile range, from the 25th to the 75th percentile, and the whiskers extend to the 5th and 95th percentiles, encompassing 90\% of the observations.
As shown in the sub-figures, for most of its iterations, a job under the Metronome scheduler requires time that is close to the Ideal scenario and significantly less than that required by the Diktyo and Default schedulers.
It is worth noting that in the Metronome scenario, low priority jobs exhibit abnormal values due to suspensions triggered by the continuous monitoring mechanism upon detecting communication contention.

Fig. \ref{8-av1}-\ref{8-av5} illustrate the average time per 1,000 iterations of high and  low priority jobs in each snapshot. If multiple low priority jobs are present in the scenario, the average time for all low priority jobs is calculated. 
In snapshot 4, Metronome successfully avoids the congested link. In contrast, although Diktyo recognizes the latency between nodes, it fails to detect the dependencies of the job's first pod. This failure results in random scheduling to congested nodes, a behavior consistent with the Default scheduler.
In snapshot 5, both Metronome and Diktyo avoid the high latency link. 
However, Diktyo still experiences significant communication contention because it is unaware of two-dimensional bandwidth resources. Moreover, although  Diktyo occasionally achieves interleaving of job traffic patterns, it cannot maintain this benefit due to communication drift.  Furthermore, the Default scheduler continues to fail in avoiding congestion. 

Metronome achieves the best performance in snapshot 2, with high priority jobs accelerated by \textbf{19.35\%, 19.50\%} compared with the Default scheduler and Diktyo, and low priority jobs accelerated by \textbf{17.63\%, 17.67\%}.

Meanwhile, when compared with other schedulers, Metronome achieves a better acceleration ratio for high priority jobs than for low priority jobs across all snapshots. Furthermore, the training speed of these high priority jobs more closely approaches the ideal time of 1,000 iterations, with a difference not exceeding 2\%.
This is due to the fact that high priority jobs maintain uninterrupted execution when contention occurs. As a result, the communication phase will be reserved, enabling high priority jobs to operate as if they are in the scenario with exclusive resources.

\begin{figure*}[htbp]
	\centering
	\captionsetup[subfloat]{
		font=footnotesize,       
		labelfont=bf,          
		textfont=normalfont,   
		justification=centering 
	}
	\centering
	\includegraphics[width=9.1cm]{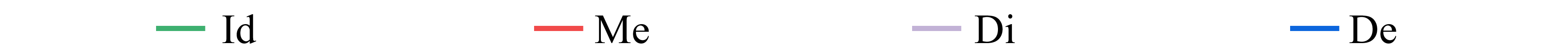}
	\par \vspace{-3mm}	
	\subfloat[S1-Loss]{\label{9-loss1}\includegraphics[width=0.19\textwidth]{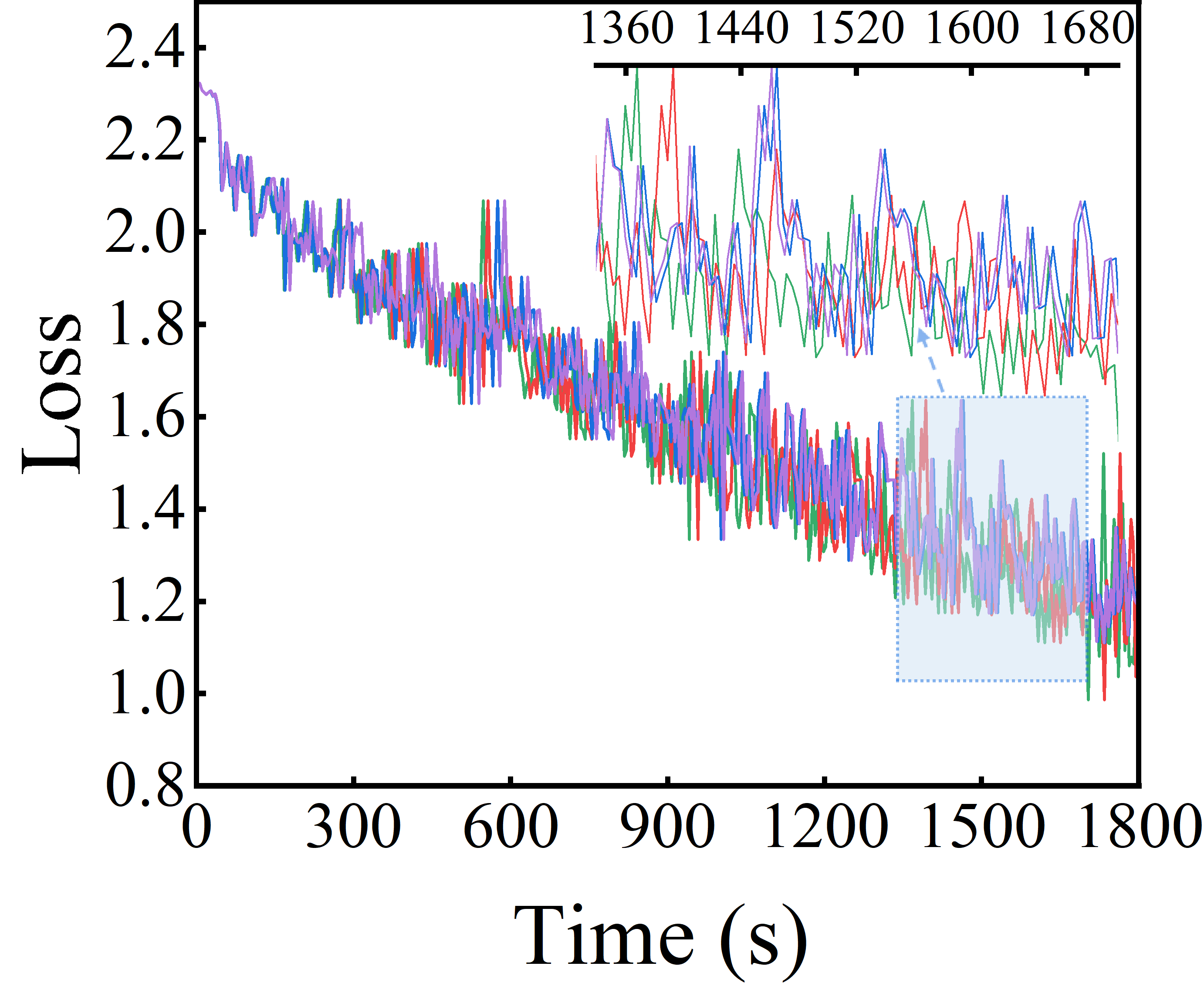}}
	\hfill
	\subfloat[S2-Loss]{\label{9-loss2}\includegraphics[width=0.19\textwidth]{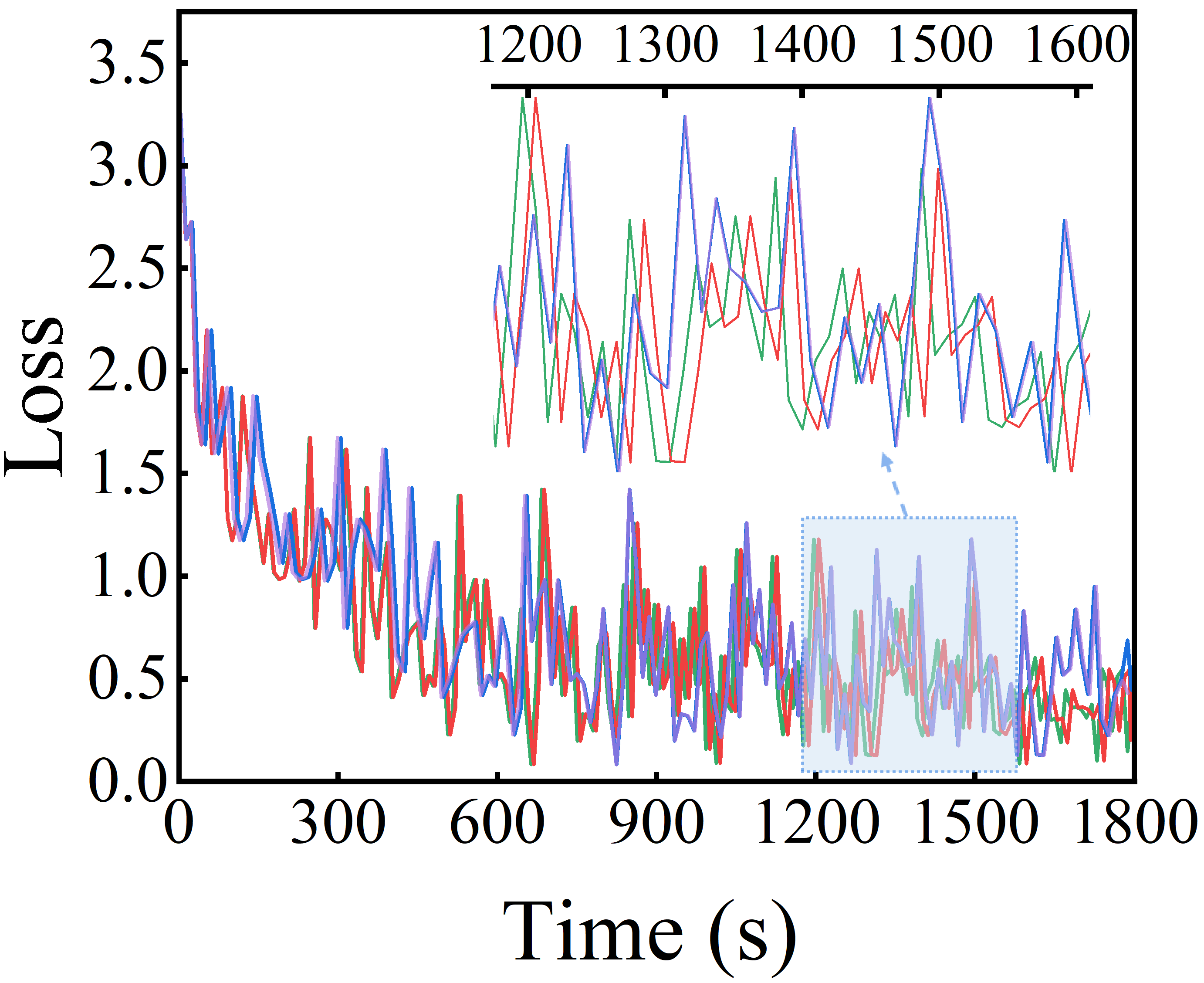}}
	\hfill
	\subfloat[S3-Loss]{\label{9-loss3}\includegraphics[width=0.19\textwidth]{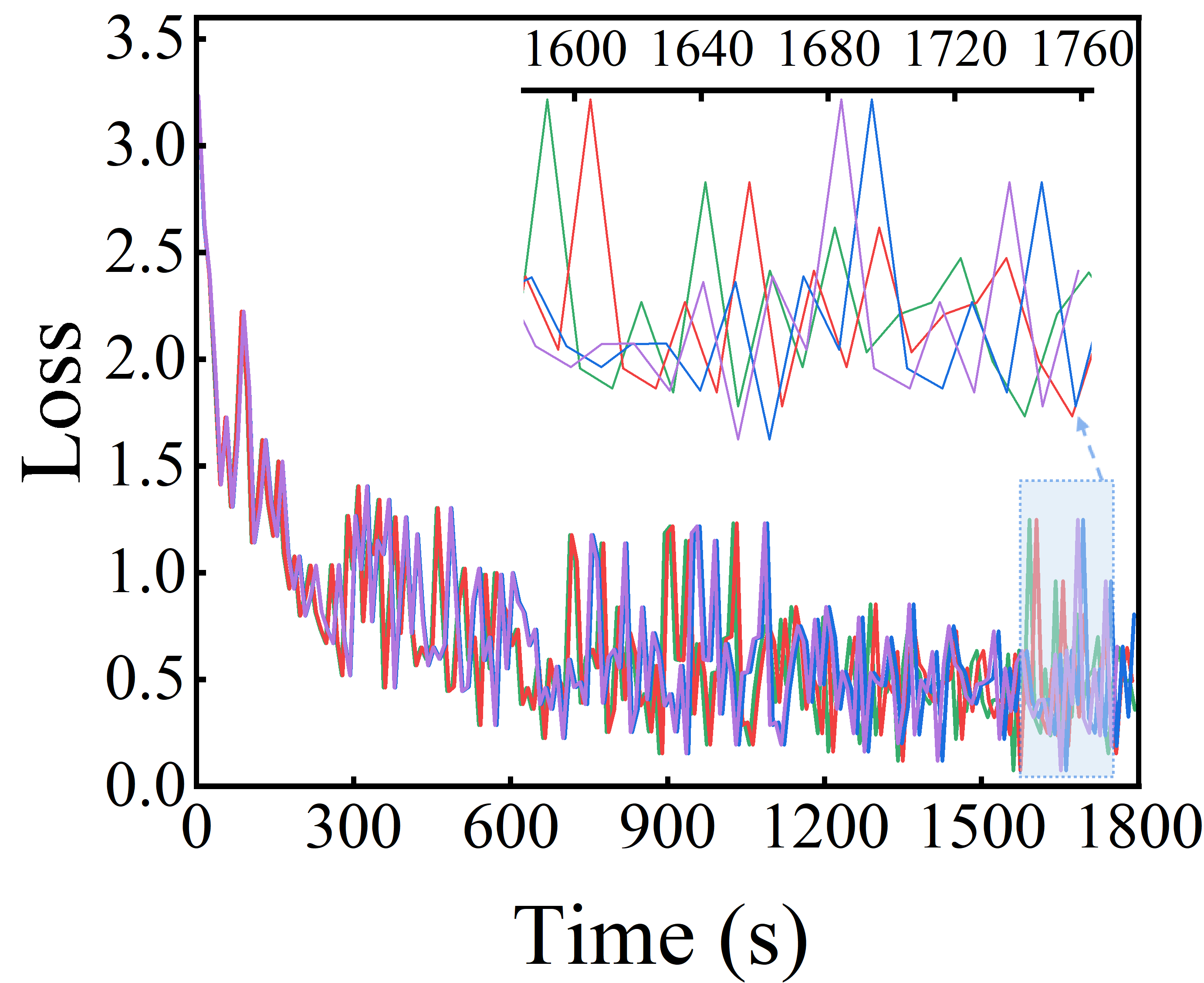}}
	\hfill
	\subfloat[S4-Loss]{\label{9-loss4}\includegraphics[width=0.19\textwidth]{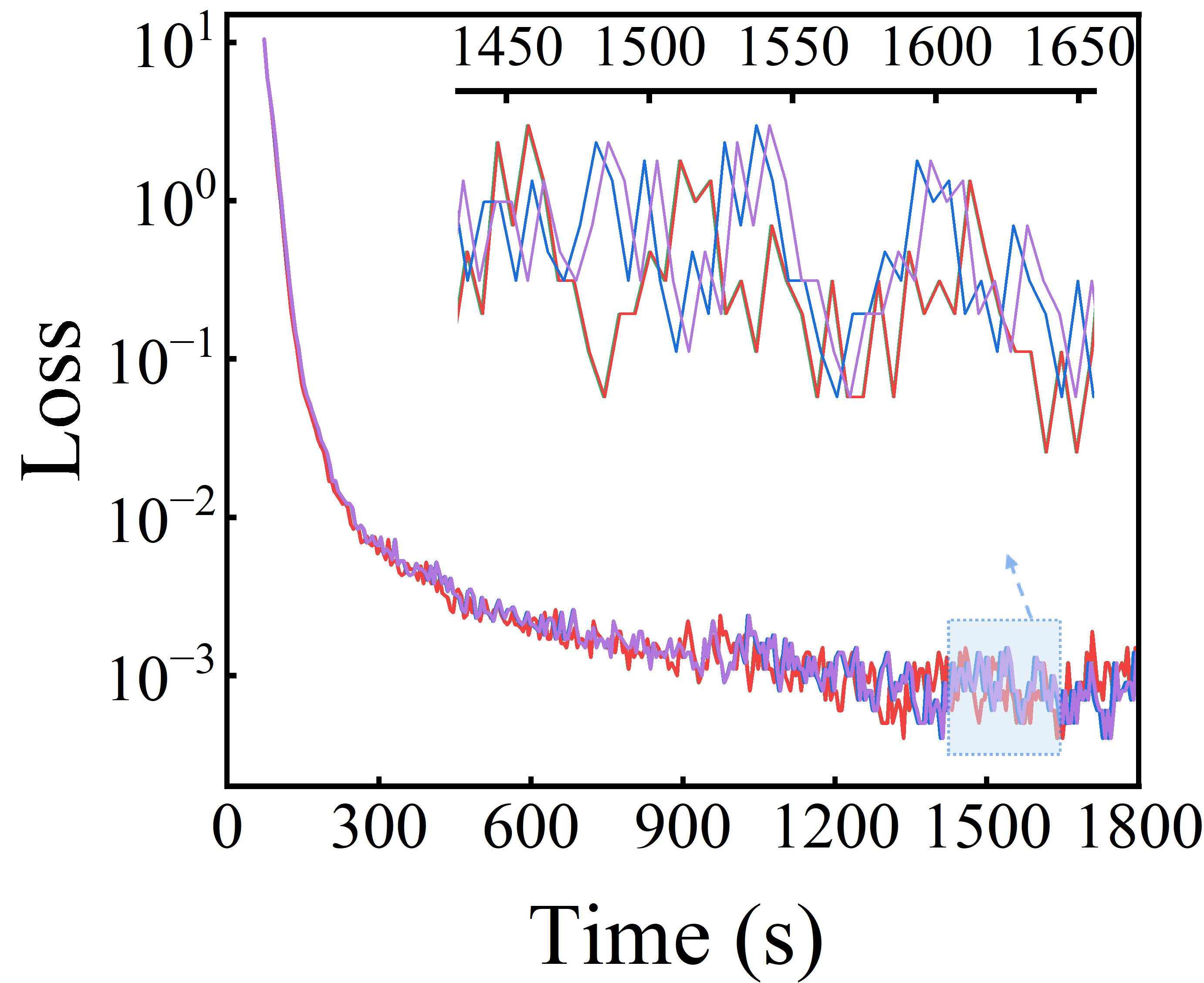}}
	\hfill
	\subfloat[S5-Loss]{\label{9-loss5}\includegraphics[width=0.19\textwidth]{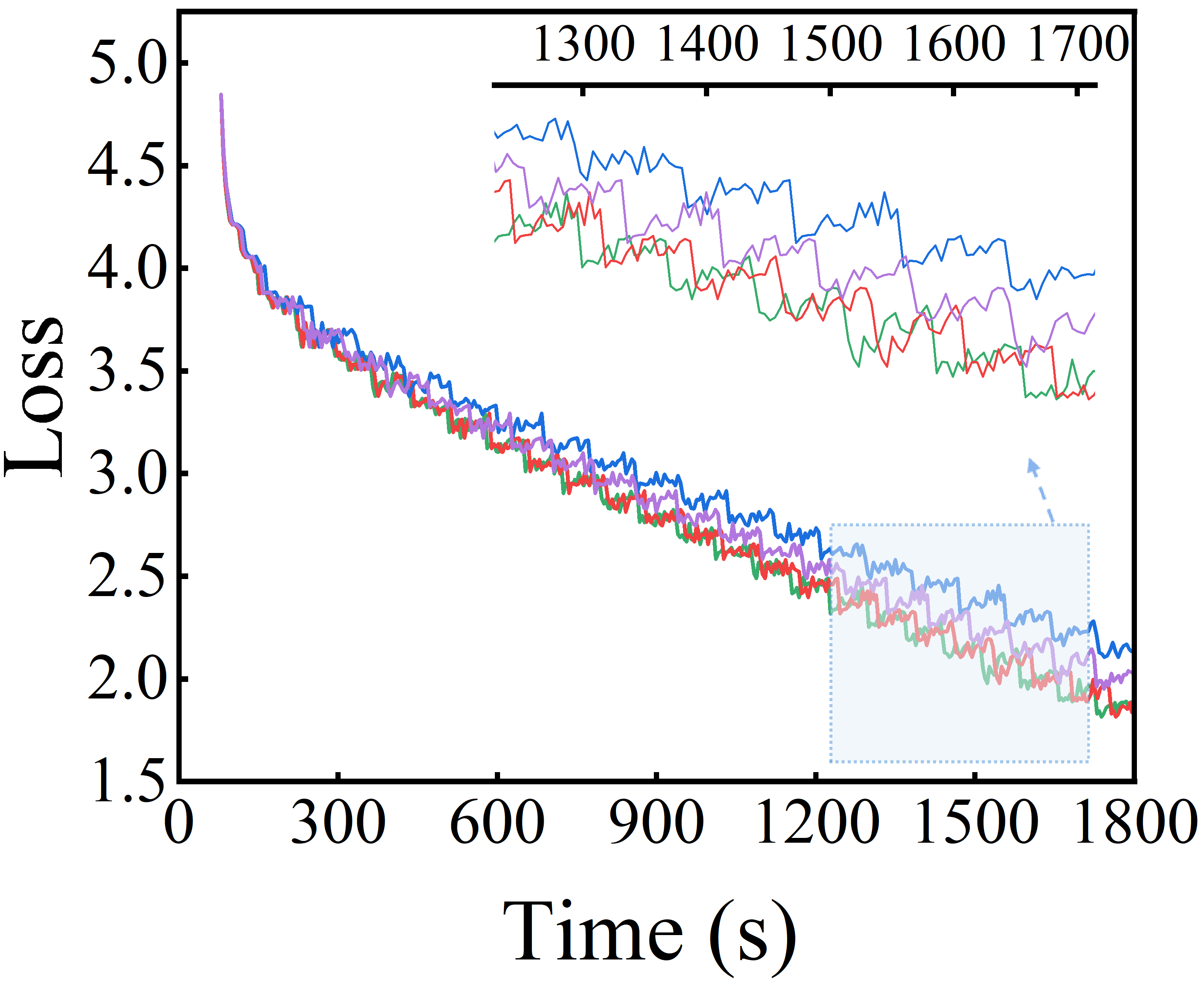}}
	\caption{Loss function curve under different schedulers.}
	\label{9-loss}
\end{figure*}

After scheduling, the loss function curve for high priority jobs is continuously plotted, as shown in Fig. \ref{9-loss1}-\ref{9-loss5}. Metronome can continually accelerate the training process through continuous monitoring, manifested as initially identical loss curves under all schedulers, with progressive divergence emerging over time. It is important to note that Metronome is not optimized for a specific model or communication mechanism, so it will not affect the performance in terms of accuracy or function loss in convergence.

In addition, Metronome improves network resource utilization versus the Default Scheduler and Diktyo, which is evidenced by increased average bandwidth utilization  (Table \ref{t5}). Metronome still achieved the best average bandwidth utilization improvement in snapshot 2. This is because the maximum job training acceleration results in the most additional synchronization traffic, which maximizes the utilization. Compared with the Default scheduler and Diktyo, Metronome improves the average bandwidth utilization by \textbf{23.10\%} and \textbf{23.20\%}, respectively.

\begin{table}[!htbp]
	\centering
	\setlength{\abovecaptionskip}{0pt}
	\setlength{\belowcaptionskip}{10pt}
	\setlength{\tabcolsep}{7.7pt} 
	\caption{Variation in average bandwidth utilization of the Metronome compared with other schedulers\label{t5}}
			\resizebox{8.8cm}{!}{
			\begin{tabular}{@{}>{\centering\arraybackslash}p{1.2cm} >{\centering\arraybackslash}p{1cm} >{\centering\arraybackslash}p{1cm} >{\centering\arraybackslash}p{1cm} >{\centering\arraybackslash}p{1cm} >{\centering\arraybackslash}p{1cm}@{}}				
			\toprule
			\textbf{Scheduler} & \textbf{S1 (\%)}  & \textbf{S2 (\%)}  & \textbf{S3 (\%)} & \textbf{S4 (\%)}  & \textbf{S5 (\%) }\\
			\midrule
			De &  4.75\;{\scriptsize ${\uparrow}$} & \textbf{23.10}\;{\textbf{\scriptsize ${\uparrow}$} }& 4.44\;{\scriptsize ${\uparrow}$}  &  5.35\;{\scriptsize ${\uparrow}$} & 16.20\;{\scriptsize ${\uparrow}$}\\
			Di &  4.87\;{\scriptsize ${\uparrow}$} & \textbf{23.20}\;{\textbf{\scriptsize ${\uparrow}$}} & 4.03\;{\scriptsize ${\uparrow}$}  &  5.21\;{\scriptsize ${\uparrow}$} & 5.30\;{\scriptsize ${\uparrow}$}\\
			Id &  2.39\;{\scriptsize ${\downarrow}$} & 1.88\;{\scriptsize ${\downarrow}$} & 1.31\;{\scriptsize ${\downarrow}$}  &  0.39\;{\scriptsize ${\downarrow}$} & 3.88\;{\scriptsize ${\downarrow}$}\\			
			\bottomrule
		\end{tabular}
	}
\end{table}

All jobs in the above snapshot are compatible. 
However, during the initial stage of tracing, an incompatibility is observed in one scenario (snapshot 0, GPT2 + GoogLeNet), as shown in Fig. \ref{5-time}.
This incompatibility arises because  the sum of their communication phases in the LCM period exceeds the length of period itself, making it impossible to interleave them without contention through TDM.
In this scenario, Metronome places the two jobs on nodes without any shared links, thereby successfully avoiding communication contention. In contrast, the Default scheduler, due to its lack of awareness of bandwidth constraints, fails to isolate the jobs, which results in contention on the shared links.

\subsubsection{Total Completion Time}
\label{4-2-2}
\begin{figure}[!htbp]
	\small
	\centering
	\includegraphics[width=8.8cm]{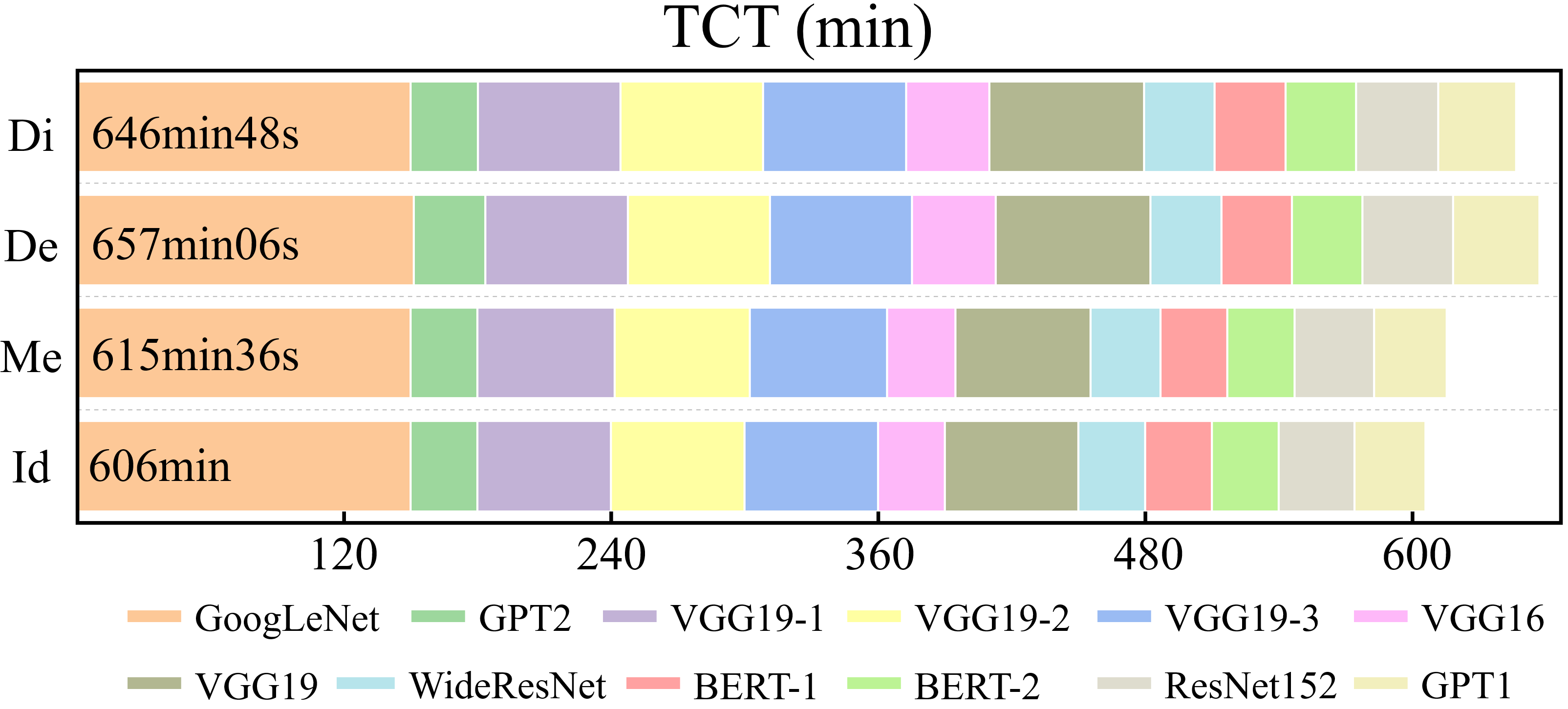}
	\caption{TCT under different schedulers.} 
	\label{10-TCT}
\end{figure}

The training jobs in the ideal trace are used as a baseline to determine the number of iterations and startup times. For each job, the estimated completion time is calculated by multiplying the number of iterations by the average iteration time and then adding the job's startup time.  
Fig. \ref{10-TCT} depicts the TCT for distributed training jobs in the trace under different schedulers.
Metronome completes all jobs 30 and 40 minutes earlier than the Default scheduler and Diktyo, respectively. Moreover, it trails the Ideal scheduler by less than 10 minutes.
Metronome can utilize the saved time to train additional jobs, improving cluster resource utilization and ultimately increasing cluster profitability.
Notably, modifying job durations in the ideal trace can lead to TCT variations, in which case Metronome might demonstrate better performance.

\subsubsection{Impact of Parameter Variations}
\label{4-2-3}
Furthermore, we change system parameters to evaluate the adaptability and the persistence of Metronome.

\begin{figure}[htbp]
	\centering
	\captionsetup[subfloat]{
		font=small,       
		labelfont=bf,            
		justification=justified 
	}
	\includegraphics[width=8.8cm]{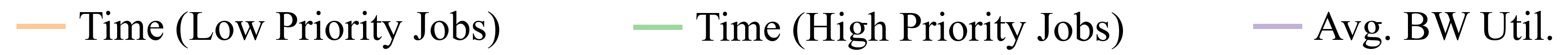}
	\par \vspace{-3mm}
	\subfloat{\label{11-bwchange-main}\includegraphics[width=8.8cm, height=2.7cm]{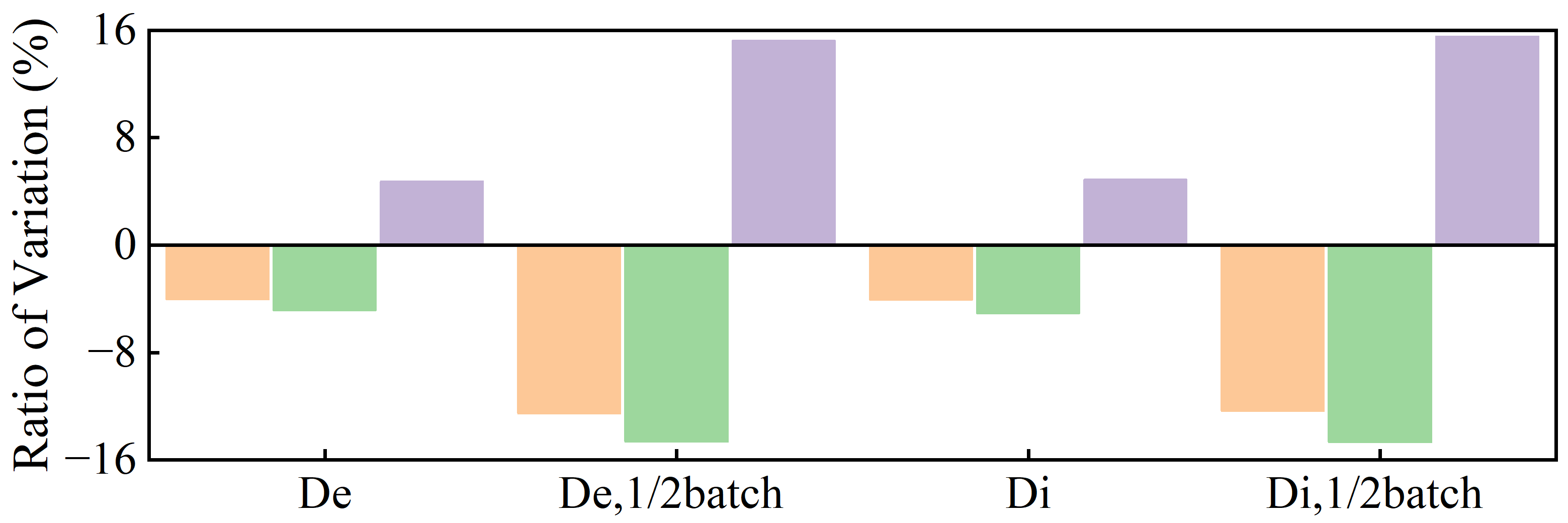}}
	\caption{Impact of changing the bandwidth requirements.}
	\label{11-bwchange}
\end{figure}

$\bullet$ \textbf{Changing the bandwidth requirement.} 
In snapshot 1, we halve the batch size of all jobs to change the communication duty cycle. Due to the continuous monitoring mechanism,  Metronome can adapt effectively to these changes.  
Fig. \ref{11-bwchange} shows the performance improvements of Metronome compared with the Default scheduler and Diktyo, in both the original scenario and the one with a reduced batch size.
In this case, a reduced batch size increases the communication duty cycle, thereby intensifying contention between jobs.
Metronome leverages continuous monitoring to achieve greater overall metric compared with prior configurations.

$\bullet$ \textbf{Changing the latency parameter.} 
By systematically varying the iPerf3 parameter to induce different levels of network latency, we quantify its impact on scheduling performance.
Metronome demonstrates its ability to avoid congested links, maintaining stable performance metrics that are largely unaffected by changes in congestion levels. 

\begin{figure}[htbp]
	\centering
	\captionsetup[subfloat]{
		font=footnotesize,      
		labelfont=bf,        
		textfont=normalfont,  
		justification=justified 
	}
	\centering
	\includegraphics[width=8.8cm]{Figures/legend-2.png}
	\par \vspace{-3mm}
	\subfloat[Performance comparison of Metronome and the Default scheduler under different latency parameter values]{\label{12-lachange1}\includegraphics[width=8.8cm, height=2.7cm]{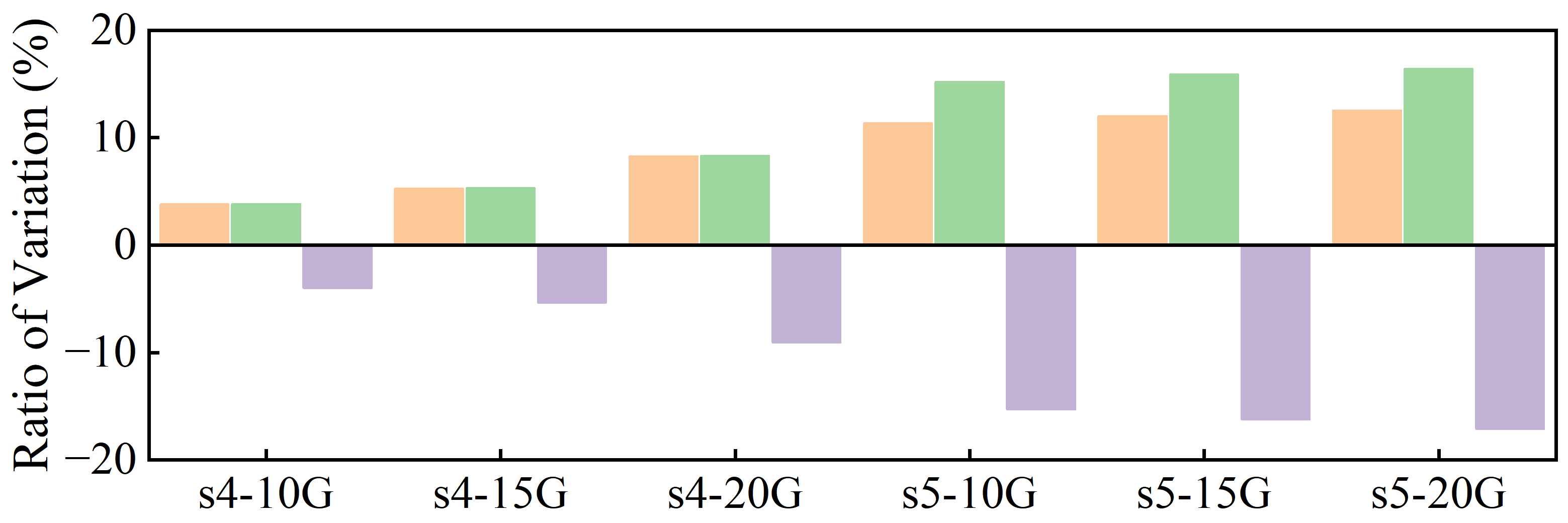}}
	\hspace{0.2cm}
	\includegraphics[width=8.8cm]{Figures/legend-2.png}
	\par \vspace{-3mm}
	\subfloat[Performance comparison of Metronome and Diktyo under different latency parameter values]{\label{12-lachange2}\includegraphics[width=8.8cm, height=2.7cm]{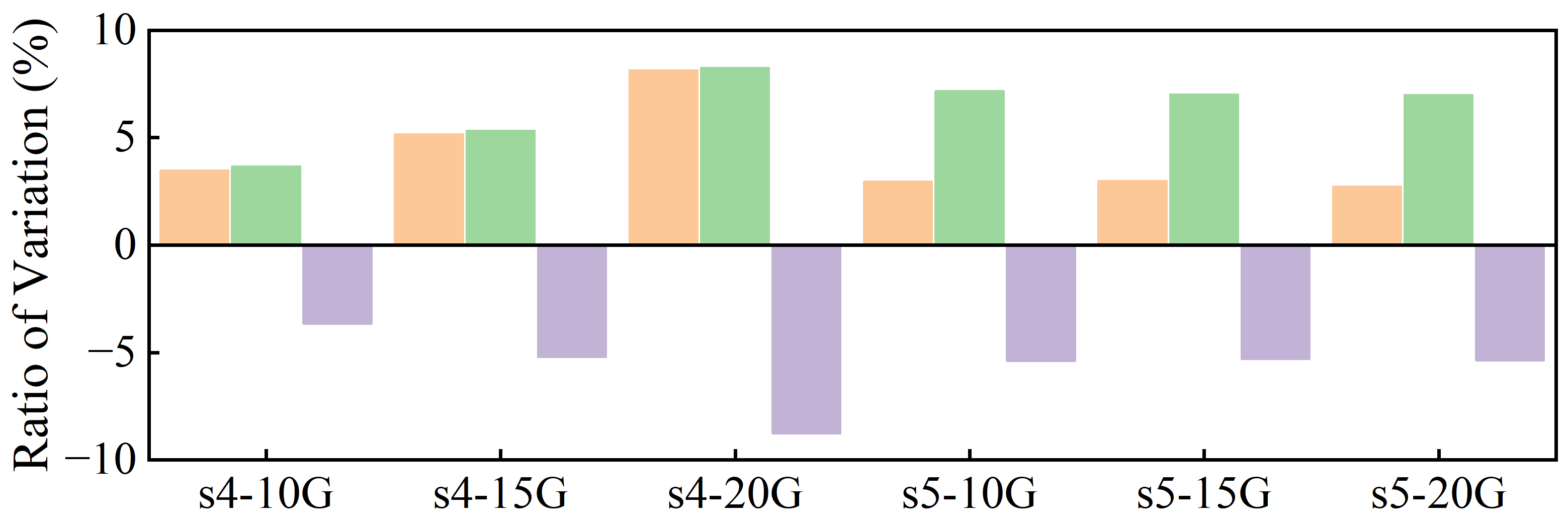}}
	\caption{Impact of changing the latency parameter.}
	\label{12-lachange}
\end{figure}

The improvement in Metronome compared with the Default scheduler across various snapshots is illustrated in Fig. \ref{12-lachange1}. Concretely, as congestion increases, jobs scheduled by the Default scheduler experience a decline in execution performance, while Metronome achieves more substantial relative gains.

The improvement in Metronome compared with the Diktyo across various snapshots is illustrated in Fig. \ref{12-lachange2}. The performance of Diktyo in snapshot 4 is comparable to that of the Default scheduler. 
In snapshot 5, Diktyo also avoids the congested links, allowing it to maintain stable performance regardless of congestion levels.

$\bullet$ \textbf{Extending the experiment observation duration.} 
The above experiments are all conducted within 0.5 hours. We further repeat the experiment in §\ref{4-2-1} for an extended duration of 5 hours. 
Using Metronome, we then compare the average time per 1,000 iterations from this long-running experiment against the metrics from the original 30-minute runs in Table \ref{t6}.
No significant differences are observed between the two sets of measurements. 
Furthermore, continuous communication contention is still observed for both the Default scheduler and Diktyo during the long-running experiment.
Therefore, the performance gains achieved by Metronome demonstrate persistence, and the same conclusion can also be drawn from the average bandwidth utilization.

\begin{table}[!htbp]
	\centering
	\setlength{\abovecaptionskip}{0pt}
	\setlength{\belowcaptionskip}{10pt}
	\setlength{\tabcolsep}{7.7pt} 
	\caption{Impact of extending the observation duration\label{t6}}
		\resizebox{8.8cm}{!}{
\begin{tabular}{w{c}{0.8cm} w{c}{1.6cm} w{c}{0cm} w{c}{1.6cm} w{c}{0cm}}
		\toprule
		\multirow{2}{*}{\textbf{Snapshot}} & \multicolumn{2}{c}{\textbf{Time (Low Priority Jobs, s)}} & \multicolumn{2}{c}{\textbf{Time (High Priority Jobs, s)}} \\
		\cmidrule(lr){2-3} \cmidrule(lr){4-5}
		& 0.5h  & 5h & 0.5h & 5h \\
		\midrule
		S1 & 426.55 & 422.28 & 422.26 & 418.67 \\
		S2 & 87.92 & 87.65 & 98.68 & 99.43 \\
		S3 & 124.25 & 124.07 & 102.76 & 102.67 \\
		S4 & 553.75 & 533.69 & 533.53 & 533.60 \\
		S5 & 112.41 & 111.87 & 429.88 & 433.56 \\
		\bottomrule
	\end{tabular}
}
\end{table}

\subsection{Ablation Studies}
\label{4-3}
To verify the necessity of each component in Metronome, we conduct ablation experiments.

\begin{figure}[htbp]
	\centering
	\captionsetup[subfloat]{
		font=footnotesize,     
		labelfont=bf,        
		textfont=normalfont,   
		justification=justified
	}
	\centering
	\includegraphics[width=8.8cm]{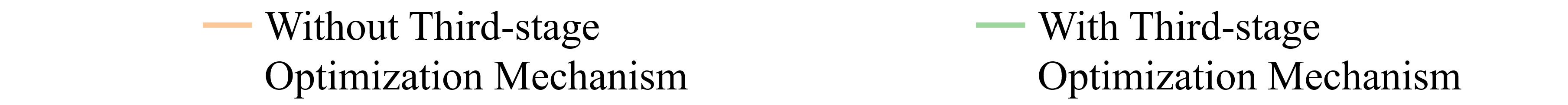}
	\par \vspace{-4mm}
	\subfloat[\footnotesize Impact of removing the third stage optimization mechanism]{\label{13-wothird}\includegraphics[width=8.8cm, height=2.7cm]{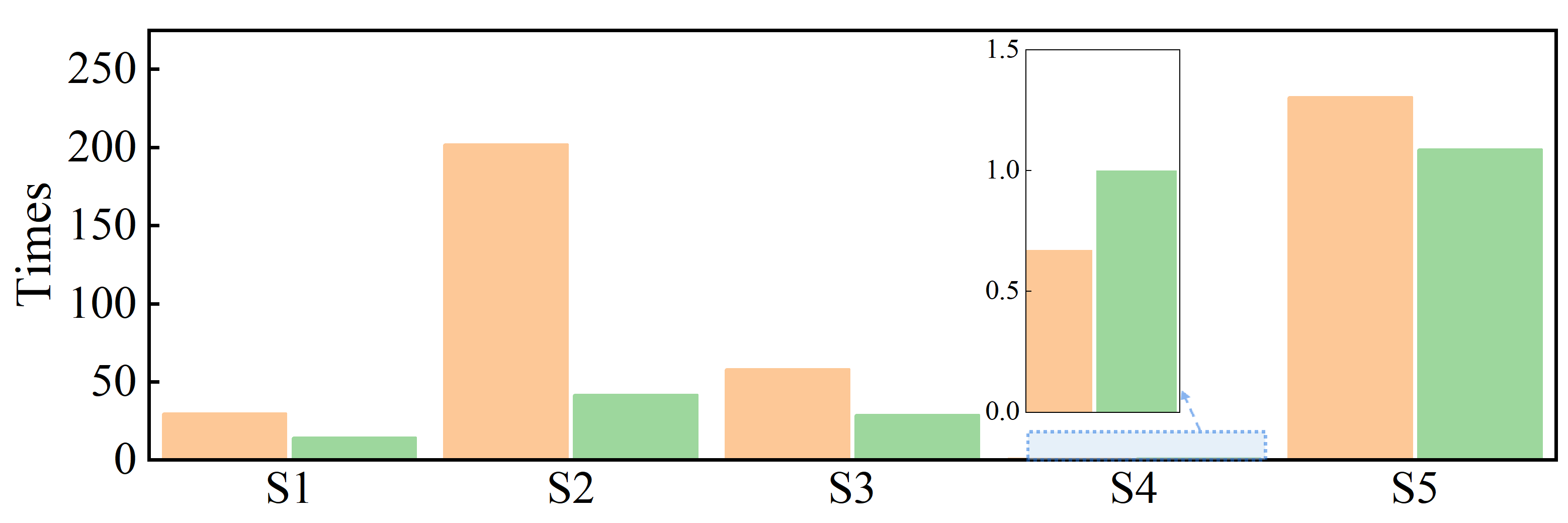}}
	\hspace{0.2cm}
	\includegraphics[width=8.8cm]{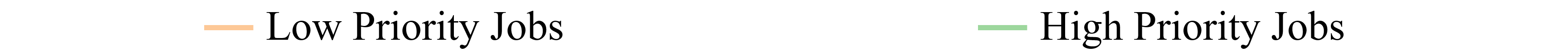}
	\par \vspace{-3.5mm}
	\subfloat[\footnotesize Impact of removing continuous monitoring mechanism]{\label{13-womon}\includegraphics[width=8.8cm, height=2.7cm]{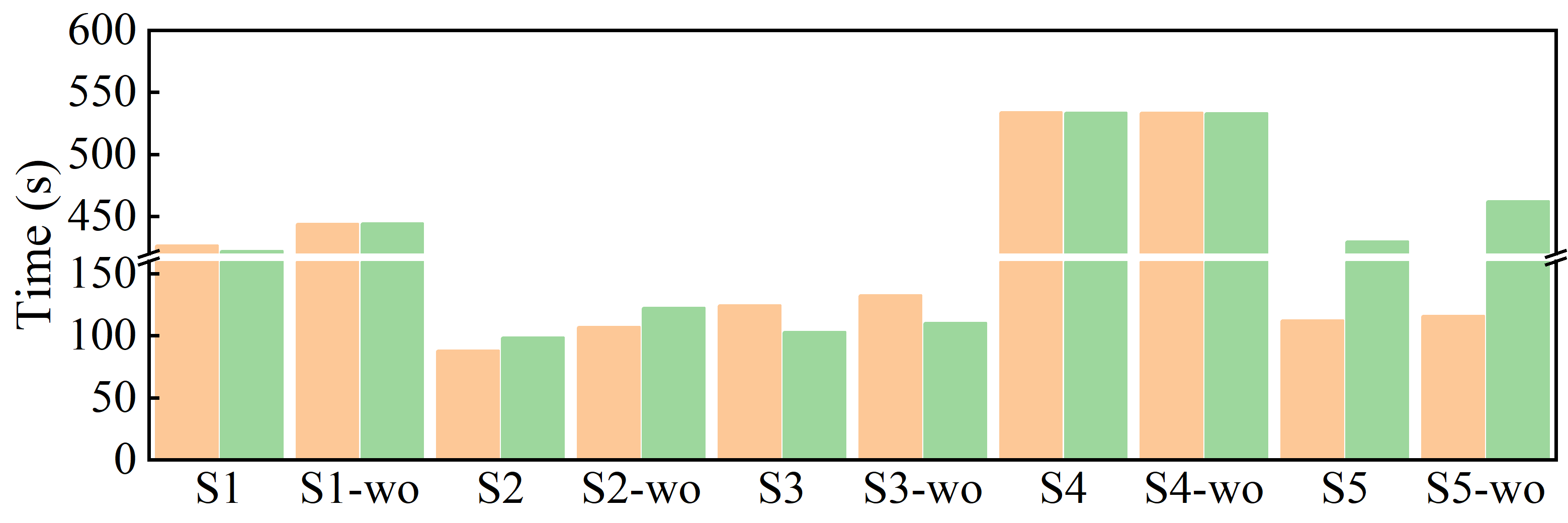}}	
	\caption{Ablation studies.}
	\label{13-wo}
\end{figure}

$\bullet$ \textbf{Removing the third stage optimization mechanism.} 
We adjust the start point of the low priority pods' communication phase to align with the end point of the previous pods' communication phase.
This approach differs from the intermediate rotation angles identified by the third stage optimization mechanism. The experimental results are shown in Fig. \ref{13-wothird}. Except for snapshot 4, where communication contention is virtually non-existent, the number of readjustment operations significantly increases compared with the intermediate rotation. This is because there are no longer cushion slots between the communication phases of pods, and even minor drift can trigger communication contention, prompting the monitoring mechanism to execute readjustment operations. The rise in readjustment operations slows down the training process of jobs and further reduces average bandwidth utilization. 
A comparison of the metrics between compact and intermediate rotation for each snapshot is provided in Table \ref{t7}.

\begin{table}[!htbp]
	\centering
	\setlength{\abovecaptionskip}{0pt}
	\setlength{\belowcaptionskip}{10pt}
	\setlength{\tabcolsep}{7.7pt} 
	\caption{Comparison of Metrics Between Metronome Configurations Without and With Third-Stage Optimization\label{t7}}
	\resizebox{8.8cm}{!}{
		
		\begin{tabular}{>{\centering\arraybackslash}p{1cm}>{\centering\arraybackslash}p{2.1cm}>{\centering\arraybackslash}p{2.1cm}>{\centering\arraybackslash}p{1.55cm}} 		
			\toprule
			\textbf{Snapshot}&\textbf{Time (Low Priority Jobs,\%)} & \textbf{Time (High Priority Jobs,\%)} &\textbf{Avg. BW Util. (\%) } \\
			\midrule
			S1 & 1.17{\scriptsize ${\uparrow}$}&0.58{\scriptsize ${\uparrow}$}&1.12{\scriptsize ${\downarrow}$} \\			
			S2 & 1.77{\scriptsize ${\uparrow}$}&3.78{\scriptsize ${\uparrow}$}&2.57{\scriptsize ${\downarrow}$}  \\			
			S3 & 1.11{\scriptsize ${\uparrow}$}&1.03{\scriptsize ${\uparrow}$}&1.08{\scriptsize ${\downarrow}$} \\			
			S4 & 0.37{\scriptsize ${\uparrow}$}&0.42{\scriptsize ${\uparrow}$}&0.22{\scriptsize ${\downarrow}$} \\			
			S5 & 0.94{\scriptsize ${\uparrow}$}&0.57{\scriptsize ${\uparrow}$}&0.97{\scriptsize ${\downarrow}$}  \\			
			\bottomrule
		\end{tabular}
	}
\end{table}

$\bullet$ \textbf{Removing the continuous monitoring mechanism.} 
After turning off the continuous monitoring mechanism (denoted as `wo' in the Fig.  \ref{13-womon}), the average time per 1,000 iterations increased, except for the snapshot 4. 
In contrast, the benefit of this mechanism is clearly demonstrated in snapshot 2, where Metronome achieves acceleration of 19.45\% and 17.74\% for high and low priority jobs, respectively.
In snapshot 4, due to the strong compatibility between the two jobs, there is basically no need for readjustment operations, making the removal of this mechanism have almost no effect. A similar situation is reflected in the average bandwidth utilization, as shown in Table \ref{t8}.

The ablation studies confirm the necessity of both the third stage optimization and the continuous monitoring mechanism, as they are integral to Metronome's design and each provides significant performance improvements.

\begin{table}[!htbp]
	\centering
	\setlength{\abovecaptionskip}{0pt}
	\setlength{\belowcaptionskip}{10pt}
	\setlength{\tabcolsep}{7.7pt} 
	\caption{	Comparison of average bandwidth utilization  Between Metronome Configurations Without and With continuous monitoring mechanism\label{t8}}
	\resizebox{8.8cm}{!}{
		\begin{tabular}{c>{\centering\arraybackslash}p{0.7cm}>{\centering\arraybackslash}p{0.7cm}>{\centering\arraybackslash}p{0.7cm}>{\centering\arraybackslash}p{0.7cm}>{\centering\arraybackslash}p{0.7cm}} 			
			\toprule
			&\textbf{S1}& \textbf{S2}&\textbf{S3}& \textbf{S4}&\textbf{S5}\\
			\midrule
			avg. BW util. (\%)  & 4.61${\downarrow}$&18.81${\downarrow}$&6.71${\downarrow}$&0.17${\downarrow}$&5.07${\downarrow}$ \\			
			\bottomrule
		\end{tabular}
	}
\end{table}		

\subsection{Threshold Selection}
\label{4-4}
We now conduct experiments on the threshold selection discussed in  §\ref{3-2} and \ref{3-3}.

$\bullet$ \textbf{\texttt{$O_T$} \& \texttt{$A_T$}.}  
For the continuous monitoring mechanism, we set \texttt{$O_T$} to 3 and 5, respectively, and \texttt{$A_T$} is randomly selected as 105\%, 110\%, and 115\%.  
Changes in the threshold directly affect the frequency of triggering readjustment operations, with the primary and most immediate impact falling on the training process of low priority jobs.
Therefore, we focus our experimental investigation on this aspect.

\begin{figure}[htbp]
	\centering
	\captionsetup[subfloat]{
		font=footnotesize,      
		labelfont=bf,      
		textfont=normalfont,
		justification=justified
	}
	\centering
	\subfloat[Flame chart for selecting the threshold value of the monitoring mechanism]{\label{14-flame}\includegraphics[width=8.8cm]{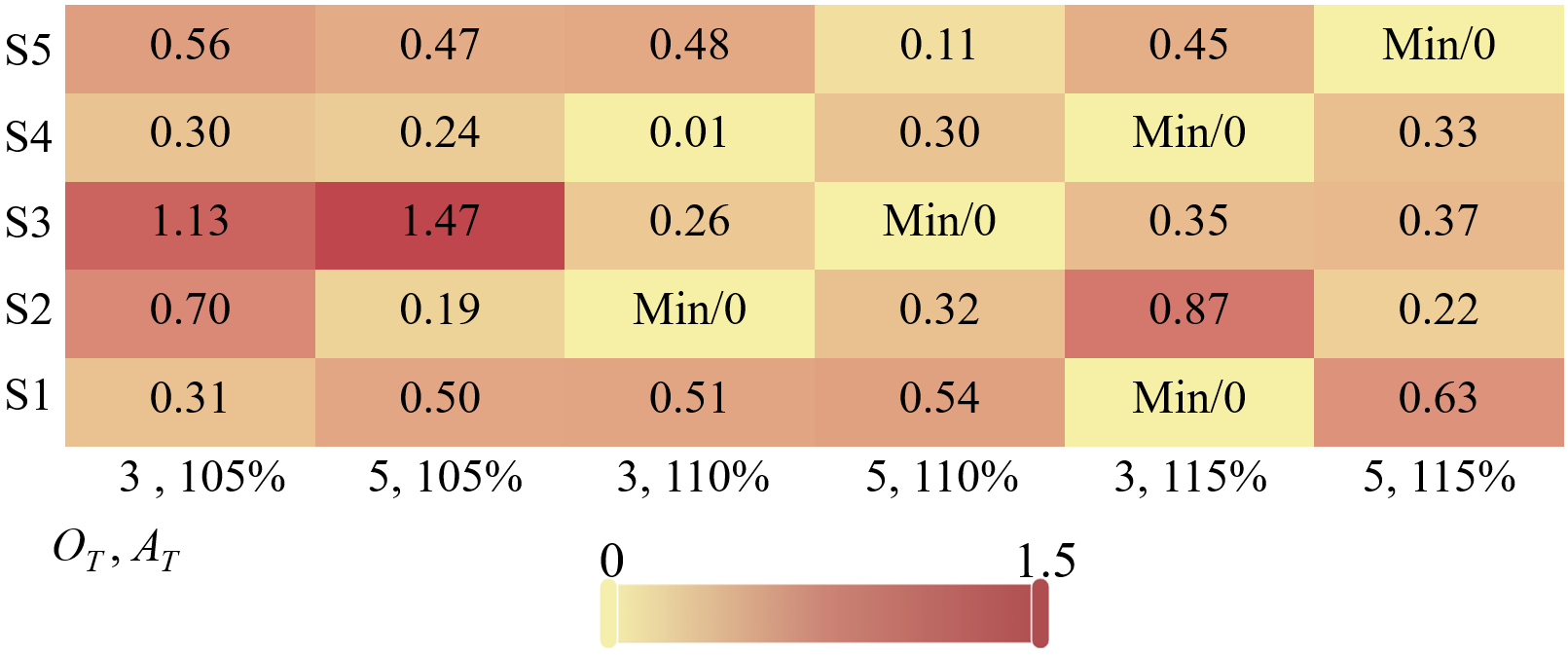}}
	\hspace{0.2cm}
	\subfloat[Impact of threshold change on the number of readjustment operations ]{\label{14-renumber}\includegraphics[width=8.8cm, height=2.7cm]{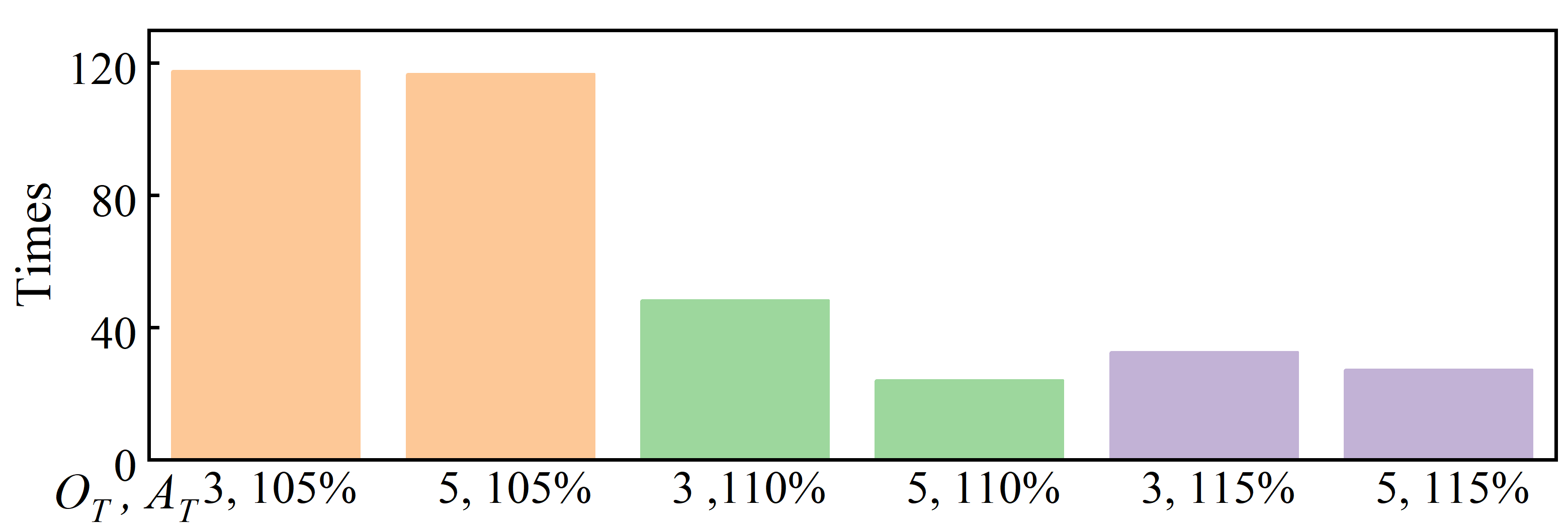}}	
	\caption{Impact of changes in monitoring mechanism thresholds.}
	\label{14-threshold}
\end{figure}

Fig.  \ref{14-flame} shows the increase in the average time per 1,000 iterations of low priority jobs compared with the best-performing threshold selection in each snapshot. Except for the case where \texttt{$A_T$} is 105\% in snapshot 3, the average time fluctuations do not exceed 1\% under other threshold selections in the flame chart. This suggests that the readjustment threshold selection of Metronome is robust within a certain range and does not require fine-tuning. 
Conversely, Fig. \ref{14-renumber} shows a significant increase in the number of readjustments at \texttt{$A_T$} = 105\%  in snapshot 3. This suggests that 105\% is a relatively low threshold value that readily triggers readjustment operations.  The same conclusion applies to the high priority job training speed and average bandwidth utilization.

$\bullet$ \textbf{\texttt{$G_T$} \& \texttt{$E_T$}.}
In snapshot 3, after we applied the period doubling operation to VGG19, the period of WideResNet is 35 ms shorter than that of VGG19.
We continuously perform idle time injection operations on the low priority job  (WideResNet)  to gradually narrow the gap between the two jobs. The LCM period is calculated as the average of WideResNet's period after injection and VGG19's period after doubling.

\begin{figure}[t]
	\centering
	\begin{minipage}[t]{0.48\columnwidth} 
		\centering
		\includegraphics[height=0.85\textwidth, width=\linewidth]{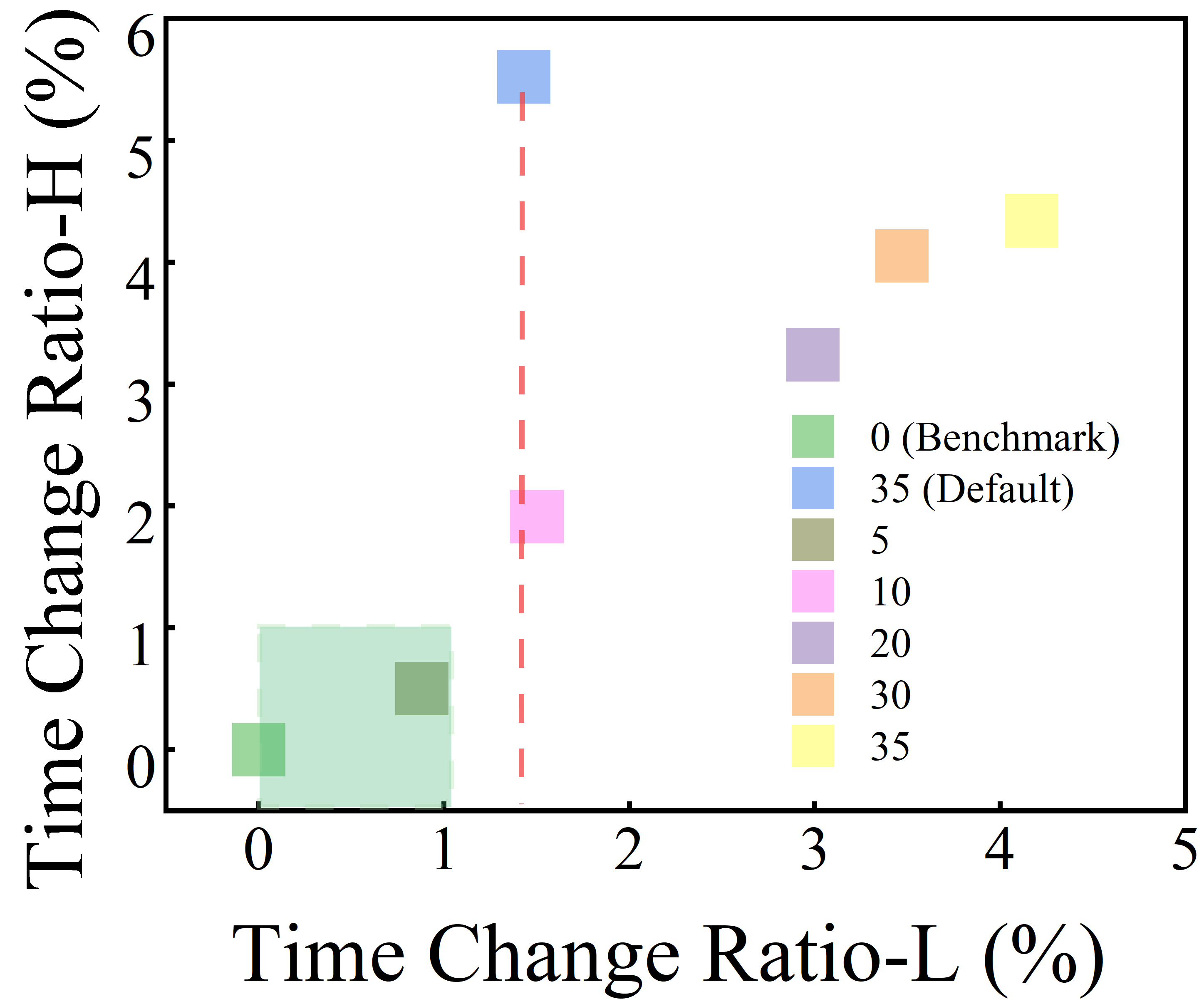} 
		\caption{ Impact of threshold selection for periodic approximation.} 
		\label{15-approximation}
	\end{minipage}
	\hfill
	\begin{minipage}[t]{0.48\columnwidth}
		\centering
		\includegraphics[height=0.85\textwidth, width=\linewidth]{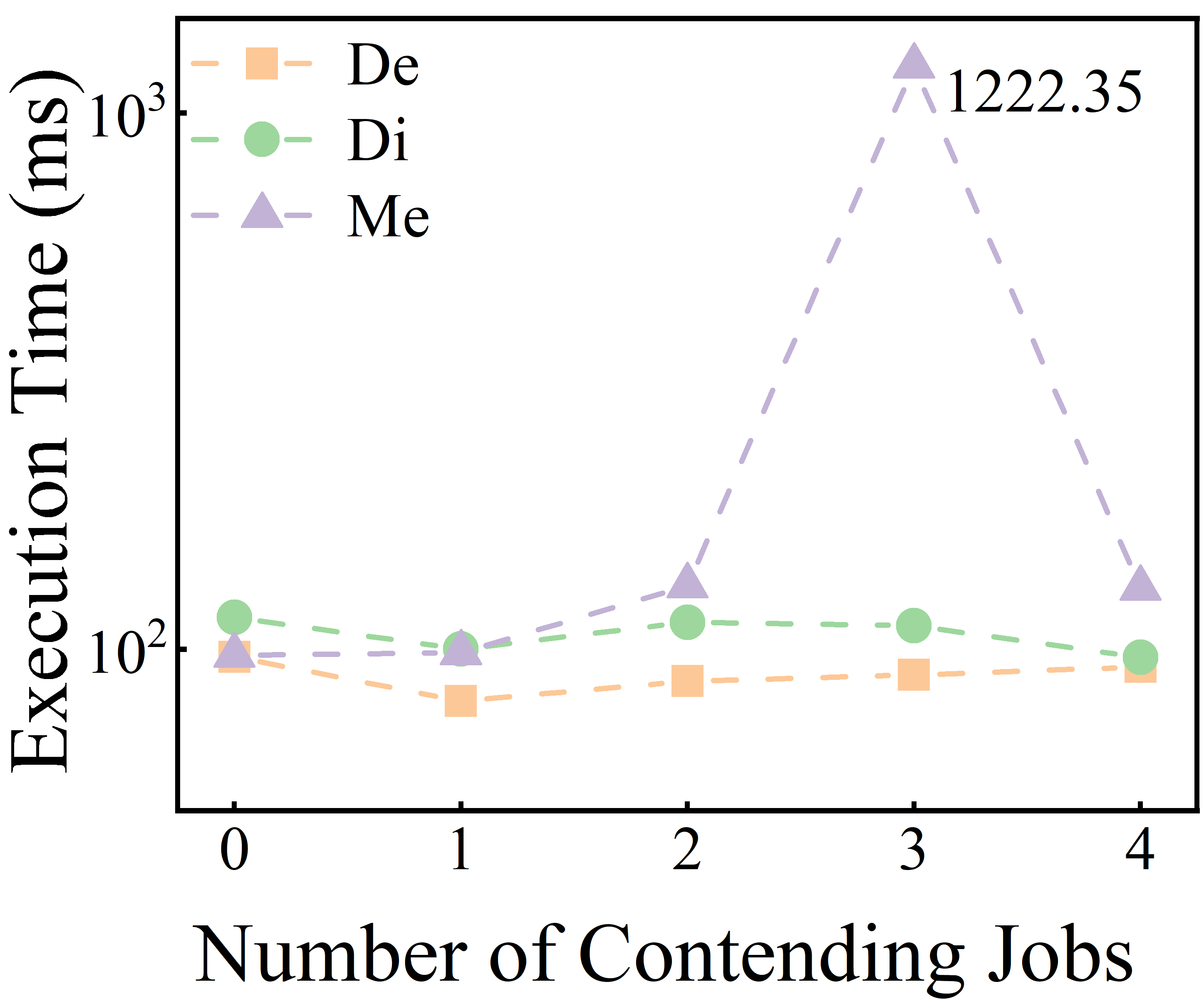}
		\caption{Execution time of schedulers.} 
		\label{16-execution}
	\end{minipage}
\end{figure}

The experimental results are shown in Fig.  \ref{15-approximation}. The legend indicates the period differences between the two jobs, including 35 ms, 30 ms, 20 ms, 10 ms, 5 ms, and 0 ms (serving as the benchmark). 
Additionally, the case of 35 ms without period injection or offset adjustment, termed the default scenario, is included for comparison.
The vertical and horizontal axes represent the performance change, relative to the benchmark, for the high priority and low priority jobs, respectively. This change is measured by the ratio of the average time per 1,000 iterations.
When the injected idle time reaches 30 ms (resulting in a period difference of 5 ms), the deviation in performance metrics remains within 1\% compared with the ideal scenario. A similar trend is observed in average bandwidth utilization. 
When the job period difference exceeds 5 ms, the difference in period exacerbates communication drift, leading to more frequent readjustment operations. Consequently, the performance metrics of low priority jobs become worse than those in the default scenario (where no period injection or offset adjustment is applied). 
However, high priority jobs do not experience any pausing, which maintains their performance advantage compared with the default results.

Therefore, we set \texttt{$G_T$} to 5 ms. Furthermore, to avoid introducing excessive overhead that would slow down the training of low priority jobs, \texttt{$E_T$} is configured to be 10\% of the job's period.

\subsection{Execution Time Analysis of Scheduler Mechanisms}
\label{4-5}
To measure the scheduler's execution time, we record the time between pod submission and the assignment of an IP address to it.
The reported value is the average obtained by repeating the experiment 100 times.

As illustrated in Fig. \ref{16-execution}, neither the Default scheduler nor Diktyo takes two-dimensional bandwidth resources into consideration.
Consequently, their scheduling complexity does not increase as the number of contending jobs sharing links grows, contrasting with Metronome.
However, since Metronome does not seek the optimal offset scheme during the scheduling phase, the longest execution time observed in the experiment is less than 1500 ms. When four jobs are already sharing links, candidate nodes may be filtered out due to GPU resource limitations, leading to a reduction in execution time.

For the stop-and-wait controller, the time required to compute the optimal offset scheme does not exceed 5 seconds when tested under the same resource quota as the scheduler. \footnote{It is worth noting that the stop-and-wait controller can be deployed on bare-metal machines with only minor code modifications and no performance degradation, although the scheduler  requires further adaptation to support platforms such as YARN \cite{yarn}.}
The extra computation time compared with the improvement in metrics is reasonable, and the computation time of the controller can be covered by the job startup time.
\section{Discussion and Limitations}
\label{5}
$\bullet$ \textbf{Sharing the network with other workloads.}
Modern large training clusters often consist of multiple networks 
\cite{alibabaHPN}, \cite{mudigere2022software}, and the computing network almost exclusively carries the traffic generated by training. On the other hand, Metronome does not allocate undeclared bandwidth resources, which can be utilized for other workloads. Even if unregulated traffic causes network congestion, Metronome can avoid congested links through dynamic latency awareness.

$\bullet$ \textbf{Improvement of the mechanism and performance verification in large-scale clusters.}
Metronome currently does not involve a rescheduling mechanism, which may struggle with scenarios where deployed jobs transition from compatible to incompatible. 
Additionally, as the number of jobs sharing links increases, Metronome may not be able to interleave communication phases across all links. 
As currently implemented, Metronome utilizes a node scoring approach that prioritizes the scheduling of jobs onto nodes experiencing lower levels of communication contention.
Additionally, a potential complementary strategy involves the intentional injection of idle time slots into low priority jobs, which reduces the communication duty cycle of such jobs and can thereby enhance job compatibility.
Finally, Metronome reduces execution time through an offline recalculation mechanism, but further algorithmic optimizations remain possible.
We leave this for future work.

\section{Related Work}
\label{6}
Our work is built on several lines of related research.

$\bullet$ \textbf{Parallel algorithms for distributed training.}
Numerous mechanisms and techniques are designed to accelerate distributed training.
Several pipeline parallelism algorithms, such as  MegaScale \cite{jiang2024megascale}, GPipe \cite{GPipe}, PiPeDream \cite{PiPeDream},  and DualPipe \cite{DualPipe} have been proposed to optimize communication between workers, with the goal of reducing bubbles in pipeline strategies. 
In network-constrained environments (e.g., edge clouds and geo-distributed data centers), approaches such as DiLoCo \cite{douillard2023diloco} and AdaSFL \cite{adasfl} reduce communication overhead by lowering update frequency, while NETSTORM \cite{netstorm} avoids the emergence of hotspots through transmission partitioning and multi-path transmission mechanisms. 
However, these algorithms focus on improving the training efficiency of individual jobs. In contrast, Metronome is orthogonal to these approaches because Metronome focuses on optimizing the operation between multiple jobs.

$\bullet$ \textbf{Network-aware approaches.}
In K8s, bandwidth can be defined via extended resources. Bandwidth awareness can be categorized into two types: capacity and dependency. The former refers to the  bandwidth capacity of links, while the latter represents the bandwidth requirements between microservices. The schedulers proposed in \cite{lai2023delay}, \cite{toka2021ultra}, use the  bandwidth capacity as the metric for node filtering, i.e., the sum of the allocated bandwidth resources cannot exceed the  bandwidth capacity of the link, which differs from the two-dimensional bandwidth resources we propose.  
Another approach \cite{lee2023optimal} uses bandwidth resources for scoring nodes to prioritize those with higher network performance.
Crucially, these capacity awareness schemes do not account for the bandwidth dependencies between microservices. 
NetMARKS \cite{wojciechowski2021netmarks} achieves compact deployment of communication-intensive pods through scheduling to facilitate intra-node communication between microservices.
Dependency awareness can be achieved by utilizing bandwidth capacity information.
Polaris \cite{pusztai2022polaris} avoids co-locating interdependent pods on node pairs with insufficient allocatable bandwidth on the link between them.
In K8s network-aware scheduling, latency is another indicator that needs to be considered. 
Diktyo \cite{diktyo} proposes a latency-aware  scheduler that favors the node with the lowest aggregated network cost to guarantee service performance.
The studies in \cite{lai2023delay}, \cite{toka2021ultra}, \cite{lee2023optimal}, and \cite{pusztai2022polaris} have all investigated scheduling using bandwidth and latency constraints.
Similarly, Metronome considers both bandwidth and latency metrics to guarantee the bandwidth requirements of the jobs while  achieving compact deployments.
A key distinction is that Metronome is specifically designed for periodic traffic jobs. By contrast, the aforementioned studies do not incorporate the interleaving of communication phases.

$\bullet$ \textbf{Interleaving scheduling.}
Gandiva \cite{Gandiva} implements GPU time-slicing to support multi-job parallelism through a job suspend-resume mechanism. To minimize overhead, it performs switching operations when GPU memory usage is at its lowest.
Muri \cite{Muri} further considers multi-dimensional resources interleaving during scheduling, but Muri's approach is only suitable for jobs that share the same set of resources. Cassini \cite{Cassini} broke this limitation by constructing an affinity graph that finds a series of time-shifts to adjust job communication phases, thereby  interleaving phase on the link between switches. However, Cassini does not consider latency and priorities indicators during scheduling and assumes that jobs do not share GPUs. Consequently, it is not well suited to the complex environment of multi-tenant clusters.
Crux \cite{cao2024crux} assigns different priorities based on traffic patterns and mitigates communication contention by routing high priority traffic through dedicated links. 
Since Crux does not consider the compatibility of jobs, the performance of low priority jobs can be significantly degraded due to delayed communication.

\section{Conclusion}
\label{7}
This paper proposes a scheduling mechanism named Metronome, designed for periodic traffic jobs. Specifically, it tackles the problem of inter-job communication contention and optimizes  cluster resource utilization in  cloud native networks.
Metronome interleaves communication phases and reserves cushion slots to mitigate contention.
It also avoids congested links during scheduling while striving to ensure compact deployment of dependent components.  
Finally, Metronome guarantees service performance by continuously monitoring deployed jobs with a stop-and-wait controller, which adjusts low priority jobs to maintain this interleaving.
Our evaluation across multiple scenarios demonstrates that Metronome reduces job completion time by up to 19.50\% while improving the average bandwidth utilization by up to 23.20\%.

\section*{References} 
\renewcommand{\refname}{}    
\footnotesize 
\bibliographystyle{IEEEtran}
\bibliography{myreference.bib}
\clearpage
\begin{IEEEbiographynophoto}{Hao Jiang}
 (Student Member, IEEE) received the B.S. degree in electronic and information engineering from Sichuan University, Chengdu, China, in 2022. He is currently pursuing the Ph.D. degree with the School of Electronic Information and Electrical Engineering, Shanghai Jiao Tong University, Shanghai, China. He is
also an intern with Pengcheng Laboratory, Shenzhen. His research interests include cloud native networks, distributed system scheduling, and distributed training.
\end{IEEEbiographynophoto}
\begin{IEEEbiographynophoto}{Meng Qin}
	(Member, IEEE) received the B.S. degree in communication engineering from the Taiyuan University of Technology, China, in 2012, and the M.S. and Ph.D. degrees in information and communication systems from Xidian University, Xi'an, China, in 2015 and 2018, respectively. He worked as a Postdoctoral Fellow with School of Electronic and Computer Engineering, Peking University and Pengcheng Laboratory, China. He is currently an assistant researcher with Department of Strategic and Advanced Interdisciplinary Research, Pengcheng Laboratory, China. His research interests include cybertwin-based cloud native networks, intelligent management and control of 6G networks, AI-aided self-organized wireless networks, edge intelligence in wireless networks.
\end{IEEEbiographynophoto}
\begin{IEEEbiographynophoto}{Ruijie Kuai}
	(Student Member, IEEE) received the B.S. degree in electronic information science and technology from Beijing University of Posts and Telecommunications, Beijing, China, in 2023. She is currently pursuing the Ph.D. degree with the School of Information and Electronic Engineering, Shanghai Jiao Tong University, Shanghai, China. She is also an intern with Pengcheng Laboratory, Shenzhen. Her research interests include cloud native networks and network-aware elastic scheduling.
\end{IEEEbiographynophoto}
\begin{IEEEbiographynophoto}{Dandan Liang}
	(Member, IEEE) received her Ph.D. degree in wireless communications from the University of Southampton, UK, in 2013. Upon completion of her PhD, she conducted research as a Research Fellow at the University of Surrey, UK. In 2017 she joined the Research Centre of Huawei Technologies in Shenzhen, China, working as Senior Engineer of wireless communications and standardization of Wi-Fi. She is currently an associated professor at Pengcheng Laboratory. She has published 20 research papers in IEEE journals and conferences, applied 68 patents and has been recognized as a CNCF Kubestronaut in 2024. Her research interests include cloud native networks and cross-layer system design.
\end{IEEEbiographynophoto}
\begin{IEEEbiographynophoto}{Yue Gao}
	(Fellow, IEEE)   received the PhD degree from the Queen Mary University of London (QMUL), U.K., in 2007. He was a lecturer, a senior lecturer, a reader, and the chair professor with QMUL and the University of Surrey, respectively. He is currently a chair professor with the School of Computer Science and the dean of the Institute of Space Internet, Fudan University, China. He has authored or coauthored 200 peer-reviewed journal and conference papers. His research interests include sparse signal processing and smart antennas and cognitive networks for mobile and satellite systems. He was a co-recipient of the EU Horizon Prize Award on
	Collaborative Spectrum Sharing in 2016 and an Engineering and Physical Sciences Research Council Fellow in 2017. He is a member of the Board of Governors and a Distinguished Speaker of IEEE Vehicular Technology Society (VTS), the chair of IEEE ComSoc Wireless Communication Technical Committee, and the past chair of IEEE ComSoc Technical Committee on Cognitive Networks. He was an editor of several IEEE Transactions and Journals, the symposia chair, and the track chair. He has other roles in the organizing committee of several IEEE ComSoC, VTS, and other conferences.
\end{IEEEbiographynophoto}
\end{document}